\documentclass[12pt]{article}
\usepackage[margin=1in]{geometry}
\usepackage{amsmath, amssymb, amsthm}
\usepackage{authblk} 
\usepackage{cuted}
\usepackage{graphicx}
\usepackage{booktabs}
\usepackage{tabularx}
\usepackage{hyperref}
\hypersetup{
	colorlinks = true,
	linkcolor  = blue!70!black,
	citecolor  = blue!70!black,
	urlcolor   = blue!70!black
}
\usepackage[authoryear,round]{natbib}
\bibpunct{(}{)}{;}{a}{}{,}
\usepackage{enumitem}
\usepackage{subcaption}
\usepackage{xcolor}
\usepackage{times}
\usepackage[section]{placeins}

\usepackage{algorithm}
\usepackage{algpseudocode}
\usepackage{float}
\usepackage{tikz}
\usetikzlibrary{calc,decorations.pathreplacing}
\usetikzlibrary{arrows.meta,shadows.blur,calc}
\makeatletter
\newenvironment{breakablealgorithm}
{\begin{center}
		\refstepcounter{algorithm}
		\renewcommand{\caption}[2][\relax]{
			\vspace{4pt}
			{\raggedright\textbf{Algorithm~\thealgorithm} ##2\par}
			\vspace{4pt}\hrule\vspace{4pt}
		}
	}
	{\vspace{4pt}\hrule\relax
\end{center}}
\makeatother

\usepackage{nomencl}
\usepackage{multicol}
\usepackage{multirow}
\makenomenclature

\renewcommand{\nompreamble}{\begin{multicols}{2}}
	\renewcommand{\nompostamble}{\end{multicols}}
\usepackage{pgffor}
\usepackage{setspace}

\title{Pooling interval-monitored step-stress accelerated life tests for reliability prediction: A failure-rate-based meta-analysis for the proportional hazards family}
\author[1]{Pranoy Palit}
\author[1]{Ayan Pal}
\author[2]{Kiran Prajapat\thanks{Corresponding author. Email: kiranprajapat92@gmail.com, kiran.prajapat@newcastle.ac.uk (Kiran Prajapat)}}

\affil[1]{\small Department of Statistics, The University of Burdwan, Burdwan, West Bengal, India}
\affil[2]{\small School of Mathematics, Statistics and Physics, Newcastle University, Newcastle upon Tyne, UK}

\date{}
\begin{document}
	
	\maketitle
	\begin{abstract}
		
		Reliability qualification often relies on several step-stress accelerated life tests conducted in separate campaigns, laboratories, or test rigs, each with its own stress profile, sample size, and inspection schedule, and with units inspected only periodically so that only failure counts between inspections are recorded. 
		Existing meta-analysis methods combine several step-stress experiments but require exactly observed failure times. Methods for interval-inspected step-stress data, on the other hand, have been developed only for a single experiment. This paper develops a failure-rate-based meta-analysis framework for multiple interval-inspected step-stress experiments within the proportional hazards family, linking the experiments through a common underlying lifetime model and a log-linear stress–life relationship. Maximum likelihood estimators and confidence intervals are derived for the model parameters and for engineering quantities of interest at normal operating conditions, including mission reliability, mean time to failure, and lifetime quantiles such as the 0.10 quantile (B10 life). For the LED data, pooling the two tests reduced the RMSE of the estimated reliability at 5 hours by about 20\% and that of the 0.10 quantile by about 66\%, compared with the better single test. A misspecification study shows that the median is robust to the choice of lifetime distribution (relative bias below 6\%), whereas tail quantiles can have a relative bias of up to 34\% that does not vanish as the sample size increases, which is critical for warranty decisions. The methodology is illustrated with simulated data and with two step-stress tests of light-emitting diodes.
		
	\end{abstract}   
	\newpage
	\newcommand{\bea}{\begin{eqnarray}}
		\newcommand{\eea}{\end{eqnarray}}
	\newcommand{\nn}{\nonumber}
	\newcommand{\bee}{\begin{eqnarray*}}
		\newcommand{\eee}{\end{eqnarray*}}
	\newcommand{\lb}{\label}
	\newcommand{\nii}{\noindent}
	\newcommand{\ii}{\indent}
	
	\newtheorem{theorem}{Theorem}[section]
	\newtheorem{corollary}{Corollary}[theorem]
	\newtheorem{proposition}{Proposition}[section]
	\newtheorem{lemma}{Lemma}[section]
	\newtheorem{remark}{Remark}[section]
	
	\nomenclature[A]{AE}{Average estimate}
	\nomenclature[A]{AIC}{Akaike information criterion}
	\nomenclature[A]{AL}{Average length}
	\nomenclature[A]{ALT}{Accelerated life testing}
	\nomenclature[A]{BIC}{Bayesian information criterion}
	\nomenclature[A]{CEM}{Cumulative exposure model}
	\nomenclature[A]{CHF}{Cumulative hazard function}
	\nomenclature[A]{CI}{Confidence interval}
	\nomenclature[A]{CP}{Coverage probability}
	\nomenclature[A]{CSALT}{Constant-stress accelerated life testing}
	\nomenclature[A]{FRM}{Failure-rate model}
	\nomenclature[A]{KHM}{Khamis-Higgins model}
	\nomenclature[A]{LED}{Light-emitting diode}
	\nomenclature[A]{MLE}{Maximum likelihood estimate}
	\nomenclature[A]{MLL}{Maximum log-likelihood}
	\nomenclature[A]{MSE}{Mean squared error}
	\nomenclature[A]{MTTF}{Mean time to failure}
	\nomenclature[A]{NOC}{Normal operating condition}
	\nomenclature[A]{PH}{Proportional hazards}
	\nomenclature[A]{RB}{Relative bias}
	\nomenclature[A]{RMSE}{Root mean squared error}
	\nomenclature[A]{RV}{Relative variability}
	\nomenclature[A]{SSALT}{Step-stress accelerated life testing}
	\nomenclature[A]{TFRM}{Tampered failure-rate model}
	\nomenclature[A]{TRV}{Tampered random variable model}
	\nomenclature[N]{$h_0(t;\alpha)$}{Baseline hazard function}
	\nomenclature[N]{$H_0(t;\alpha)$}{Baseline cumulative hazard function}
	\nomenclature[N]{$H_k(t)$}{cumulative hazard function for experiment $k$}
	\nomenclature[N]{$I(\boldsymbol{\theta})$}{Fisher information matrix}
	\nomenclature[N]{$I_k$}{Number of inspection intervals in experiment $k$}
	\nomenclature[N]{$K$}{Number of independent SSALT experiments}
	\nomenclature[N]{$k$}{Index of each experiment}
	\nomenclature[N]{$N$}{Total number of test units}
	\nomenclature[N]{$N_k$}{Number of test units in experiment $k$}
	\nomenclature[N]{$n_{k,i}$}{Number of failures in interval $i$ of experiment $k$}
	\nomenclature[N]{$p_k$}{Multinomial probability vector for $k$-th experiment}
	\nomenclature[N]{$q$}{Quantile level}
	\nomenclature[N]{$R_0(t;\alpha)$}{Baseline reliability function}
	\nomenclature[N]{$R_{\text{NOC}}(t)$}{Reliability function at time $t$ under NOC}
	\nomenclature[N]{$R_k(t)$}{Reliability function for experiment $k$}
	\nomenclature[N]{$s_{\text{NOC}}$}{Stress level at NOC}
	\nomenclature[N]{$s_H$}{Highest stress level considered}
	\nomenclature[N]{$S_k$}{Stress path of experiment $k$}
	\nomenclature[N]{$t$}{Time}
	\nomenclature[N]{$t_M$}{Mission time}
	\nomenclature[N]{$t_{q,\text{NOC}} $}{$q$-th lifetime quantile at NOC}
	\nomenclature[N]{$\tau_{k,i}$}{$i$-th inspection time in experiment $k$}
	\nomenclature[N]{$x_{k,i}$}{Standardized stress level in interval $i$ of experiment $k$}
	\nomenclature[N]{$\boldsymbol{\theta}$}{Vector of model parameters}
	\nomenclature[N]{$\widehat{\boldsymbol{\theta}}$}{MLE of $\boldsymbol{\theta}$}
	\nomenclature[N]{$\hat{t}^{(\mathrm{mis})}_{q,l}$}{Estimated $q$-th quantile under the misspecified model}
	\nomenclature[N]{$\hat{t}^{(\mathrm{true})}_{q,l}$}{Estimated $q$-th quantile under the true model}
	\nomenclature[N]{$\bar{\widehat{t}}^{(\mathrm{mis})}_{q}$}{Average estimated $q$-th quantile under the misspecified model}
	\nomenclature[N]{$\bar{\widehat{t}}^{(\mathrm{true})}_{q}$}{Average estimated $q$-th quantile under the true model}
	\printnomenclature[2cm]
	
	\section{Introduction}
	\label{sec:intro}
	
	Modern products such as light-emitting diodes (LEDs), power electronics, and battery cells are designed to operate for years under normal operating conditions (NOC), so that very few failures can be observed within a practical testing period. Accelerated life testing (ALT) addresses this by exposing units to elevated stress, such as temperature, voltage, or load, and extrapolating the resulting failure information to NOC \cite{nelson1978,meeker1998,bagdonavicius2001accelerated}. In constant-stress ALT, a separate group of units is required for each stress level, and the groups tested at lower levels may yield few failures within the available time. Step-stress ALT (SSALT) avoids this by increasing the stress on the same units at pre-specified times, and therefore yields failures within a shorter test duration and with fewer units in total \cite{Nelson1980,kateri2008}. 
	
	Two features of industrial SSALT practice are, however, rarely accommodated together by existing methods. First, units are frequently inspected only periodically rather than monitored continuously; for example, the light output of LEDs is measured at fixed intervals, battery capacity is checked after a given number of cycles, and semiconductor devices in high-temperature operating life tests are tested electrically only at pre-specified inspection points. The data then consist only of the numbers of failures between successive inspections. Second, reliability qualification seldom rests on a single test: several SSALT experiments on the same product are often conducted in different campaigns, laboratories, or test rigs, each with its own stress path, sample size, and inspection schedule. If the data from different sources were obtained under an identical step-stress plan, they could be merged into a single sample, and any unobserved differences between the sources could be accommodated through a finite mixture model, as in the heterogeneous SSALT models of \cite{lu2025simple,palit2026finite}. When the experiments differ in their stress levels, stress-change times, or inspection schedules, however, such merging is not possible, and the experiments must be combined in a way that respects their individual designs. Analyzing these experiments separately wastes information and can lead to imprecise estimates of the quantities that drive engineering decisions, such as lower lifetime quantiles (e.g., the 0.10 quantile, known in reliability engineering as the B10 life) used for warranty periods, the reliability at a mission time used for qualification, and the mean time to failure (MTTF) used in maintenance and spare-parts planning.
	
	Statistical inference for SSALT requires a model linking the lifetime distributions at successive stress levels. The cumulative exposure model (CEM) \cite{sedyakin1966} and the tampered failure-rate model (TFRM) \cite{bhattacharyya1989} are the most widely used; these and other models are reviewed in \cite{Kateri2017,kundu2017}. In this paper, we adopt the failure-rate-based model (FRM) of \cite{kateri2015}, a generalization of the TFRM that specifies the hazard function directly at each stress level. The FRM is a natural choice when lifetimes belong to the proportional hazards (PH) family, since stress then acts multiplicatively on the hazard, and it leads to a tractable likelihood for interval-monitored data. For exponential lifetimes, the FRM and the CEM coincide.
	
	Meta-analysis of step-stress experiments was introduced by \cite{katerimeta2009} for exponential lifetimes under Type-II censoring and extended to experiments with time constraints by \cite{katerimeta2010}. \cite{bedbur2015} generalized it to arbitrary baseline distributions under repeated Type-II censoring, \cite{samanta2021} considered Weibull lifetimes under the generalized  \cite{khamis1996} model, and \cite{prajapat2024} studied exponential lifetimes under Type-I hybrid censoring. More recently, \cite{tung2026} proposed unified accelerated life testing for one-shot devices with Weibull lifetimes, which combines the principles of constant-stress and step-stress testing by subjecting several test groups to different stress-change patterns, and showed that analyzing such groups jointly yields more accurate reliability estimates. Apart from the one-shot setting, however, these methods require exactly observed failure times. Interval-monitored SSALT, in contrast to the continuously monitored setting 
	, has been studied in \cite{lee2012interval,bobotas2015interval,balakrishnan2026a}, but only for a single experiment. To the best of our knowledge, no method is available for combining several step-stress experiments when failures are observed only through periodic inspections.
	
	This gap is illustrated by the LED data analyzed in Section~\ref{sec:data_analysis} \cite{shyur1999,samanta2024led}. Two tests were run, one starting at 7V and the other at 7.5V, and both were later stepped up to 8V. The second test never operated at the use condition of 7V, so on its own it says little about the LEDs in use. Combined with the first test, however, it can contribute substantially to the estimation of reliability at NOC.
	
	The main contributions of this paper are as follows. \vspace{-0.65em}
	\begin{itemize}
		\item We propose a meta-analysis framework for multiple independent interval-monitored SSALT experiments under the FRM, in which the experiments may differ in sample size, stress path, and inspection schedule, and the lifetime distribution may be any member of the PH family. The experiments are linked through a common log-linear stress-life relationship. \vspace{-0.65em}
		
		\item We derive maximum likelihood estimators and asymptotic confidence intervals for the model parameters and for reliability characteristics at NOC: the reliability at a mission time, the MTTF, and important lifetime quantiles. \vspace{-0.65em}
		
		\item Through simulation, we assess the performance of the estimators and examine the consequences of misspecifying the lifetime distribution, showing that the tail quantiles, such as the 0.10 and 0.95 quantiles, are far more sensitive to this choice than the median. \vspace{-0.65em}
		
		\item Using the LED data, we quantify the gain in precision obtained by pooling the two tests compared with analyzing each test on its own. \vspace{-0.65em}
	\end{itemize}
	
	The remainder of the paper is organized as follows. Section~\ref{sec:model_description} introduces the statistical model, and Sections~\ref{sec:estimation} and~\ref{sec:CI} develop point and interval estimation. Section~\ref{sec:simulations} reports the simulation study, and Section~\ref{sec:misspecification} examines model misspecification. Section~\ref{sec:data_analysis} analyzes the LED data and simulated data, and Section~\ref{sec:conclusion} concludes the paper.
	
	\section{Model description}
	\label{sec:model_description}
	
	Let us consider the framework of  $K$ independent interval-monitored step-stress accelerated life testing (SSALT) experiments in which  the $k$-th experiment ($k = 1, 2, \dots, K$) allocates $N_k$ identical test units to an initial stress level $s_{k,1}$. The pre-fixed inspection times are denoted by
	$
	0=\tau_{k,0}<\tau_{k,1}<\cdots<\tau_{k,I_k}=t_c,
	$
	where $I_k$ denotes the total number of inspection times and $t_c$ specifies  the termination time of the $k$-th experiment. Each experiment is conducted under a sequence of stress levels, where the initial stress level may equal or exceed the normal operating condition (NOC), while the subsequent stress levels exceed it, i.e.,
	$s_{k,1}\ge s_{\text{NOC}}$ and $s_{k,i}>s_{\text{NOC}}$ for $i=2,\ldots,I_k$. 
	The corresponding stress path for the $k$-th experiment is defined as
	\[
	\mathcal{S}_k=
	\left\{
	(\tau_{k,1},s_{k,1}),
	(\tau_{k,2},s_{k,2}),
	\ldots,
	(\tau_{k,I_k},s_{k,I_k})
	\right\}.
	\]
	Here, $s_{k,i}$ denotes the stress level applied during the interval $(\tau_{k,i-1},\tau_{k,i}]$. \vspace{0.65em}
	
	\begin{figure}[!ht]
		\centering
		\resizebox{1\linewidth}{!}{
			\begin{tikzpicture}[
    font=\large,
    >={Stealth[length=6pt,width=4pt]},
    axis/.style={->, thick},
    interval/.style={decorate, decoration={brace, amplitude=4pt, mirror}, thin},
    stress node/.style={circle, fill, inner sep=2pt},
    open node/.style={circle, draw, fill=white, inner sep=2pt, thick},
    box/.style={draw, rounded corners, align=center, thick, inner sep=6pt, text width=2.2cm},
    c1/.style={blue!80!black},
    c2/.style={red!90!black},
    c3/.style={green!50!black}
]

\node[draw, rounded corners, thick, inner sep=8pt, fill=white, align=center, font=\large] at (7, 4.0) {
    \begin{tikzpicture}[baseline=-0.5ex]
        \draw[c1, thick] (0,0) -- (0.8,0);
    \end{tikzpicture} Experiment $k=1$ \hspace{0.5cm}
    \begin{tikzpicture}[baseline=-0.5ex]
        \draw[c2, thick] (0,0) -- (0.8,0);
    \end{tikzpicture} Experiment $k=2$ \hspace{0.5cm}
    \begin{tikzpicture}[baseline=-0.5ex]
        \draw[c3, thick] (0,0) -- (0.8,0);
    \end{tikzpicture} Experiment $k=K$
};

\begin{scope}[yshift=0cm, c1]
    \draw[thick, black] (0,0) -- (10.0,0);
    \draw[thick, black, dotted] (10.0,0) -- (11.5,0);
    \draw[axis, black] (11.5,0) -- (15.0,0) node[right=0.8cm, yshift=6pt, font=\large] {Time};
    \draw[axis, black] (0,0) -- (0,2.5) node[above, align=center, font=\large] {Stress};

    \draw[black] (-0.1, 0.4) -- (0.1, 0.4) node[left=0.18cm, font=\large, text=black] {$s_{1,1}$};
    \draw[black] (-0.1, 1.0) -- (0.1, 1.0) node[left=0.18cm, font=\large, text=black] {$s_{1,2}$};
    \node[left=0.25cm, text=black] at (0, 1.5) {$\vdots$};
    \draw[black] (-0.1, 2.0) -- (0.1, 2.0) node[left=0.18cm, font=\large, text=black] {$s_{1,I_1}$};
    \node[below=3pt, font=\large, text=black] at (0,0) {$\mathbf{0}$};

    \draw[c1, thick] 
        (0,0.4) node[stress node, fill=blue!80!black] {} 
        -- (4.5,0.4) node[midway, above] {} node[open node, draw=blue!80!black] (p11) {}
        -- (4.5,1.0)
        -- (9.0,1.0) node[midway, above] {} node[open node, draw=blue!80!black] (p12) {}
        -- (10.0,1.0); 
        
    \node[c1] at (10.75, 1.5) {$\dots$};
    
    \draw[dashed, black] (11.5,-0.18) node[anchor=north east, font=\large] {$\tau_{1,I_1-1}$} -- (11.5,2.0);
    \draw[c1, thick] (11.5,2.0) -- (13.5,2.0) node[open node, draw=blue!80!black] (p1end) {};
    
    \draw[dashed, thick, black] (13.5, 2.0) -- (13.5, -2.0);
    \node[fill=white, inner sep=3pt, font=\large, anchor=north west, text=black] at (13.5,-0.18) {$\tau_{1,I_1} = t_c$};

    \draw[dashed, black] (p11) -- (4.5,0) node[below=3pt, font=\large] {$\tau_{1,1}$};
    \draw[dashed, black] (p12) -- (9.0,0) node[below=3pt, font=\large] {$\tau_{1,2}$};

    \draw[interval, black] (0,-0.75) -- (4.5,-0.75);
    \draw[interval, black] (4.5,-0.75) -- (9.0,-0.75);
    \draw[interval, black] (11.5,-0.75) -- (13.5,-0.75);
    \draw[interval, black] (13.5,-0.75) -- (14.8,-0.75);

  \node[text=blue!80!black, left, font=\large] at (-0.5, -1.65) {No. of failures};
    \node[text=blue!80!black, font=\large] at (2.25, -1.65) {$n_{1,1}$};
    \node[text=blue!80!black, font=\large] at (6.75, -1.65) {$n_{1,2}$};
    \node[text=blue!80!black, font=\large] at (10.75, -1.65) {$\dots$};
    \node[text=blue!80!black, font=\large] at (12.5, -1.65) {$n_{1,I_1}$};
    \node[text=blue!80!black, font=\large] at (14.35, -1.65) {$n_{1,I_1+1}$};
\end{scope}

\begin{scope}[yshift=-5.4cm, c2]
    \draw[thick, black] (0,0) -- (10.0,0);
    \draw[thick, black, dotted] (10.0,0) -- (11.5,0);
    \draw[axis, black] (11.5,0) -- (15.0,0) node[right=0.8cm, yshift=6pt, font=\large] {Time};
    \draw[axis, black] (0,0) -- (0,2.5) node[above, align=center, font=\large] {Stress};

    \draw[black] (-0.1, 0.4) -- (0.1, 0.4) node[left=0.18cm, font=\large, text=black] {$s_{2,1}$};
    \draw[black] (-0.1, 0.9) -- (0.1, 0.9) node[left=0.18cm, font=\large, text=black] {$s_{2,2}$};
    \node[left=0.25cm, text=black] at (0, 1.4) {$\vdots$};
    \draw[black] (-0.1, 1.9) -- (0.1, 1.9) node[left=0.18cm, font=\large, text=black] {$s_{2,I_2}$};
    \node[below=3pt, font=\large, text=black] at (0,0) {$\mathbf{0}$};

    \draw[c2, thick] 
        (0,0.4) node[stress node, fill=red!90!black] {} 
        -- (3.0,0.4) node[midway, above] {} node[open node, draw=red!90!black] (p21) {}
        -- (3.0,0.9)
        -- (6.0,0.9) node[midway, above] {} node[open node, draw=red!90!black] (p22) {}
        -- (6.0,1.4)
        -- (8.5,1.4) node[midway, above] {} node[open node, draw=red!90!black] (p23) {}
        -- (10.0,1.4); 
        
    \node[c2] at (10.75, 1.65) {$\dots$};
    
    \draw[dashed, black] (11.5,-0.18) node[anchor=north east, font=\large] {$\tau_{2,I_2-1}$} -- (11.5,1.9);
    \draw[c2, thick] (11.5,1.9) -- (13.5,1.9) node[open node, draw=red!90!black] (p2end) {};
    
    \draw[dashed, thick, black] (13.5, 1.9) -- (13.5, -2.0);
    \node[fill=white, inner sep=3pt, font=\large, anchor=north west, text=black] at (13.5,-0.18) {$\tau_{2,I_2} = t_c$};

    \draw[dashed, black] (p21) -- (3.0,0) node[below=3pt, font=\large] {$\tau_{2,1}$};
    \draw[dashed, black] (p22) -- (6.0,0) node[below=3pt, font=\large] {$\tau_{2,2}$};
    \draw[dashed, black] (p23) -- (8.5,0) node[below=3pt, font=\large] {$\tau_{2,3}$};

    \draw[interval, black] (0,-0.75) -- (3.0,-0.75);
    \draw[interval, black] (3.0,-0.75) -- (6.0,-0.75);
    \draw[interval, black] (6.0,-0.75) -- (8.5,-0.75);
    \draw[interval, black] (11.5,-0.75) -- (13.5,-0.75);
    \draw[interval, black] (13.5,-0.75) -- (14.8,-0.75);

   \node[text=red!90!black, left, font=\large] at (-0.5, -1.65) {No. of failures};
    \node[text=red!90!black, font=\large] at (1.5, -1.65) {$n_{2,1}$};
    \node[text=red!90!black, font=\large] at (4.5, -1.65) {$n_{2,2}$};
    \node[text=red!90!black, font=\large] at (7.25, -1.65) {$n_{2,3}$};
    \node[text=red!90!black, font=\large] at (10.75, -1.65) {$\dots$};
    \node[text=red!90!black, font=\large] at (12.5, -1.65) {$n_{2,I_2}$};
    \node[text=red!90!black, font=\large] at (14.35, -1.65) {$n_{2,I_2+1}$};
\end{scope}

\node[font=\Huge] at (6, -8.0) {$\vdots$};

\begin{scope}[yshift=-10.8cm, c3]
    \draw[thick, black] (0,0) -- (10.0,0);
    \draw[thick, black, dotted] (10.0,0) -- (11.5,0);
    \draw[axis, black] (11.5,0) -- (15.0,0) node[right=0.8cm, yshift=6pt, font=\large] {Time};
    \draw[axis, black] (0,0) -- (0,2.5) node[above, align=center, font=\large] {Stress};

    \draw[black] (-0.1, 0.4) -- (0.1, 0.4) node[left=0.18cm, font=\large, text=black] {$s_{K,1}$};
    \draw[black] (-0.1, 1.2) -- (0.1, 1.2) node[left=0.18cm, font=\large, text=black] {$s_{K,2}$};
    \node[left=0.25cm, text=black] at (0, 1.6) {$\vdots$};
    \draw[black] (-0.1, 2.0) -- (0.1, 2.0) node[left=0.18cm, font=\large, text=black] {$s_{K,I_K}$};
    \node[below=3pt, font=\large, text=black] at (0,0) {$\mathbf{0}$};

    \draw[c3, thick] 
        (0,0.4) node[stress node, fill=green!50!black] {} 
        -- (3.5,0.4) node[midway, above] {} node[open node, draw=green!50!black] (p31) {}
        -- (3.5,1.2)
        -- (7.5,1.2) node[midway, above] {} node[open node, draw=green!50!black] (p32) {}
        -- (10.0,1.2); 
        
    \node[c3] at (10.75, 1.6) {$\dots$};
    
    \draw[dashed, black] (11.5,-0.18) node[anchor=north east, font=\large] {$\tau_{K,I_K-1}$} -- (11.5,2.0);
    \draw[c3, thick] (11.5,2.0) -- (13.5,2.0) node[open node, draw=green!50!black] (p3end) {};
    
    \draw[dashed, thick, black] (13.5, 2.0) -- (13.5, -2.0);
    \node[fill=white, inner sep=3pt, font=\large, anchor=north west, text=black] at (13.5,-0.18) {$\tau_{K,I_K} = t_c$};

    \draw[dashed, black] (p31) -- (3.5,0) node[below=3pt, font=\large] {$\tau_{K,1}$};
    \draw[dashed, black] (p32) -- (7.5,0) node[below=3pt, font=\large] {$\tau_{K,2}$};

    \draw[interval, black] (0,-0.75) -- (3.5,-0.75);
    \draw[interval, black] (3.5,-0.75) -- (7.5,-0.75);
    \draw[interval, black] (11.5,-0.75) -- (13.5,-0.75);
    \draw[interval, black] (13.5,-0.75) -- (14.8,-0.75);

\node[text=green!50!black, left, font=\large] at (-0.5, -1.65) {No. of failures};
    \node[text=green!50!black, font=\large] at (1.75, -1.65) {$n_{K,1}$};
    \node[text=green!50!black, font=\large] at (5.5, -1.65) {$n_{K,2}$};
    \node[text=green!50!black, font=\large] at (10.75, -1.65) {$\dots$};
    \node[text=green!50!black, font=\large] at (12.5, -1.65) {$n_{K,I_K}$};
    \node[text=green!50!black, font=\large] at (14.5, -1.65) {$n_{K,I_K+1}$};
\end{scope}

\end{tikzpicture} } 
		
		\caption{K step-stress experiments with interval-monitoring.}
		\label{fig:K_experiments}
	\end{figure}
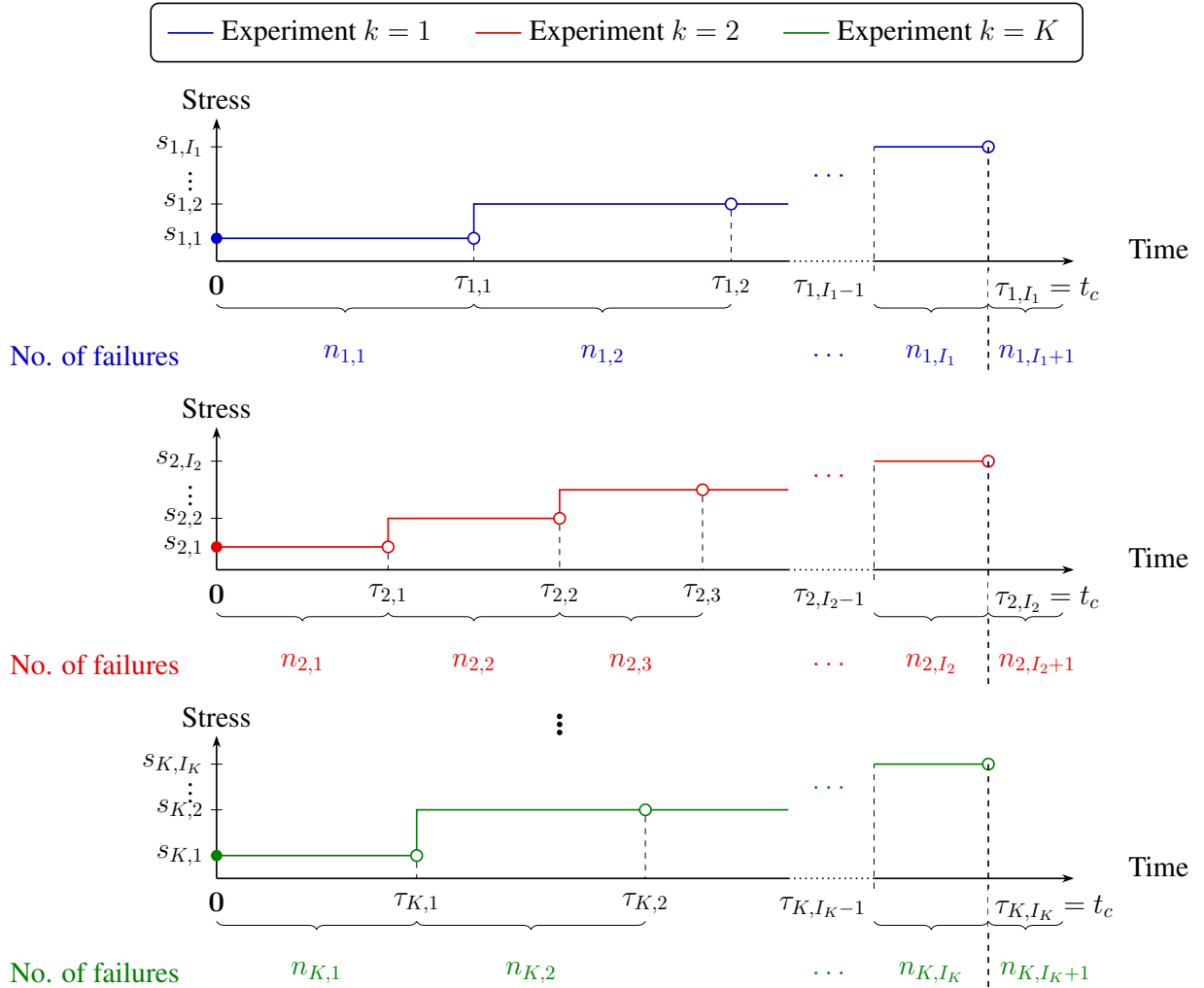
	Since the experiments are interval monitored (see Figure \ref{fig:K_experiments}), the exact failure times are  unobserved and  only the failure counts within successive inspection intervals are recorded. Let $n_{k,i}$ denote the number of failed units  within the interval $(\tau_{k,i-1}, \tau_{k,i}]$ for $i = 1, 2, \dots, I_k$, and let $n_{k,I_k+1}$ represent  the number of units surviving beyond the termination time $t_c$. Consequently, the total sample size for the $k$-th experiment satisfies
	$\displaystyle \sum_{i=1}^{I_k+1} n_{k,i} = N_k.
	$
	Accordingly, the complete observed data from the $K$ independent experiments are given by
	\[
	\mathcal{D}
	=
	\left\{
	n_{k,i},\;
	k=1,\ldots,K,\;
	i=1,\ldots,I_k, n_{k,I_k+1}
	\right\}.
	\]
	
	Moreover, the time-to-failure  distribution of each test unit is assumed to belong to the proportional hazards (PH) class of distributions. A class of distributions is referred to as the PH class of distributions with parameters $\alpha>0$ and $\lambda>0$ if its reliability  function can be expressed as
	\begin{equation}
		R(t;\alpha,\lambda)=
		\begin{cases}
			\left[R_0(t;\alpha)\right]^{\lambda}, & \text{if} \:\: t>0,\\
			1, & \text{otherwise},
		\end{cases}
		\label{eq:PH}
	\end{equation}
	where $R_0(t;\alpha)$ denotes an absolutely continuous baseline reliability  function assumed to be specified completely, except the unknown parameter, $\alpha$ and $H_0(t;\alpha)=-\ln R_0(t;\alpha)$ defines  the associated cumulative hazard function (CHF). A broad class of time-to-failure distributions can be obtained by specifying different functional forms of  $R_0(t;\alpha)$.  Some notable members of this class are presented 
	in Table  \ref{tab:models_phc}. Here, the notations  C, I, D, BT, and UBT denote constant, increasing, decreasing, bathtub-shaped, and upside-down bathtub-shaped failure-rate patterns  respectively.

	\begin{table}[!ht]
		\centering
		\caption{Some notable time-to-failure models belonging to the PH class of distributions}
		\label{tab:models_phc}
		\scriptsize
		\setlength{\tabcolsep}{3pt}
		\renewcommand{\arraystretch}{1.3}
		
		\begin{tabular}{lccccccc}
			\hline
			Distribution
			& $H_0(t;\alpha)$
			& $R_0(t;\alpha)$
			& $R_{PH}(t;\alpha,\lambda)$
			& Failure-rate
			& Shape
			& Scale
			& (Type)
			\\
			\hline
			
			Exponential
			& $t$
			& $e^{-t}$
			& $e^{-\lambda t}$
			& $\lambda$
			& $-$
			& $\lambda$
			& (C)
			\\
			
			Weibull
			& $t^{\alpha}$
			& $e^{-t^{\alpha}}$
			& $e^{-\lambda t^{\alpha}}$
			& $\lambda\alpha t^{\alpha-1}$
			& $\alpha$
			& $\lambda$
			& (I, D)
			\\
			
			Pareto
			& $\ln(1+t)$
			& $(1+t)^{-1}$
			& $(1+t)^{-\lambda}$
			& $\dfrac{\lambda}{1+t}$
			& $\lambda$
			& $-$
			& (D)
			\\
			
			Burr XII
			& $\ln(1+t^{\alpha})$
			& $(1+t^{\alpha})^{-1}$
			& $(1+t^{\alpha})^{-\lambda}$
			& $\dfrac{\lambda\alpha t^{\alpha-1}}
			{1+t^{\alpha}}$
			& $\alpha,\lambda$
			& $-$
			& (D, UBT)
			\\
			
			Gompertz
			& $e^{\alpha t}-1$
			& $\exp[-(e^{\alpha t}-1)]$
			& $\exp[-\lambda(e^{\alpha t}-1)]$
			& $\lambda\alpha e^{\alpha t}$
			& $\alpha$
			& $\lambda$
			& (I)
			\\
			
			Lomax
			& $\ln(1+\alpha t)$
			& $(1+\alpha t)^{-1}$
			& $(1+\alpha t)^{-\lambda}$
			& $\dfrac{\lambda\alpha}{1+\alpha t}$
			& $\lambda$
			& $\alpha$
			& (D)
			\\
			Chen
			& $e^{t^{\alpha}}-1$
			& $\exp\!\left[-(e^{t^{\alpha}}-1)\right]$
			& $\exp\!\left[-\lambda(e^{t^{\alpha}}-1)\right]$
			& $\lambda\alpha t^{\alpha-1}e^{t^{\alpha}}$
			& $\alpha$
			& $\lambda$
			& (I, BT)
			\\
			\hline
		\end{tabular}
	\end{table}
	
	In SSALT experiments, the cumulative exposure model (CEM) and the failure rate model (FRM) have been widely adopted in order to establish the relationship between the stress-specific time-to-failure distributions. The CEM assumes that the remaining lifetime of a test unit depends solely on the cumulative exposure accumulated during the testing process. In this study, we adopt the flexible FRM proposed by \cite{kateri2015}, which establishes the relationship between the  stress-specific lifetime distributions through their hazard (failure rate) functions. Accordingly, the overall failure-rate function corresponding to the $k$-th experiment is given by
	\begin{equation}
		h_k(t)=
		\begin{cases}
			\lambda_{k,1}h_0(t;\alpha), & 0<t\le\tau_{k,1},\\
			\lambda_{k,i}h_0(t;\alpha), & \tau_{k,i-1}<t\le\tau_{k,i},
		\end{cases}
		\qquad i=2,\ldots,I_k,\quad k=1,\ldots,K,
		\label{eq:h_k(t)}
	\end{equation}
	\noindent where $h_0(t;\alpha)$ denotes the baseline failure rate function (completely specified except the unknown parameter $\alpha$) and $\lambda_{k,i}>0$ defines the shape/scale parameter corresponding to the $i$-th stress-specific time-to-failure distribution.  For $i=2,\ldots,I_k; \: k=1,\ldots,K,$  one can obtain the CHF and the reliability function corresponding to the $k$-th experiment by integrating the  failure rate function over time and are obtained  as:
	\begin{align}
		H_k(t)
		&=
		\begin{cases}
			\lambda_{k,1}H_0(t;\alpha), & 0<t\le \tau_{k,1},\\
			\displaystyle
			\sum_{j=1}^{i-1}\lambda_{k,j}
			\left[
			H_0(\tau_{k,j};\alpha)-H_0(\tau_{k,j-1};\alpha)
			\right]
			+\lambda_{k,i}
			\left[
			H_0(t;\alpha)-H_0(\tau_{k,i-1};\alpha)
			\right], & \tau_{k,i-1}<t\le\tau_{k,i};
		\end{cases}
		\label{eq:H_k(t)}
	\end{align}
	and 
	\begin{align}
		R_k(t)=e^{-H_k(t)}
		=
		\begin{cases}
			R_{k,1}(t), & 0<t\le\tau_{k,1},\\
			R_{k,i}(t), & \tau_{k,i-1}<t\le\tau_{k,i},
		\end{cases}
		\label{eq:R_k(t)}
	\end{align}
	respectively, 
	\noindent where 
	\[
	R_{k,i}(t)=e^{-H_{k,i}(t)}, \quad i=1, 2,\ldots,I_k.
	\]
	Additionally, standardized stress levels are employed to characterize the relationship between the applied stress levels and the time-to-failure  distribution. Let $s_{k,i}\in[s_{\text{NOC}},s_H]$ denote the $i$-th stress level of the $k$-th experiment, where $s_{\text{NOC}}$ denotes the NOC and $s_H$ is the maximum stress level recommended by the practitioner for the experiments. The corresponding standardized stress level is defined as 
	\begin{equation}
		x_{k,i}=\frac{s_{k,i}-s_{\text{NOC}}}{s_H-s_{\text{NOC}}}, \qquad x_{k,i}\in[0,1].
	\end{equation}
	It is hence noteworthy that  $x_{k,i}=0$ corresponds to the NOC, while  $x_{k,i}=1$ corresponds to the highest accelerated stress level. The parameter $\lambda_{k,i}$ associated with the $i$-th stress level is assumed to depend log-linearly on the standardized stress level according to the following relationship
	\begin{equation}
		\ln(\lambda_{k,i})=\beta_0+\beta_1x_{k,i},
	\end{equation}
	where $\beta_0\in\mathbb{R}$ is the intercept term and $\beta_1>0$ is the slope parameter describing the effect of the standardized stress level. Equivalently, $
	\lambda_{k,i}=\exp(\beta_0+\beta_1x_{k,i})$  ensures that $\lambda_{k,i}$ is a positive quantity  and it increases with increase in the standardized stress level. Consequently, the parameter vector of the proposed model is represented as
	$\boldsymbol{\theta}=(\alpha,\beta_0,\beta_1).$
	
	\section{Maximum likelihood estimation}
	\label{sec:estimation}
	
	In this section, we discuss the point estimation procedures for the proposed meta-analysis model in association with the interval-monitored  SSALT experiments.  The method of maximum likelihood is employed to obtain the point estimates of the  unknown model parameters. Subsequently, the asymptotically normal distributions of the resulting MLEs  are used in Section~\ref{sec:CI} to construct CIs of the model parameters and of the reliability characteristics at NOC.  
	
	\subsection{Model parameters}
	\label{subsec:est_par}
	For the $k$-th experiment, the vector of observed failure counts $\mathcal D_k=(n_{k,1},n_{k,2},\ldots,n_{k,I_k},n_{k,I_k+1})$ follows a multinomial distribution with $N_k$ trials and cell probabilities $R_{k,i-1}-R_{k,i}$, $i=1,\ldots,I_k$, and $R_{k,I_k}$, where $R_{k,0}=1$.  
	Consequently, the likelihood function (up to a multiplicative constant) is given by
	\begin{equation}
		L_k(\boldsymbol{\theta}|\mathcal D_k)
		=
		\prod_{i=1}^{I_k}
		\left(R_{k,i-1}-R_{k,i}\right)^{n_{k,i}}
		R_{k,I_k}^{\,n_{k,I_k+1}},
		\label{eq:L_k(t)}
	\end{equation}
	where $\boldsymbol{\theta}=(\alpha, \beta_0,\beta_1),$ and $R_{k,i} \equiv R_k(\tau_{k,i}) $ denotes the reliability at the inspection time $\tau_{k,i}$.
	Since the $K$ interval-monitored SSALT experiments are assumed to be independent, the overall likelihood is obtained by combining the multinomial likelihoods from all experiments and  can be expressed as: 
	\begin{equation}
		L(\boldsymbol{\theta}|\mathcal D) =\prod_{k=1}^{K}
		\left[\prod_{i=1}^{I_k}\{R_{k,i-1}-R_{k,i}\}^{n_{k,i}}\right] R_{k,I_k}^{\,n_{k,I_k+1}}
		\label{eq:Likelihood}
	\end{equation} 
	The corresponding log-likelihood function is then given by
	\begin{equation}
		\ell(\boldsymbol{\theta}|\mathcal D) = \sum_{k=1}^{K}\sum_{i=1}^{I_k} n_{k,i}\ln\big(R_{k,i-1}-R_{k,i}\big)
		+ \sum_{k=1}^{K} n_{k,I_k+1}\ln R_{k,I_k}.
		\label{eq:loglikelihood}
	\end{equation}
	Differentiating the log-likelihood function with respect to the unknown
	parameters $\beta_0$, $\beta_1$, and $\alpha$ yields the following
	likelihood equations:
	\begin{align}
		\frac{\partial \ell}{\partial \beta_0}
		&=
		\sum_{k=1}^{K}\sum_{i=1}^{I_k}
		\frac{n_{k,i}}{D_{k,i}}
		\left(
		R_{k,i}H_{k,i}
		-
		R_{k,i-1}H_{k,i-1}
		\right)
		-
		\sum_{k=1}^{K}
		n_{k,I_k+1}H_{k,I_k}
		=0,
		\\
		\frac{\partial\ell}{\partial\beta_1}
		&=
		\sum_{k=1}^{K}\sum_{i=1}^{I_k}
		\frac{n_{k,i}}{D_{k,i}}
		\left(
		R_{k,i}H_{k,i}B_{k,i}
		-
		R_{k,i-1}H_{k,i-1}B_{k,i-1}
		\right)
		-
		\sum_{k=1}^{K}
		n_{k,I_k+1}H_{k,I_k}B_{k,I_k}
		=0,
	\end{align}
	and
	\begin{align}
		\frac{\partial\ell}{\partial\alpha}
		&=
		\sum_{k=1}^{K}\sum_{i=1}^{I_k}
		\frac{n_{k,i}}{D_{k,i}}
		\left(
		R_{k,i}H_{k,i}C_{k,i}
		-
		R_{k,i-1}H_{k,i-1}C_{k,i-1}
		\right)
		-
		\sum_{k=1}^{K}
		n_{k,I_k+1}H_{k,I_k}C_{k,I_k}
		=0,
	\end{align}
	where, for $i=1,\ldots,I_k$ (with $H_{k,0}=0$), 
	\begin{align*}
		D_{k,i}
		&=
		R_{k,i-1}-R_{k,i},
		\\
		H_{k,i}
		&=
		\sum_{j=1}^{i}
		\lambda_{k,j}
		\left[
		H_0(\tau_{k,j};\alpha)
		-
		H_0(\tau_{k,j-1};\alpha)
		\right],
		\\
		B_{k,i}
		&=
		\frac{1}{H_{k,i}}
		\sum_{j=1}^{i}
		x_{k,j}\lambda_{k,j}
		\left[
		H_0(\tau_{k,j};\alpha)
		-
		H_0(\tau_{k,j-1};\alpha)
		\right],
		\\
		C_{k,i}
		&=
		\frac{1}{H_{k,i}}
		\sum_{j=1}^{i}
		\lambda_{k,j}
		\left[
		\frac{\partial H_0(\tau_{k,j};\alpha)}{\partial\alpha}
		-
		\frac{\partial H_0(\tau_{k,j-1};\alpha)}{\partial\alpha}
		\right].
	\end{align*}
	\noindent Since the likelihood equations corresponding to $\boldsymbol{\theta}=(\alpha,\beta_0,\beta_1)$
	are nonlinear and do not admit closed-form solutions, one may resort to  iterative numerical procedures in order to obtain their MLEs.
	
	\subsection{Reliability characteristics at the NOC}
	\label{subsec:est_lifetime_char}
	
	Some well-known metrics available in reliability literature for assessing lifetime behavior  under the NOC  are (i) reliability at a specified mission time, (ii)  mean time to failure (MTTF), and (iii) quantile time-to-failure; we refer to these  collectively as the reliability characteristics at NOC. In the proposed framework, these quantities are evaluated at the NOC, corresponding to the standardized stress level $x=0$. Consequently, the CHF is obtained as:
	\[
	H_{\text{NOC}} (t;\alpha,\beta_0)
	=
	\exp(\beta_0)H_0(t;\alpha),
	\]
	where $H_0(t;\alpha)$ denotes the baseline CHF.
	
	\paragraph{Reliability at a specified mission time:}The reliability function at a specified mission time $t_M>0$ under the NOC is defined as
	\begin{align}
		R_{\text{NOC}}(t_M)=
		\exp\left\{-H_{\text{NOC}} (t_M;\alpha,\beta_0)\right\}
		=
		\exp\left\{
		-\exp(\beta_0)H_0(t_M;\alpha)
		\right\}.
		\label{eq:noc_reliability}
	\end{align}
	
	\paragraph{Mean time to failure:}
	The mean time to failure (MTTF) (whenever it
	exists) under the NOC (denoted as $\mu_\text{NOC}$) is defined by
	\begin{equation}
		\mu_\text{NOC}
		=
		\int_0^\infty R_{\text{NOC}}(t)\,dt
		=
		\int_0^\infty
		\exp\left\{
		-\exp(\beta_0)H_0(t;\alpha)
		\right\}\,dt.
		\label{eq:noc_mttf}
	\end{equation}
	For some members of the  PH class of distributions, the integral in Equation~\eqref{eq:noc_mttf} admits a closed-form expression, while  for other members, it may be evaluated employing suitable numerical procedures.
	
	\paragraph{Quantile time-to-failure:} 
	For $0<q<1,$ the $q$-th quantile of the time-to-failure distribution under NOC (denoted by $t_{q,\text{NOC}} $)  is defined as the solution of $R_{\text{NOC}}(t_{q,\text{NOC}} )=1-q$, or equivalently,
	\begin{equation}
		H_{\text{NOC}}(t_{q,\text{NOC}} ;\alpha,\beta_0) = -\ln(1-q).
	\end{equation}
	Whenever the inverse of the baseline CHF exists, the $q$-th quantile can be expressed as:
	\begin{equation}
		t_{q,\text{NOC}}  = H_0^{-1}\big(-\ln(1-q)\exp(-\beta_0);\,\alpha\big).
		\label{eq:noc_quantile}
	\end{equation}
	In particular, the median time-to-failure  is obtained by setting $q=0.5$ and is given by:
	\begin{equation}
		t_{0.5,\text{NOC}} 
		=
		H_0^{-1}
		\left(
		\ln(2)\exp(-\beta_0);\alpha
		\right).
		\label{eq:noc_median}
	\end{equation}
	
	\noindent Employing  the invariance property, the MLEs of the reliability characteristics \(R_{\text{NOC}}(t_M)\), \(\mu_{\text{NOC}}\), \(t_{q,\text{NOC}} \), and \(t_{0.5,\text{NOC}} \) are obtained by direct substitution of \(\hat{\boldsymbol{\theta}}=(\hat{\beta}_0,\hat{\alpha},\hat{\beta}_1)\) for \(\boldsymbol{\theta}\) in their corresponding expressions in
	\eqref{eq:noc_reliability}-\eqref{eq:noc_median}.
	
	\subsection{Some special cases}
	\label{subsec:est_special_cases}
	
	For a given member of the PH family, all quantities in the likelihood equations and in the Fisher information, such as $H_{k,i}$, $R_{k,i}$, $D_{k,i}$, $B_{k,i}$, and $C_{k,i}$, follow from the baseline CHF $H_0(t;\alpha)$. The model parameters are therefore estimated by maximizing the log-likelihood \eqref{eq:loglikelihood} with the corresponding $H_0(t;\alpha)$; for the exponential distribution, $\alpha=1$ and only $(\beta_0,\beta_1)$ are estimated. Since the likelihood equations have no closed-form solution, the MLEs are obtained numerically, with starting values chosen from contour plots of the (profile) log-likelihood. The log-likelihood functions, the likelihood equations, and the estimation procedure for the exponential, Weibull, and Burr XII distributions, considered in the simulation study, are given in detail in Section~S4 of the Supplementary Material. The Chen distribution, used in Sections~\ref{sec:misspecification} and~\ref{sec:data_analysis}, is handled in the same way. Table~\ref{tab:special_cases} lists $H_0(t;\alpha)$ and the resulting reliability characteristics at NOC for these four models.
	
	\begin{table}[H]
		\centering
		\caption{Baseline CHF and reliability characteristics at NOC for four members of the PH family. The MTTF is finite for the exponential, Weibull, and Chen distributions, and for the Burr XII distribution if $\alpha e^{\beta_0}>1$.}
		\label{tab:special_cases}
		\scriptsize
		\setlength{\tabcolsep}{5pt}
		\begin{tabular}{@{}lcccc@{}}
			\toprule
			Distribution & $H_0(t;\alpha)$ & $R_{\text{NOC}}(t_M)$ & $\mu_{\text{NOC}}$ & $t_{0.5,\text{NOC}}$ \\
			\midrule
			Exponential ($\alpha=1$) & $t$ & $e^{-e^{\beta_0}t_M}$ & $e^{-\beta_0}$ & $(\ln 2)\,e^{-\beta_0}$ \\
			Weibull & $t^{\alpha}$ & $e^{-e^{\beta_0}t_M^{\alpha}}$ & $e^{-\beta_0/\alpha}\,\Gamma\!\left(1+\frac{1}{\alpha}\right)$ & $\left(\ln 2\; e^{-\beta_0}\right)^{1/\alpha}$ \\
			Burr XII & $\ln\left(1+t^{\alpha}\right)$ & $\left(1+t_M^{\alpha}\right)^{-e^{\beta_0}}$ & $\dfrac{\Gamma\!\left(1+\frac{1}{\alpha}\right)\Gamma\!\left(e^{\beta_0}-\frac{1}{\alpha}\right)}{\Gamma\!\left(e^{\beta_0}\right)}$ & $\left(2^{e^{-\beta_0}}-1\right)^{1/\alpha}$ \\
			Chen & $e^{t^{\alpha}}-1$ & $e^{-e^{\beta_0}\left(e^{t_M^{\alpha}}-1\right)}$ & $\displaystyle\int_0^{\infty} e^{-e^{\beta_0}\left(e^{t^{\alpha}}-1\right)}\,dt$ & $\left[\ln\!\left(1+\ln 2\; e^{-\beta_0}\right)\right]^{1/\alpha}$ \\
			\bottomrule
		\end{tabular}
	\end{table}
	
	\section{Interval estimation}
	\label{sec:CI}
	
	\subsection{Model parameters}
	\label{subsec:CI_par}
	
	Since the MLEs of the model parameters do not
	admit closed-form expressions, their exact sampling distributions, and hence the exact CIs are difficult to obtain. Hence, we resort to construct approximate CIs based on the asymptotic properties of the MLEs. Under the usual regularity conditions for ML estimation, $\widehat{\boldsymbol{\theta}}=(\widehat{\beta}_0,\widehat{\alpha},\widehat{\beta}_1)$ is
	asymptotically multivariate normally distributed as the total sample size $N$ grows. In particular, $\widehat{\boldsymbol{\theta}}
	\overset{}{\sim}
	N_3\left\{
	\boldsymbol{\theta},
	\mathcal{I}^{-1}(\boldsymbol{\theta})
	\right\}$, where
	the Fisher information matrix $\mathcal{I}(\boldsymbol{\theta})$ is defined as the negative expectation of the Hessian matrix of the log-likelihood function $\ell(\boldsymbol{\theta}|\mathcal D)$ in  \eqref{eq:loglikelihood}. Assuming  $\boldsymbol{\theta}=(\theta_1,\theta_2,\theta_3)=(\beta_0,\alpha,\beta_1),$  we thus have 
	\begin{equation*}
		\mathcal{I}(\boldsymbol{\theta})
		=
		\Bigg(\Bigg(-E\!\left[
		\frac{\partial^{2}\ell(\boldsymbol{\theta})}
		{\partial\theta_p\,\partial\theta_q}
		\right]\Bigg)\Bigg),
	\end{equation*}
	which can further be simplified as
	\begin{equation}
		\begin{aligned}
			\mathcal{I}_{pq}(\boldsymbol{\theta}) 
			=&
			\sum_{k=1}^{K}\sum_{i=1}^{I_k}
			\left(
			\frac{\partial D_{k,i}}{\partial\theta_p}
			\right)
			\left(
			\frac{\partial D_{k,i}}{\partial\theta_q}
			\right)
			\frac{E(n_{k,i})}{D_{k,i}^{2}}
			\\
			&+
			\sum_{k=1}^{K}
			\left(
			\frac{\partial R_{k,I_k}}{\partial\theta_p}
			\right)
			\left(
			\frac{\partial R_{k,I_k}}{\partial\theta_q}
			\right)
			\frac{E(n_{k,I_k+1})}{R_{k,I_k}^{2}}.
		\end{aligned}
	\end{equation}
	Since $E(n_{k,i})=N_kD_{k,i}$ and
	$E(n_{k,I_k+1})=N_kR_{k,I_k},
	$
	the above expression simplifies to
	\begin{equation}
		\mathcal{I}_{pq}(\boldsymbol{\theta}) 
		=
		\sum_{k=1}^{K}
		N_k
		\left[
		\sum_{i=1}^{I_k}
		\frac{1}{D_{k,i}}
		\left(
		\frac{\partial D_{k,i}}{\partial\theta_p}
		\right)
		\left(
		\frac{\partial D_{k,i}}{\partial\theta_q}
		\right)
		+
		\frac{1}{R_{k,I_k}}
		\left(
		\frac{\partial R_{k,I_k}}{\partial\theta_p}
		\right)
		\left(
		\frac{\partial R_{k,I_k}}{\partial\theta_q}
		\right)
		\right].
	\end{equation}
	For notational convenience, let us define
	\begin{align}
		U_{k,i}^{(\beta_0)}
		&=
		\frac{\partial D_{k,i}}{\partial\beta_0}
		=
		R_{k,i}H_{k,i}
		-
		R_{k,i-1}H_{k,i-1},
		\\
		U_{k,i}^{(\alpha)}
		&=
		\frac{\partial D_{k,i}}{\partial\alpha}
		=
		R_{k,i}H_{k,i}C_{k,i}
		-
		R_{k,i-1}H_{k,i-1}C_{k,i-1},
		\\
		U_{k,i}^{(\beta_1)}
		&=
		\frac{\partial D_{k,i}}{\partial\beta_1}
		=
		R_{k,i}H_{k,i}B_{k,i}
		-
		R_{k,i-1}H_{k,i-1}B_{k,i-1},
	\end{align}
	and
	\begin{align}
		V_{k}^{(\beta_0)}
		&=
		\frac{\partial R_{k,I_k}}{\partial\beta_0}
		=
		-
		R_{k,I_k}H_{k,I_k},\\	
		V_{k}^{(\alpha)}
		&=
		\frac{\partial R_{k,I_k}}{\partial\alpha}
		=
		-
		R_{k,I_k}H_{k,I_k}C_{k,I_k},
	\end{align}
	\begin{align}
		V_{k}^{(\beta_1)}
		&=
		\frac{\partial R_{k,I_k}}{\partial\beta_1}
		=
		-
		R_{k,I_k}H_{k,I_k}B_{k,I_k}.
	\end{align}
	Subsequently, the Fisher information matrix can be expressed as a $3\times 3$ matrix as follows:
	\[
	\mathcal{I}(\boldsymbol{\theta})
	=
	\begin{pmatrix}
		\mathcal{I}_{11} & \mathcal{I}_{12}\\
		\mathcal{I}_{12}^{T} & \mathcal{I}_{22}
	\end{pmatrix},
	\]
	
	where
	\begin{gather*}
		\mathcal{I}_{11}
		=
		\begin{pmatrix}
			\mathcal{I}_{\beta_0\beta_0} & \mathcal{I}_{\beta_0\alpha}\\
			\mathcal{I}_{\beta_0\alpha} & \mathcal{I}_{\alpha\alpha}
		\end{pmatrix}_{2\times2},
		\\
		\mathcal{I}_{12}
		=
		\begin{pmatrix}
			\mathcal{I}_{\beta_0\beta_1}\\
			\mathcal{I}_{\alpha\beta_1}
		\end{pmatrix}_{2\times1},
		\qquad
		\mathcal{I}_{22}=\mathcal{I}_{\beta_1\beta_1},
	\end{gather*}
	and, for $p,q\in\{\beta_0,\alpha,\beta_1\}$,
	\begin{equation}
		\mathcal{I}_{pq}
		=
		\sum_{k=1}^{K}N_k
		\left[
		\sum_{i=1}^{I_k}
		\frac{U_{k,i}^{(p)}U_{k,i}^{(q)}}{D_{k,i}}
		+
		\frac{V_k^{(p)}V_k^{(q)}}{R_{k,I_k}}
		\right].
		\label{eq:fisher_elements}
	\end{equation}
	
	The asymptotic variance-covariance matrix of the MLE is $\operatorname{Var}(\widehat{\boldsymbol{\theta}})=\mathcal{I}^{-1}(\boldsymbol{\theta})$, and its estimate is $\widehat{\boldsymbol{\Sigma}}=\mathcal{I}^{-1}(\widehat{\boldsymbol{\theta}})$. Here, the $i$-th diagonal element $\widehat{\Sigma}_{ii}$ represents the  estimated asymptotic variance of $\widehat{\theta}_i$, and an approximate $100(1-\gamma)\%$ CI for $\theta_i$ is given by 
	\[
	\widehat{\theta}_i
	\pm
	z_{\gamma/2}\sqrt{ \widehat{\Sigma}_{ii}},
	\qquad i=1,2,3,
	\]
	where $z_{\gamma/2}$ denotes the upper $\gamma/2$ percentile of the
	standard normal distribution. For the reliability characteristics,
	the corresponding asymptotic confidence intervals are obtained by
	applying the delta method to their respective functions of the model
	parameters.

	\subsection{Reliability characteristics at the NOC}
	\label{subsec:CI_lifetime_char}
	
	The asymptotic confidence intervals described above are based on the asymptotic
	normality of the maximum likelihood estimator. However,
	the reliability characteristics defined in Subsection \ref{subsec:est_lifetime_char} are subject to natural parameter-space restrictions. In particular, the reliability
	function takes values in $(0,1)$, whereas the mean time to failure (MTTF) and the lifetime quantiles are strictly positive. Consequently, the direct
	normal-approximation intervals may extend beyond their admissible ranges.
	To account for these restrictions, suitable transformations are employed to
	construct more appropriate confidence intervals for the reliability and lifetime
	characteristics based on the MLE.
	
	For $\psi\in\{R_{\text{NOC}}(t_M),\,\mu_{\text{NOC}},\,t_{q,\text{NOC}} \}$,  since $\psi$ depends on $\boldsymbol{\theta}$ only through $(\beta_0,\alpha)$ (only through $\beta_0$ for the exponential model), the gradient vector is defined as
	\[
	P_{\psi}
	=
	\left(
	\frac{\partial\psi}{\partial\beta_0},
	\frac{\partial\psi}{\partial\alpha}
	\right).
	\]
	Consequently, the asymptotic variance of $\widehat{\psi}$ obtained
	using the delta method is expressed as
	\[
	V_{\widehat{\psi}}
	=
	P_{\psi}
	\left[
	\mathcal{I}_{11}-\mathcal{I}_{12}\mathcal{I}_{22}^{-1}\mathcal{I}_{12}^{\top}
	\right]^{-1}
	P_{\psi}^{\top},
	\] 
	where $\left[\mathcal{I}_{11}-\mathcal{I}_{12}\mathcal{I}_{22}^{-1}\mathcal{I}_{12}^{\top}\right]^{-1}$ is the $(\beta_0,\alpha)$ block of $\mathcal{I}^{-1}(\boldsymbol{\theta})$, evaluated at $\widehat{\boldsymbol{\theta}}$. The standard error of $\widehat{\psi}$ is then $\operatorname{se}(\widehat{\psi})=\sqrt{V_{\widehat{\psi}}}$. 
	
	To construct confidence intervals that respect the parameter-space
	restrictions, logarithmic transformations are used for the MTTF and
	the lifetime quantiles, whereas a logit transformation is adopted for
	the reliability function. Specifically, consider
	\[
	\phi_{\mu}(\mu_{\text{NOC}})=\ln(\mu_{\text{NOC}}),
	\quad
	\phi_R\{R_{\text{NOC}}(t_M)\}
	=
	\ln\left\{\frac{R_{\text{NOC}}(t_M)}{1-R_{\text{NOC}}(t_M)}\right\}, \text{and}
	\quad
	\phi_q( t_{q,\text{NOC}} )=\ln( t_{q,\text{NOC}} ).
	\]
	Applying the delta method on the transformed scales and subsequently
	back-transforming the resulting limits gives the approximate
	$100(1-\gamma)\%$ log-transformed 
	confidence interval for the MTTF and lifetime quantile respectively as 
	\begin{align}
		& \text{CI}_{\gamma}(\mu_{\text{NOC}})
		=
		\left[
		\widehat{\mu}_{\text{NOC}}
		\exp\left\{
		-\frac{z_{1-\gamma/2}
			\operatorname{se}(\widehat{\mu}_{\text{NOC}})}
		{\widehat{\mu}_{\text{NOC}}}
		\right\},
		\;
		\widehat{\mu}_{\text{NOC}}
		\exp\left\{
		\frac{z_{1-\gamma/2}
			\operatorname{se}(\widehat{\mu}_{\text{NOC}})}
		{\widehat{\mu}_{\text{NOC}}}
		\right\}
		\right], &\\
		&\text{CI}_{\gamma}( t_{q,\text{NOC}} )
		=
		\left[
		\widehat{t}_{q,\text{NOC}}
		\exp\left\{
		-\frac{z_{1-\gamma/2}
			\operatorname{se}(\widehat{t}_{q,\text{NOC}})}
		{\widehat{t}_{q,\text{NOC}}}
		\right\},
		\;
		\widehat{t}_{q,\text{NOC}}
		\exp\left\{
		\frac{z_{1-\gamma/2}
			\operatorname{se}(\widehat{t}_{q,\text{NOC}})}
		{\widehat{t}_{q,\text{NOC}}}
		\right\}
		\right].&
		\label{eq:quantile_ci}
	\end{align}
	Here, $\operatorname{se}(\widehat{\mu}_{\text{NOC}})$, and
	$\operatorname{se}(\widehat{ t}_{q,\text{NOC}} )$ denote the estimated standard errors of the MTTF and the $q$-th lifetime quantile, respectively. For $q=0.5$, $t_{q,\text{NOC}} $ represents the median lifetime, with the corresponding confidence interval obtained from \eqref{eq:quantile_ci}.
	
	For the reliability at the specified mission time $t_M$, a logit
	transformation is adopted to accommodate its natural range $(0,1)$. Define
	\[
	S_R=
	\exp\left\{
	\frac{z_{1-\gamma/2}
		\operatorname{se}\{\widehat{R}_{\text{NOC}}(t_M)\}}
	{\widehat{R}_{\text{NOC}}(t_M)\{1-\widehat{R}_{\text{NOC}}(t_M)\}}
	\right\}.
	\]
	Back-transforming the corresponding confidence limits from the logit scale gives the approximate $100(1-\gamma)\%$ confidence interval for
	$R_{\text{NOC}}(t_M)$ as
	\begin{equation}
		\text{CI}_{\gamma}\{R_{\text{NOC}}(t_M)\}
		=
		\left[
		\frac{\widehat{R}_{\text{NOC}}(t_M)}
		{\widehat{R}_{\text{NOC}}(t_M)+
			[1-\widehat{R}_{\text{NOC}}(t_M)]S_R},
		\;
		\frac{\widehat{R}_{\text{NOC}}(t_M)}
		{\widehat{R}_{\text{NOC}}(t_M)+
			[1-\widehat{R}_{\text{NOC}}(t_M)]/S_R}
		\right].
	\end{equation}
	
	Here, $\operatorname{se}\{\widehat{R}_{NOC}(t_M)\}$ denotes the estimated standard error of the reliability function at the specified mission time $t_M$. This transformation ensures that the resulting confidence limits remain within $(0,1)$.
	
	\FloatBarrier
	\section{Simulation studies}
	\label{sec:simulations}
	
	In this section, an extensive simulation study is conducted to investigate the finite-sample performance of the proposed inferential methods. To illustrate the proposed meta-analysis framework of $K=3$ independent interval-monitored step-stress experiments (each involving $I_k=3$ inspection intervals),  we consider three important members of the PH family, namely (i) Exponential, (ii) Weibull, and (iii) Burr XII distributions as the time-to-failure distributions, with varying   sample sizes and experimental configurations (presented in Table  \ref{tab:simulation_design_all}). The total sample size $N=\sum_{k=1}^{K}N_k$ is allocated equally across the three experiments, i.e., $N_k=N/3$. Under the chosen values of $\beta_1$, the failure intensity at the highest stress level is about $e^{\beta_1}\approx 20$ times that at NOC for the exponential and Weibull models and about $1.6$ times for the Burr XII model, so that both strongly and weakly accelerated tests are represented.
	
	\begin{table}[H]
		\centering
		\footnotesize
		\caption{Simulation design and parameter settings for the Exponential,
			Weibull, and Burr XII distributions.}
		\label{tab:simulation_design_all}

		\setlength{\tabcolsep}{5pt}
		
		\begin{tabular}{@{}llccc@{}}
			\toprule
			Parameters & Notation
			& Exponential & Weibull & Burr XII \\
			\midrule
			
			No. of experiments
			& $K$
			& $3$ & $3$ & $3$ \\
			
			No. of inspection intervals
			& $I_k$
			& $3$ & $3$ & $3$ \\
			
			Total sample size 
			& $N$
			& $N\in\{150,300,450\}$
			& $N\in\{150,300,450\}$
			& $N\in\{150,300,450\}$  \\
			
			True parameter vector
			& $\boldsymbol{\theta}=(\beta_0,\alpha,\beta_1)$
			& $(-2,\,1,\,3)$
			& $(-3,\,2,\,3)$
			& $(-0.1,\,3.5,\,0.5)$ \\
			
			Mission time
			& $t_M$
			& $3.5$ & $3.5$ & $1.3$ \\
			
			\midrule
			\multicolumn{5}{c}{$k=1$} \\[-1mm]
			
			Stress levels
			& $x_{1,i},  i=1,2,3$
			& $(0.20,\,0.55,\,1.00)$
			& $(0.20,\,0.55,\,1.00)$
			& $(0.10,\,0.55,\,1.00)$ \\
			
			Inspection times
			& $\tau_{1,i},  i=1,2,3$
			& $(0.50,\,1.00,\,1.50)$
			& $(0.50,\,1.00,\,1.50)$
			& $(0.50,\,1.00,\,1.50)$ \\
			
			\midrule
			\multicolumn{5}{c}{$k=2$} \\[-1mm]
			
			Stress levels
			& $x_{2,i},  i=1,2,3$
			& $(0.25,\,0.60,\,1.00)$
			& $(0.25,\,0.60,\,1.00)$
			& $(0.15,\,0.60,\,1.00)$ \\
			
			Inspection times
			& $\tau_{2,i},  i=1,2,3$
			& $(1.00,\,1.50,\,2.00)$
			& $(1.00,\,1.50,\,2.00)$
			& $(1.00,\,1.50,\,2.00)$ \\
			
			\midrule
			\multicolumn{5}{c}{$k=3$} \\[-1mm]
			
			Stress levels
			& $x_{3,i},  i=1,2,3$
			& $(0.30,\,0.65,\,1.00)$
			& $(0.30,\,0.65,\,1.00)$
			& $(0.20,\,0.65,\,1.00)$ \\
			
			Inspection times
			& $\tau_{3,i},  i=1,2,3$
			& $(2.00,\,2.50,\,3.00)$
			& $(2.00,\,2.50,\,3.00)$
			& $(1.50,\,2.00,\,2.50)$ \\
			
			
			
			
			\bottomrule
		\end{tabular}
	\end{table}
	For each parameter configuration, $B=1000$ independent interval-monitored SSALT datasets are generated from each of the corresponding time-to-failure distributions using Algorithm  \ref{algo:data_generation}. Information from all the three experiments are  combined through the associated  likelihood function, and the point estimates of the model parameters are obtained  employing the ML principle. The performance of the point estimators is assessed in terms of the average estimate
	(AE) and  root mean squared error (RMSE), while that of the interval estimators is evaluated in terms of coverage probability (CP), and average length (AL) of the corresponding 95\% CIs. In addition to the model parameters, we report the important reliability characteristics at NOC. The associated point estimates are obtained using the invariance property of MLEs, while the CIs is computed  using the delta method with appropriate transformations as discussed in Section  \ref{subsec:CI_lifetime_char}.
	\begin{breakablealgorithm}
		\caption{Generating interval-monitored SSALT data for meta-analysis.}
		\label{algo:data_generation}
		\begin{algorithmic}[1]
			\State {\textbf{Input:}}  Number of experiments $K$, parameter values $\alpha, \beta_0, \beta_1$, inspection intervals $I_k$, sample sizes $N_k$, standardized stress levels $x_{k,i}$, and inspection times $\tau_{k,i}$.
			\For{$k = 1, 2, \dots, K$}
			\State Initialize baseline cumulative hazard $H_{k,0} = 0$ and reliability $R_{k,0} = 1$.
			\For{$i = 1, 2, \dots, I_k$}
			\State Given the stress levels, determine $\lambda_{k,i} = \exp(\beta_0 + \beta_1 x_{k,i})$.
			\If{$i = 1$}
			\State Set $H_{k,i} = \lambda_{k,i} H_0(\tau_{k,i}; \alpha)$.
			\Else
			\State Set $H_{k,i} = H_{k,i-1} + \lambda_{k,i} [H_0(\tau_{k,i}; \alpha) - H_0(\tau_{k,i-1}; \alpha)]$.
			\EndIf
			\State Determine the reliability $R_{k,i} = \exp(-H_{k,i})$.
			\State Obtain the interval failure probability $D_{k,i} = R_{k,i-1} - R_{k,i}$. 
			\EndFor
			\State Set $D_{k,I_k+1} = R_{k,I_k}$ and determine the probability vector 
			
			\quad$\boldsymbol{p}_k = (D_{k,1}, D_{k,2}, \dots, D_{k,I_k+1})$.
			\State Associated with it, generate from $\text{Multinomial}(N_k; \boldsymbol{p}_k)$ to obtain the failure counts $(n_{k,1}, \newline n_{k,2}, \dots, n_{k,I_k+1})$.
			\EndFor
			\State \textbf{Output:} The complete observed interval-monitored SSALT data 
			
			$\mathcal{D}
			=
			\left\{
			n_{k,i},\;
			k=1,\ldots,K,\;
			i=1,\ldots,I_k, n_{k,I_k+1}
			\right\}.
			$
		\end{algorithmic}
	\end{breakablealgorithm}
	
	\begin{table*}[!t]
		\centering
		\caption{Simulation results under the Exponential, Weibull, and Burr XII
			distributions for different sample sizes, $N \in\{150,300,450\}$. The table reports the average estimate (AE), root mean squared error (RMSE),
			coverage probability (CP), and average length (AL) of the confidence
			intervals.}
		\scriptsize
		\label{tab:simulation_results_all}
		\setlength{\tabcolsep}{2.8pt}
		\renewcommand{\arraystretch}{1.25}
		
		\resizebox{\textwidth}{!}{%
			\begin{tabular}{llc|cccc|cccc|cccc}
				\toprule
				
				\multicolumn{15}{c}{Model Parameters}\\
				\midrule
				
				Distribution
				& Parameter & True
				& \multicolumn{4}{c|}{$N=150$}
				& \multicolumn{4}{c|}{$N=300$}
				& \multicolumn{4}{c}{$N=450$} \\
				
				\cmidrule(lr){4-7}
				\cmidrule(lr){8-11}
				\cmidrule(lr){12-15}
				
				& &
				& AE & RMSE & CP & AL
				& AE & RMSE & CP & AL
				& AE & RMSE & CP & AL \\
				
				\midrule

				\multirow{3}{*}{Exponential}
				& $\beta_0$ & $-2.000$
				& $-2.006$ & $0.215$ & $0.944$ & $0.812$
				& $-2.003$ & $0.148$ & $0.960$ & $0.573$
				& $-2.007$ & $0.118$ & $0.951$ & $0.468$ \\
				
				& $\beta_1$ & $3.000$
				& $3.007$ & $0.303$ & $0.942$ & $1.148$
				& $3.006$ & $0.210$ & $0.946$ & $0.809$
				& $3.009$ & $0.166$ & $0.951$ & $0.660$ \\
				
				\midrule

				\multirow{3}{*}{Weibull}
				& $\beta_0$ & $-3.000$
				& $-3.052$ & $0.271$ & $0.954$ & $1.067$
				& $-3.015$ & $0.186$ & $0.958$ & $0.747$
				& $-3.021$ & $0.154$ & $0.954$ & $0.609$ \\
				
				& $\alpha$ & $2.000$
				& $2.031$ & $0.283$ & $0.951$ & $1.073$
				& $2.026$ & $0.206$ & $0.932$ & $0.756$
				& $2.004$ & $0.156$ & $0.945$ & $0.614$ \\
				
				& $\beta_1$ & $3.000$
				& $3.047$ & $0.366$ & $0.957$ & $1.482$
				& $2.998$ & $0.258$ & $0.962$ & $1.042$
				& $3.026$ & $0.215$ & $0.953$ & $0.852$ \\
				
				\midrule
				
				\multirow{3}{*}{Burr XII}
				& $\beta_0$ & $-0.100$
				& $-0.107$ & $0.160$ & $0.951$ & $0.633$
				& $-0.103$ & $0.115$ & $0.952$ & $0.446$
				& $-0.105$ & $0.095$ & $0.950$ & $0.364$ \\
				
				& $\alpha$ & $3.500$
				& $3.581$ & $0.594$ & $0.971$ & $2.301$
				& $3.535$ & $0.413$ & $0.952$ & $1.599$
				& $3.532$ & $0.339$ & $0.949$ & $1.303$ \\
				
				& $\beta_1$ & $0.500$
				& $0.515$ & $0.367$ & $0.954$ & $1.445$
				& $0.508$ & $0.269$ & $0.952$ & $1.017$
				& $0.511$ & $0.224$ & $0.934$ & $0.828$ \\
				
				
				\midrule
				\multicolumn{15}{c}{reliability characteristics}\\
				\midrule
				
				Distribution
				& Measure & True
				& \multicolumn{4}{c|}{$N=150$}
				& \multicolumn{4}{c|}{$N=300$}
				& \multicolumn{4}{c}{$N=450$} \\
				
				\cmidrule(lr){4-7}
				\cmidrule(lr){8-11}
				\cmidrule(lr){12-15}
				
				& &
				& AE & RMSE & CP & AL
				& AE & RMSE & CP & AL
				& AE & RMSE & CP & AL \\
				
				\midrule

				\multirow{3}{*}{Exponential}
				& $R_{\text{NOC}}(t_M)$ & $0.623$
				& $0.621$ & $0.062$ & $0.950$ & $0.233$
				& $0.622$ & $0.043$ & $0.961$ & $0.166$
				& $0.624$ & $0.035$ & $0.951$ & $0.136$ \\
				
				& MTTF & $7.389$
				& $7.611$ & $1.709$ & $0.944$ & $6.406$
				& $7.497$ & $1.152$ & $0.960$ & $4.371$
				& $7.492$ & $0.901$ & $0.951$ & $3.547$ \\
				
				& Median lifetime & $5.122$
				& $5.275$ & $1.184$ & $0.944$ & $4.441$
				& $5.196$ & $0.798$ & $0.960$ & $3.030$
				& $5.193$ & $0.624$ & $0.951$ & $2.458$ \\
				
				\midrule

				\multirow{3}{*}{Weibull}
				& $R_{\text{NOC}}(t_M)$ & $0.543$
				& $0.541$ & $0.113$ & $0.960$ & $0.417$
				& $0.534$ & $0.083$ & $0.965$ & $0.311$
				& $0.546$ & $0.065$ & $0.952$ & $0.255$ \\
				
				& MTTF & $3.972$
				& $4.184$ & $1.012$ & $0.956$ & $3.842$
				& $4.018$ & $0.611$ & $0.932$ & $2.376$
				& $4.062$ & $0.503$ & $0.956$ & $1.956$ \\
				
				& Median lifetime & $3.731$
				& $3.896$ & $0.809$ & $0.952$ & $3.118$
				& $3.764$ & $0.505$ & $0.939$ & $1.978$
				& $3.804$ & $0.418$ & $0.953$ & $1.633$ \\
				
				\midrule

				\multirow{3}{*}{Burr XII}
				& $R_{\text{NOC}}(t_M)$ & $0.321$
				& $0.321$ & $0.065$ & $0.954$ & $0.256$
				& $0.321$ & $0.048$ & $0.947$ & $0.184$
				& $0.322$ & $0.040$ & $0.948$ & $0.151$ \\
				
				& MTTF & $1.223$
				& $1.262$ & $0.202$ & $0.944$ & $0.748$
				& $1.243$ & $0.126$ & $0.944$ & $0.464$
				& $1.237$ & $0.100$ & $0.950$ & $0.366$ \\
				
				& Median lifetime & $1.041$
				& $1.049$ & $0.071$ & $0.955$ & $0.283$
				& $1.045$ & $0.051$ & $0.954$ & $0.196$
				& $1.045$ & $0.042$ & $0.949$ & $0.159$ \\
				
				\bottomrule
			\end{tabular}%
		}
	\end{table*}
	
\begin{figure}[!t]
		\centering
		\includegraphics[width=0.85\linewidth]{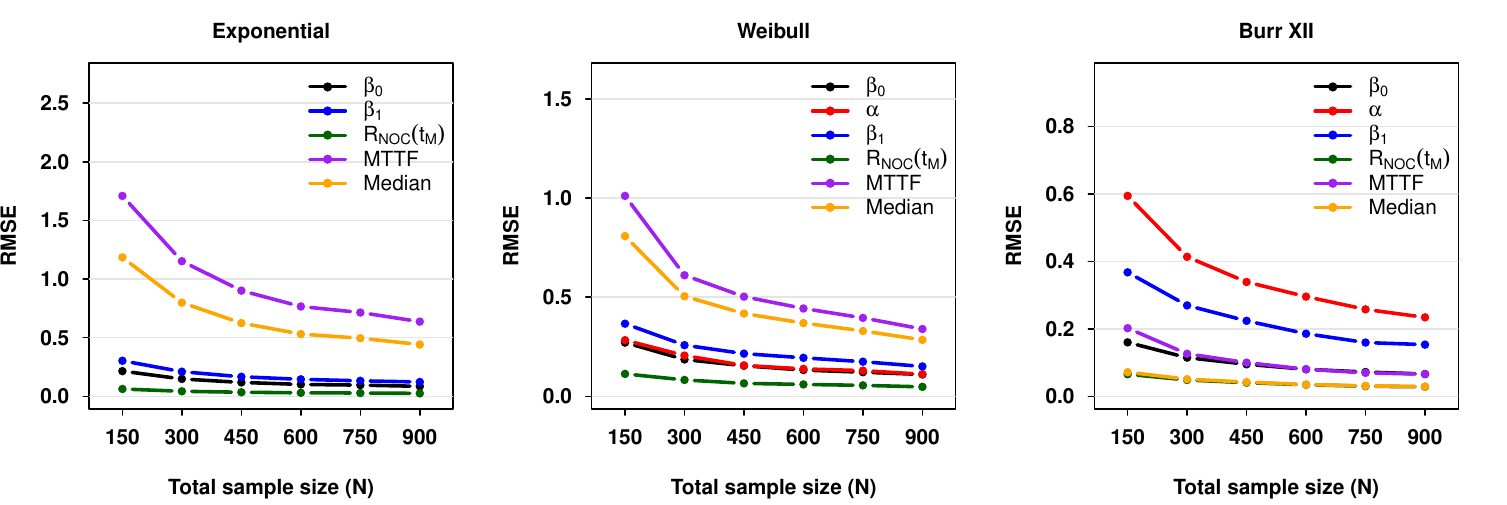}
		
		\vspace{2mm}
		\includegraphics[width=0.85\linewidth]{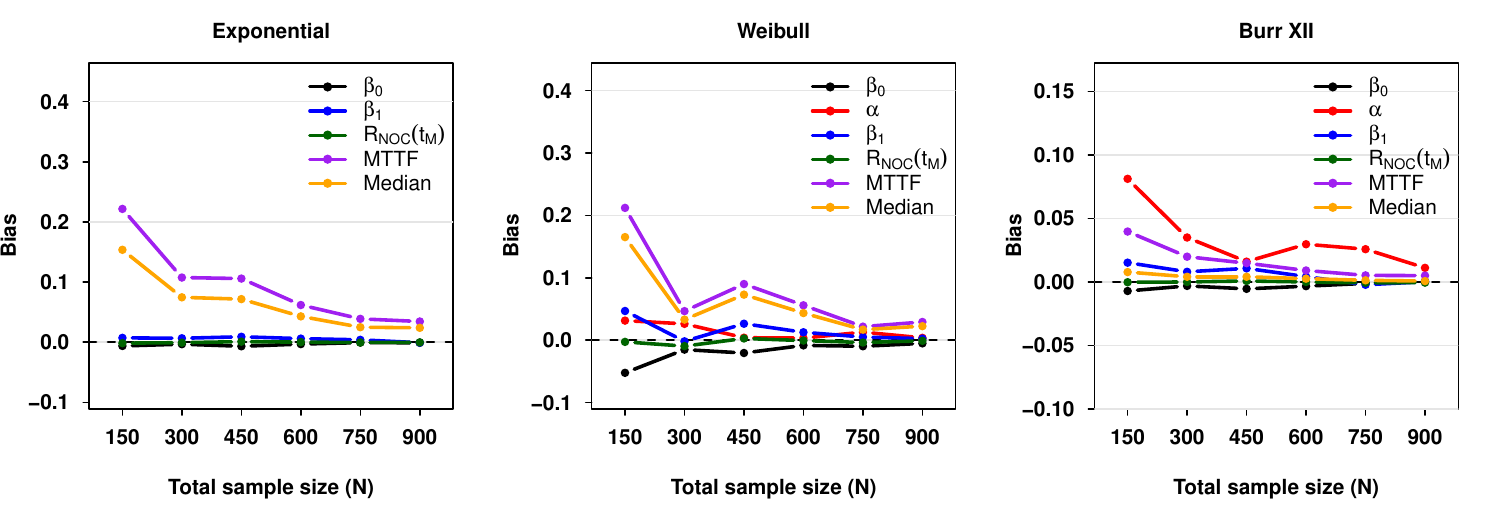}
		
		\caption{RMSE (top row) and Bias (bottom row) of the proposed estimators under varying total sample sizes for the Exponential, Weibull, and Burr XII models.}
		\label{fig:rmse_bias}
	\end{figure}

	The numerical results in Table  \ref{tab:simulation_results_all} indicate satisfactory performance of the estimation procedure across all considered sample sizes ($N=150$, $300$, and $450$) assuming each of the  three time-to-failure  distributions. For both the model parameters and the derived reliability characteristics, the AEs remain consistently close to their respective true values. Furthermore, as the total sample size increases, the RMSEs exhibit a steady decline, reflecting the expected gain in estimation precision. Figure \ref{fig:rmse_bias} provides further insight into the behavior of the RMSEs and biases across the considered sample sizes and lifetime distributions. Regarding interval estimation, the CPs of the asymptotic 95\% CIs are maintained close to the nominal level, the Monte Carlo band for the CP is approximately $[0.936,\,0.964]$, and nearly all CPs fall within it. Consistent with asymptotic theory, the associated ALs of these intervals decrease systematically as the amount of information increases due to an increase in $N$. This reduction in RMSE and interval length is clearly observable not only for the model parameters but also for the nonlinear functions like MTTF and median lifetime. A slight upward bias of the MTTF estimator is observed  at $N=150$ for both the exponential and the Weibull model, which diminishes as $N$ increases.
	
	From a practical point of view, the precision of the reliability characteristics at NOC depends strongly on the sample size. For the exponential and Weibull models, with $50$ units per test ($N=150$), the 95\% CI of the MTTF is almost as wide as the MTTF itself, whereas with $150$ units per test ($N=450$) its width falls to about half of the MTTF. 
	
	\begin{figure}[!t]
		\centering
		\subfloat[$N=150$]{\includegraphics[width=0.5\textwidth, height=0.45\textwidth]{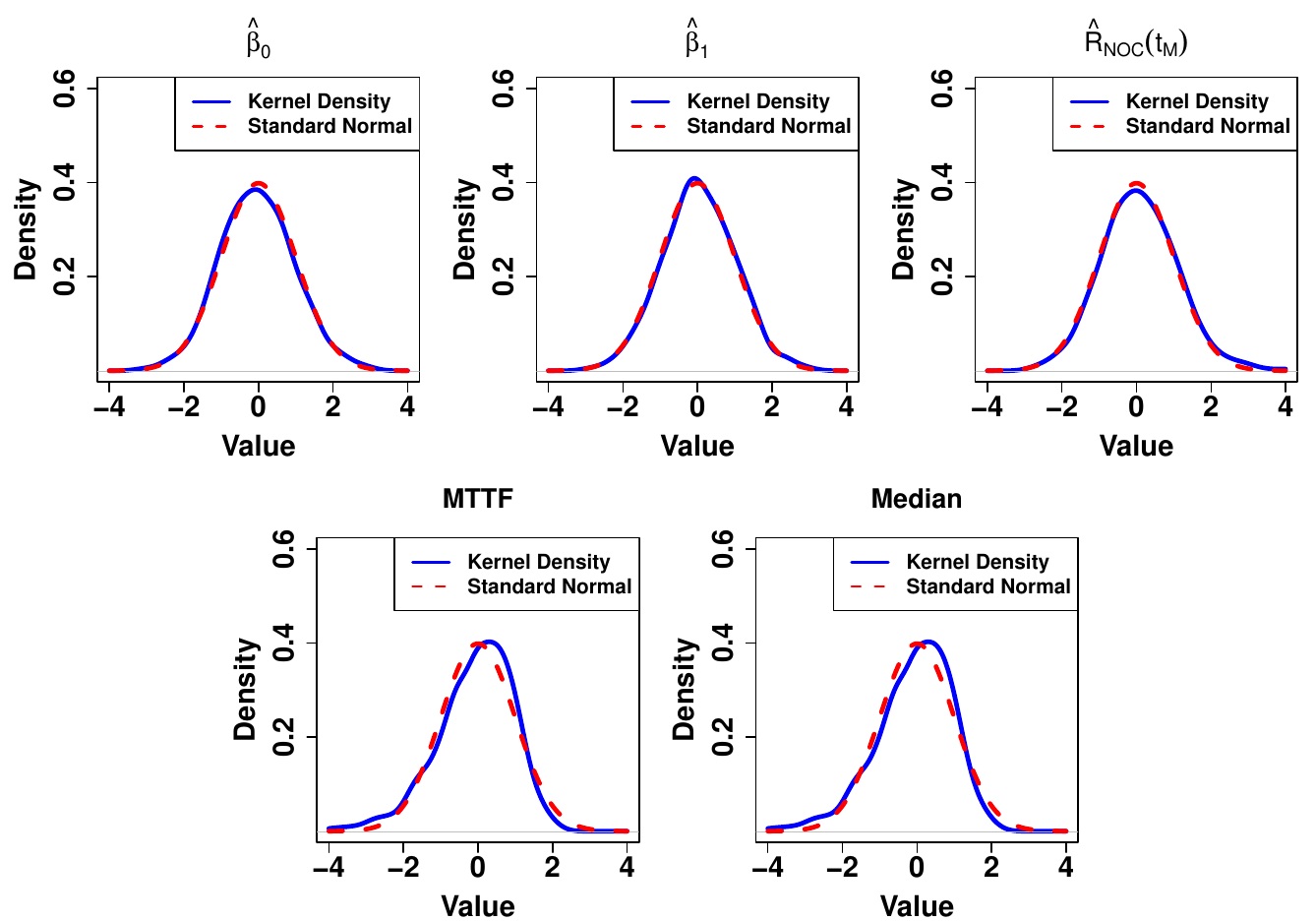}\label{fig:expo_kernel_a}}
		\hfil
		\subfloat[$N=300$]{\includegraphics[width=0.5\textwidth, height=0.45\textwidth]{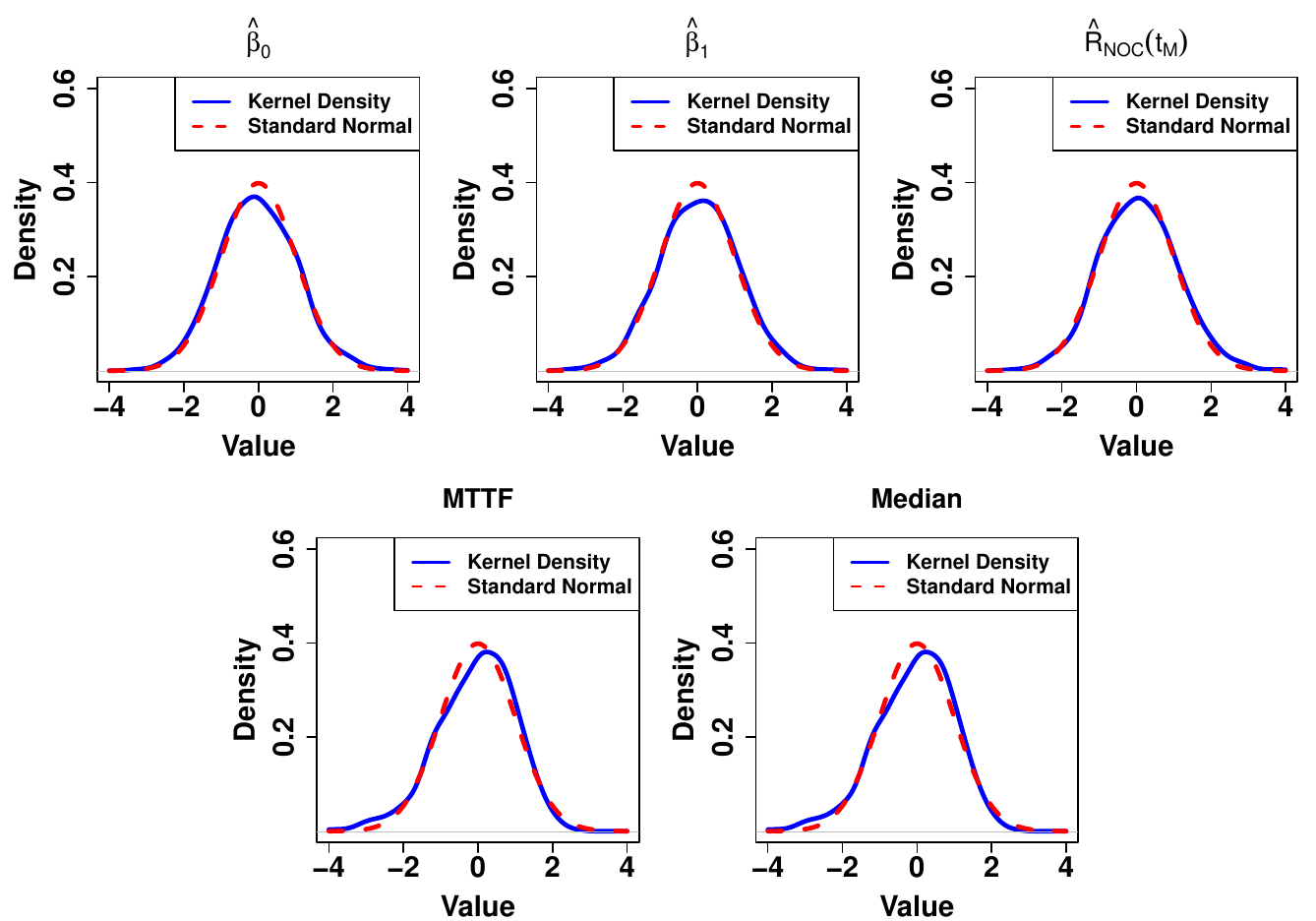}\label{fig:expo_kernel_b}}
		\caption{Kernel density plots of the standardized estimators with the standard normal curve under the exponential distribution for sample sizes (a) $N=150$ and (b) $N=300$.}
		\label{fig:expo_kernel}
	\end{figure}
	
	\begin{figure}[!t]
		\centering
		\subfloat[$N=150$]{\includegraphics[width=0.5\textwidth, height=0.45\textwidth]{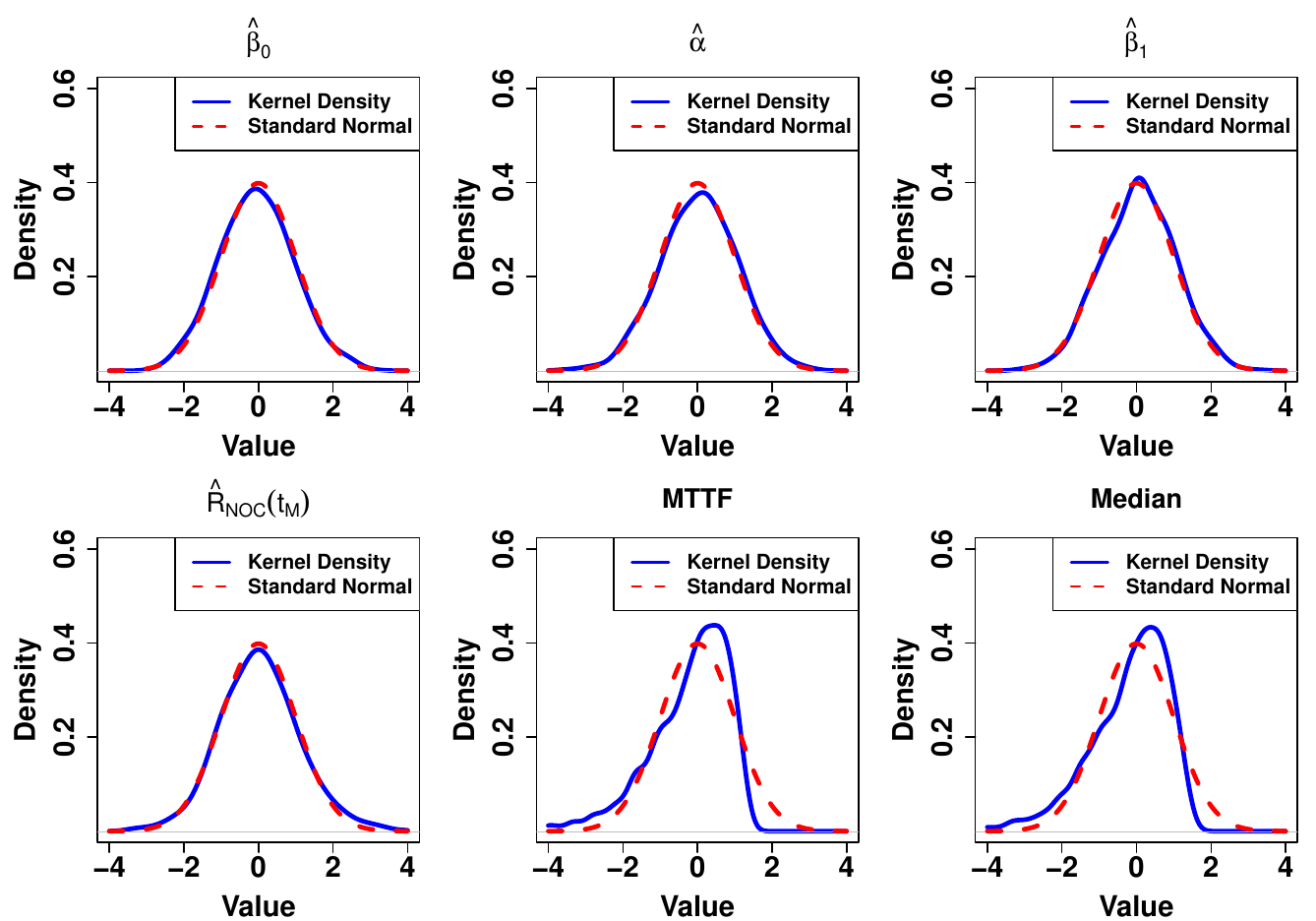}\label{fig:weibull_kernel_a}}
		\hfil
		\subfloat[$N=300$]{\includegraphics[width=0.5\textwidth, height=0.45\textwidth]{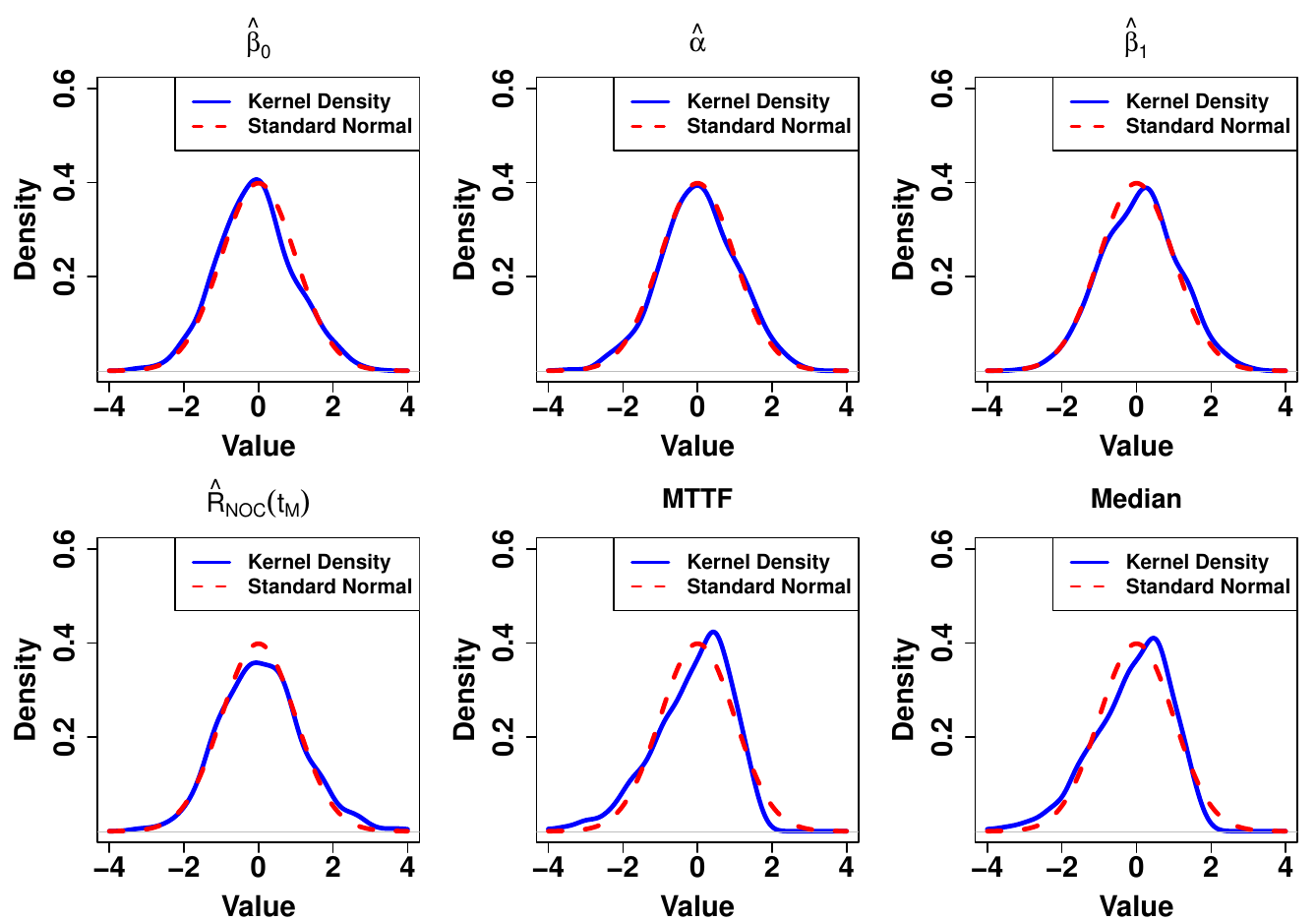}\label{fig:weibull_kernel_b}}
		\caption{Kernel density plots of the standardized estimators with the standard normal curve under the Weibull distribution for sample sizes (a) $N=150$ and (b) $N=300$.}
		\label{fig:weibull_kernel}
	\end{figure}
	
	The normal approximation to the sampling distributions of the MLEs  is further examined in Figures \ref{fig:expo_kernel} and \ref{fig:weibull_kernel} (corresponding to exponential and Weibull time-to-failure distributions)  through the kernel density estimates of the standardized estimators of \(\hat{\beta}_0\), \(\hat{\alpha}\) (Weibull only), \(\hat{\beta}_1\), \(\hat{R}_{\text{NOC}}(t_M)\), MTTF, and the median lifetime, together with the standard normal density. The standardized estimator is defined as $Z{\hat{\eta}} = \frac{\hat{\eta}-\eta} {\widehat{\operatorname{SE}}(\hat{\eta})},$ where $\hat{\eta}$ denotes the estimator of the corresponding model parameter or reliability characteristic $\eta$, and $\widehat{\operatorname{SE}}(\hat{\eta})$ is its estimated standard error. For \(N=150\), the empirical densities of the standardized model-parameter estimators are  reasonably close to the standard normal curve, although some deviations are observed particularly for the nonlinear reliability characteristics. The standardized MTTF and median lifetime estimators exhibit comparatively greater deviations from normality at the smaller sample size, which is consistent with their nonlinear dependence on the underlying model parameters. The agreement improves noticeably when the total sample size is increased to \(N=300\). The comparatively close agreement observed for the standardized estimator of $\alpha$ is also consistent with its estimation being based on information pooled across the three experiments (Figure \ref{fig:weibull_kernel}), since the Weibull shape parameter is common to the experiments. The corresponding plots under the Burr XII distribution are  provided in Figure S1 of the Supplementary Material and the plots show a similar pattern. The results therefore suggest that the proposed inferential procedure achieves a satisfactory normal approximation for moderate sample sizes, with the approximation improving as the sample size increases. In practice, this supports the use of the asymptotic CIs for the model parameters, while for the reliability characteristics at small sample sizes, the transformed intervals of Section~\ref{subsec:CI_lifetime_char} are preferable.
	\FloatBarrier
	
	\FloatBarrier
	\section{Model misspecification analysis}
	\label{sec:misspecification}
	
	In this section, we investigate the consequences of model misspecification within the proposed framework. The  interval-monitored  SSALT data are generated from a certain  time-to-failure  distribution (referred to as the true model), whereas a different time-to-failure  distribution is fitted to the generated data (often  referred to as the misspecified model). The objective is then to assess the effect of such misspecification on the estimation of important characteristics of the unit's lifetime distribution. Four combinations of the true and misspecified models are considered: (a) Weibull-Burr XII, (b) Burr XII-Weibull, (c) Weibull-Chen, and (d) Chen-Weibull, where the first distribution in each pair denotes
	the true model and the second denotes the misspecified model. The Weibull distribution is chosen as the reference because it is the default lifetime model in many applications, whereas the Burr XII and Chen distributions allow upside-down bathtub-shaped and bathtub-shaped failure rates, respectively, which the Weibull model cannot capture.
	
	Following the approach of \cite{Khakifirooz2020}, the $q$-th quantile measure  under NOC is adopted as a characteristic measure of the unit's lifetime
	distribution. The analysis considers the lower, median, and upper
	quantiles corresponding to $q=0.10$, $0.50$, and $0.95$, respectively. For each simulation replication, the $q$-th quantile is obtained under both the true and misspecified models using \eqref{eq:noc_quantile}. Let $\widehat{t}_{q,l}^{(\mathrm{true})}$ and $\widehat{t}_{q,l}^{(\mathrm{mis})}$ denote the estimated $q$-th
	quantiles under the true and misspecified models, respectively, in the $l$-th simulation replication, $l=1,\ldots,B$. For each model, the average estimated quantile over the replications is defined as
	\begin{equation*}
		\overline{t}_{q}^{(\mathrm{true})}
		=
		\frac{1}{B}\sum_{l=1}^{B}
		\widehat{t}_{q,l}^{(\mathrm{true})},
		\qquad
		\overline{t}_{q}^{(\mathrm{mis})}
		=
		\frac{1}{B}\sum_{l=1}^{B}
		\widehat{t}_{q,l}^{(\mathrm{mis})}.
	\end{equation*}
	The empirical relative bias (RB), empirical relative variability (RV), and mean squared error (MSE) are used to assess the effect of model misspecification and are defined as
		\begin{align*}
			\text{RB} &=
			\frac{
				\overline{t}_{q}^{(\mathrm{mis})}
				-
				\overline{t}_{q}^{(\mathrm{true})}
			}{
				\overline{t}_{q}^{(\mathrm{true})}
			}, \\
			\text{RV} &=
			\frac{
				\displaystyle\sum_{l=1}^{B}
				\left[
				\widehat{t}_{q,l}^{(\mathrm{mis})}
				-
				\overline{t}_{q}^{(\mathrm{true})}
				\right]^2
			}{
				\displaystyle\sum_{l=1}^{B}
				\left[
				\widehat{t}_{q,l}^{(\mathrm{true})}
				-
				\overline{t}_{q}^{(\mathrm{true})}
				\right]^2
			}, \text{~and}
			\\
			\text{MSE} &=
			\frac{1}{B}
			\sum_{l=1}^{B}
			\left[
			\widehat{t}_{q,l}^{(\mathrm{mis})}
			-
			t_q^{(\mathrm{true})}
			\right]^2.
		\end{align*}
	\begin{table}[!t]
		\centering
		\caption{Simulation Results for the Estimated Quantiles Under Model Misspecification at $q=0.10$, $0.50$, and $0.95$, With True Parameter Values $\beta_0=-0.2$, $\alpha=2.5$, and $\beta_1=0.5$}
		\label{tab:misspecification}
		\renewcommand{\arraystretch}{1.15}
		\setlength{\tabcolsep}{3.5pt}
		\resizebox{\textwidth}{!}{%
			\begin{tabular}{@{}llc|ccccc|ccccc|ccccc@{}}
				\toprule
				\multirow{2}{*}{True model} &
				\multirow{2}{*}{Fitted model} &
				\multirow{2}{*}{$N$} &
				\multicolumn{5}{c|}{$q=0.10$} &
				\multicolumn{5}{c|}{$q=0.50$} &
				\multicolumn{5}{c}{$q=0.95$} \\
				\cmidrule(lr){4-8}\cmidrule(lr){9-13}\cmidrule(l){14-18}
				& &
				& $t_q^{(\mathrm{true})}$ & $\overline{\hat{t}}_q^{(\mathrm{mis})}$ & MSE & RB & RV
				& $t_q^{(\mathrm{true})}$ & $\overline{\hat{t}}_q^{(\mathrm{mis})}$ & MSE & RB & RV
				& $t_q^{(\mathrm{true})}$ & $\overline{\hat{t}}_q^{(\mathrm{mis})}$ & MSE & RB & RV \\
				\midrule
				\multirow{2}{*}{Weibull} & \multirow{2}{*}{Burr XII}
				& 150
				& 0.440 & 0.510 & 0.008 & 0.123 & 2.069
				& 0.936 & 0.924 & 0.006 & $-0.015$ & 0.638
				& 1.680 & 1.998 & 0.318 & 0.188 & 4.448 \\
				& & 300
				& 0.440 & 0.503 & 0.005 & 0.132 & 3.566
				& 0.936 & 0.919 & 0.003 & $-0.017$ & 0.645
				& 1.680 & 1.950 & 0.155 & 0.161 & 4.880 \\
				\midrule
				\multirow{2}{*}{Burr XII} & \multirow{2}{*}{Weibull}
				& 150
				& 0.452 & 0.331 & 0.018 & $-0.273$ & 4.218
				& 1.121 & 1.213 & 0.039 & 0.058 & 1.499
				& 4.277 & 3.388 & 1.705 & $-0.335$ & 0.393 \\
				& & 300
				& 0.452 & 0.330 & 0.017 & $-0.274$ & 7.808
				& 1.121 & 1.190 & 0.017 & 0.057 & 1.806
				& 4.277 & 3.321 & 1.321 & $-0.291$ & 0.775 \\
				\midrule
				\multirow{2}{*}{Weibull} & \multirow{2}{*}{Chen}
				& 150
				& 0.440 & 0.357 & 0.011 & $-0.208$ & 4.159
				& 0.936 & 0.986 & 0.020 & 0.044 & 2.176
				& 1.680 & 1.678 & 0.075 & 0.001 & 1.076 \\
				& & 300
				& 0.440 & 0.349 & 0.010 & $-0.219$ & 8.162
				& 0.936 & 0.988 & 0.011 & 0.051 & 2.668
				& 1.680 & 1.701 & 0.040 & 0.013 & 1.207 \\
				\midrule
				\multirow{2}{*}{Chen} & \multirow{2}{*}{Weibull}
				& 150
				& 0.430 & 0.451 & 0.003 & 0.038 & 0.647
				& 0.822 & 0.821 & 0.006 & $-0.016$ & 0.973
				& 1.188 & 1.322 & 0.084 & 0.091 & 1.303 \\
				& & 300
				& 0.430 & 0.450 & 0.002 & 0.039 & 0.731
				& 0.822 & 0.813 & 0.003 & $-0.019$ & 0.980
				& 1.188 & 1.291 & 0.030 & 0.077 & 2.975 \\
				\bottomrule
			\end{tabular}%
		}
	\end{table}
	For numerical illustration, we consider $\beta_0=-0.2$, $\alpha=2.5$, and $\beta_1=0.5$, with sample sizes $N=150$ and $300$ and $B=1000$ replications. Three experiments with different
	stress paths $x_{k,i}$ and inspection-time schedules $\tau_{k,i}$ are considered. The respective stress paths are $(0.20,0.55,1.00)$, $(0.25,0.60,1.00)$,
	and $(0.30,0.65,1.00)$, while the corresponding inspection times are $(0.5,1.0,1.5)$, $(1.0,1.5,2.0)$, and $(2.0,2.5,3.0)$.
	The quantiles are evaluated at $q=0.10$, $0.50$, and $0.95$ under the model-misspecification settings described above. The corresponding true and estimated quantiles of the fitted model, together with the MSE, empirical RB, and empirical RV, are reported in Table  \ref{tab:misspecification}.
	
	\begin{figure}[!t]
		\centering
		
		\includegraphics[width=1\linewidth]{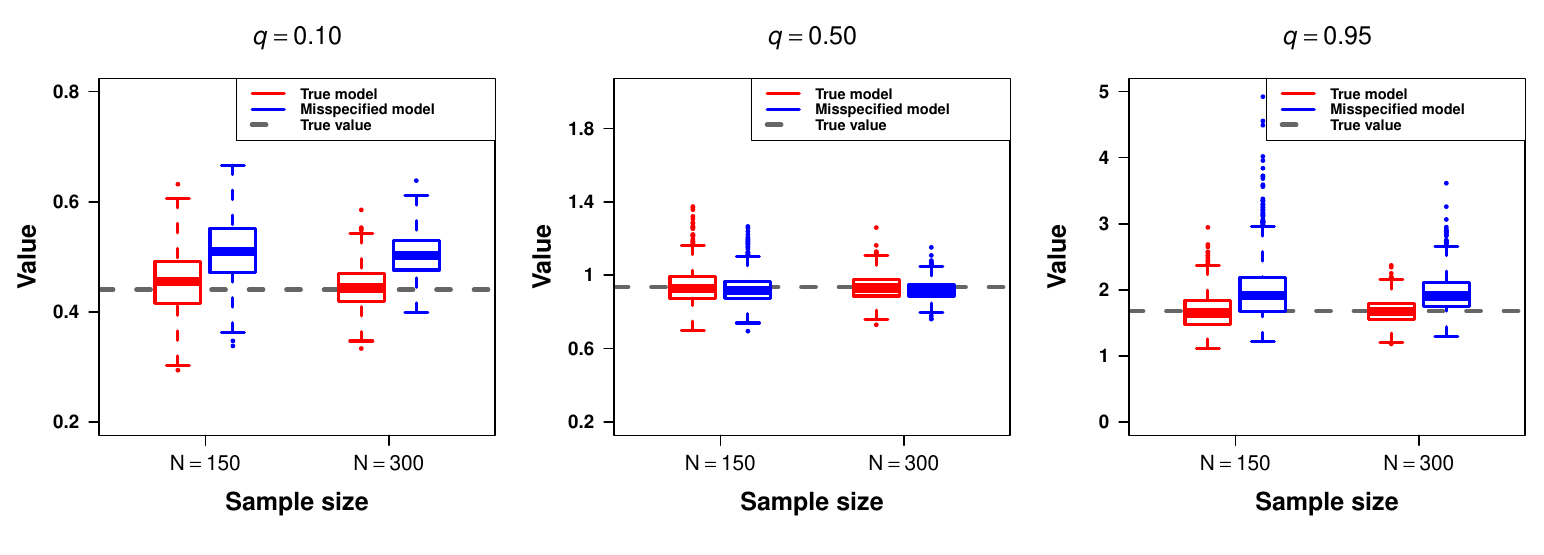}
		
		\includegraphics[width=1\linewidth]{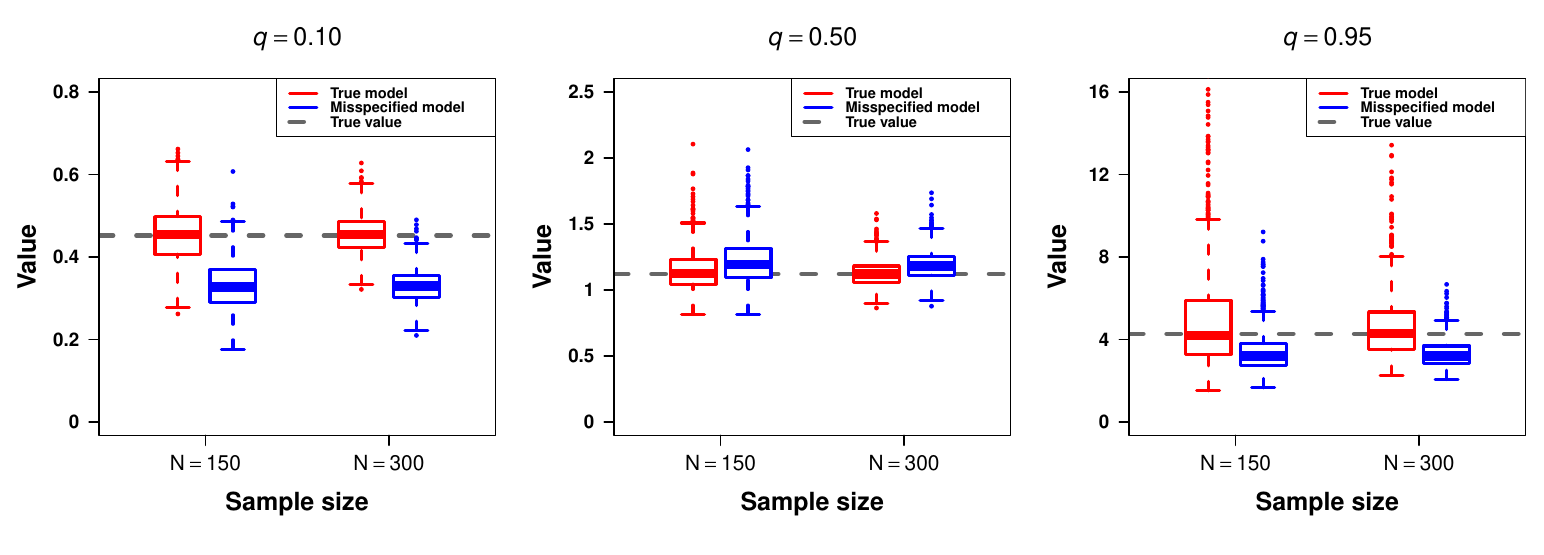}
		\caption{Boxplots of the estimated $q$-th quantiles at $q=0.10$, $0.50$, and $0.95$ for sample sizes $N=150$ and $N=300$: (a) Weibull as the true model and Burr XII as the misspecified model (top); (b) Burr XII as the true model and Weibull as the misspecified model (bottom).}
		\label{fig:Boxplots_1}
	\end{figure}
	
	\begin{figure}[!t]
		\centering
		
		\includegraphics[width=1\linewidth]{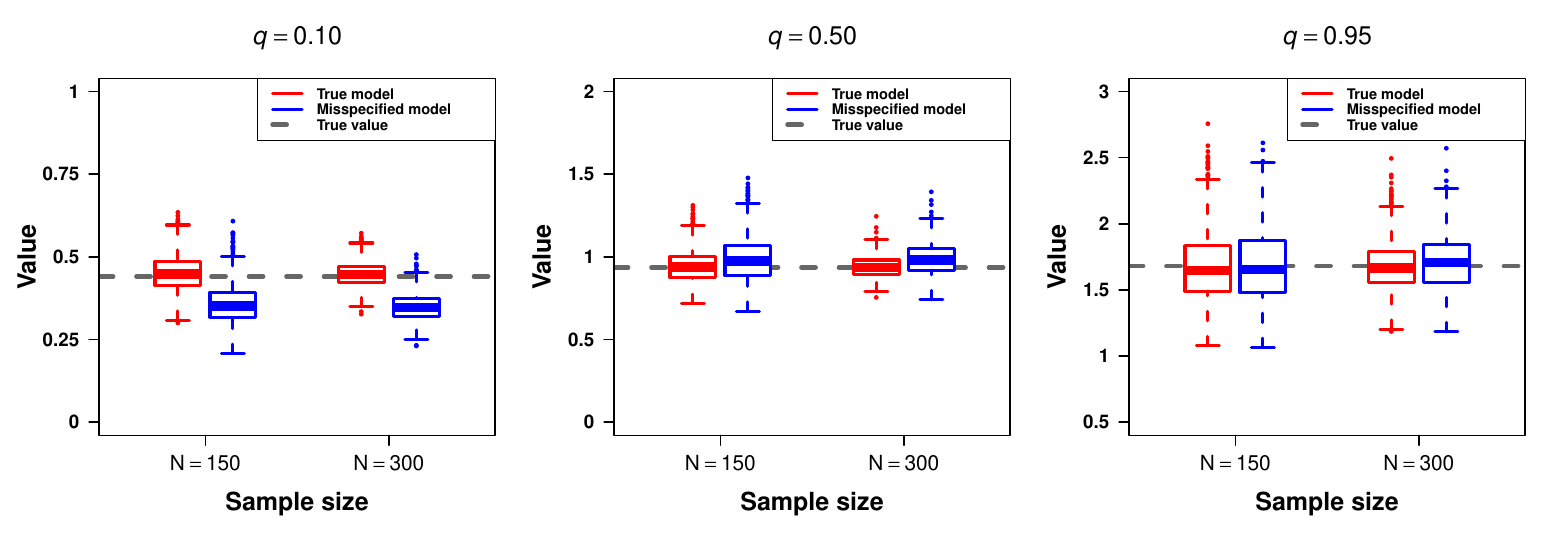}
		
		\includegraphics[width=1\linewidth]{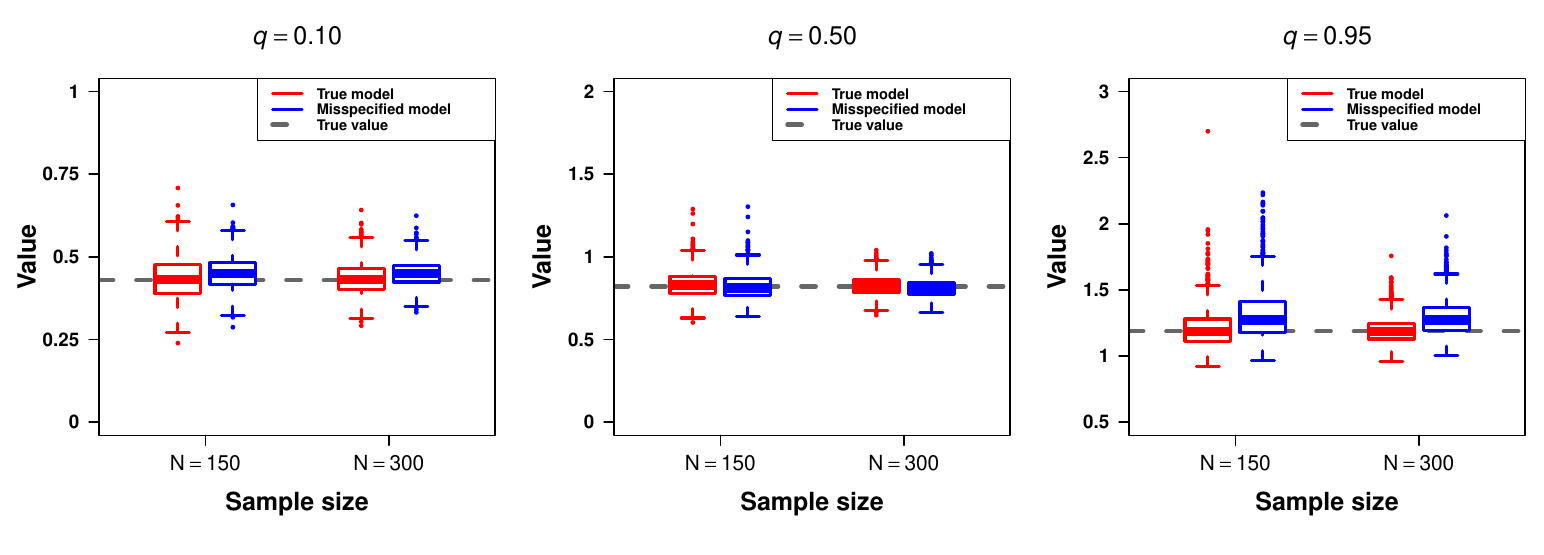}
		\caption{Boxplots of the estimated $q$-th quantiles at $q=0.10$, $0.50$, and $0.95$ for sample sizes $N=150$ and $N=300$: (a) Weibull as the true model and Chen as the misspecified model (top); (b) Chen as the true model and Weibull as the misspecified model (bottom).}
		
		\label{fig:Boxplots_2}
	\end{figure}

	Several important patterns emerge from the quantile-based model-misspecification analysis. The median is hardly affected by misspecification: its relative bias stays within about $6\%$ in all four settings, whereas in the tails it reaches about $34\%$. When the Weibull distribution is the true model and the Burr XII distribution is fitted, the lower and upper quantiles tend to be overestimated, whereas the median remains relatively close to the true value. For instance, at $N=150$, the estimated $q=0.95$ quantile is $1.998$, compared with the true value $1.680$. In practice, such overestimation at the upper tail could project an overly optimistic view of product longevity and affect warranty or maintenance planning. For example, a warranty period based on the $0.10$ quantile would be about $12\%$ longer than under the correctly specified model. Conversely, fitting a Weibull model to true Burr XII data results in underestimation at both the lower and upper quantiles, with the discrepancy more evident in the upper tail (e.g., an estimated $q=0.95$ of $3.388$ against the true $4.277$ at $N=150$). This may result in shorter estimated product lifetimes and consequently influence decisions related to replacement and maintenance. In this case, a warranty period based on the $0.10$ quantile would be about $27\%$ shorter than under the correctly specified model. For the Weibull-Chen combination, the main discrepancy is observed at the lower quantile, where the misspecified Chen model underestimates the true value, while the estimates at the upper quantile remain very close to the corresponding true value, with an RB of only $0.001$ at $q=0.95$ for $N=150$. In contrast, when the Chen distribution is the true model and the Weibull distribution is fitted, the estimated quantiles remain relatively close to their true values across the considered quantiles, indicating comparatively limited distortion. The boxplots in Figures \ref{fig:Boxplots_1} and~\ref{fig:Boxplots_2} further illustrate these differences through the location and spread of the estimated quantiles relative to their true values.
	
	The RV values need to be read with some care. Because the numerator of RV measures deviations from the mean of the correctly specified estimates, it picks up the bias as well as the spread, so it is largest where the bias is large; the maximum, $8.162$, occurs for the Weibull-Chen case at $q=0.10$ and $N=300$. Although increasing the sample size generally reduces the MSE and improves estimation precision, it does not necessarily remove the effects of an incorrect distributional assumption. Indeed, the RB hardly changes when $N$ increases from $150$ to $300$. Overall, the results emphasize that the consequences of model misspecification vary across the lifetime distribution, and both the accuracy and variability of the estimated quantiles should be considered when selecting an appropriate lifetime model under the proposed meta-analysis framework. In particular, if a decision rests on a tail quantile, such as a warranty period based on the $0.10$ quantile, it is worth comparing several candidate distributions, for example through AIC, BIC, and goodness-of-fit tests, before reporting the result. 
	
	\FloatBarrier
	\section{Data analysis}
	\label{sec:data_analysis}
	
	\FloatBarrier
	
	\subsection{Real LED data illustration}
	\label{subsec:real_data}
	
	To demonstrate the proposed meta-analysis framework, we consider the LED failure-time data arising from two simultaneously conducted SSALT  experiments. The experiments were designed to investigate the effect of loading voltage on the lifetime distribution of light-emitting diodes (LEDs). In each experiment, \(30\) LEDs were tested under a controlled temperature of \(80^\circ\mathrm{C}\) and a relative
	humidity of \(70\%\mathrm{RH}\). The first set of LEDs was initially subjected to a voltage of \(7\mathrm{V}\), whereas the second set was initially subjected to \(7.5\mathrm{V}\). In both the  experiments, the voltage was increased to \(8\mathrm{V}\) at \(100\) minutes. The failure times of the individual test units were recorded, and each experiment was terminated after the \(28\)-th failure under a Type-II censoring scheme (see \cite{shyur1999,samanta2024led}). The complete failure-time data
	are presented in Table  \ref{tab:LED_failure_times}.
	
	\begin{table}[!t]
		\centering
		\caption{Failure-time data of LEDs obtained from the two SSALT experiments.}
		\label{tab:LED_failure_times}
		\footnotesize
		\setlength{\tabcolsep}{4pt}
		\renewcommand{\arraystretch}{1.2}
		\begin{tabular}{@{}cc>{\raggedright\arraybackslash}p{0.4\columnwidth}@{}}
			\toprule
			Experiment & Stress level & Observed failure times (min) \\
			\midrule
			I  & \(S_1(7\mathrm{V})\) &
			2, 2, 3, 5, 8, 8, 23, 26, 29, 35, 94 \\
			& \(S_2(8\mathrm{V})\) &
			189, 218, 233, 256, 364, 379, 414, 424, 542, 550, 672, 742, 830, 972, 1041, 1289, 1340 \\
			\midrule
			II & \(S_1(7.5\mathrm{V})\) &
			2, 2, 2, 2, 2, 2, 3, 3, 3, 5, 6, 7, 9, 43, 43, 97 \\
			& \(S_2(8\mathrm{V})\) &
			158, 235, 269, 269, 410, 500, 500, 550, 732, 970, 972, 1137 \\
			\bottomrule
		\end{tabular}
	\end{table}
	
	For the purpose of our analysis, the recorded failure times are divided by \(60\) to convert the time scale from minutes to hours.  Since voltage is the applied stress, the standardized stress levels are computed on the logarithmic scale, in line with the inverse power law, i.e., $x_{k,i}=(\ln s_{k,i}-\ln s_{\text{NOC}})/(\ln s_H-\ln s_{\text{NOC}})$. With \(s_{\text{NOC}}=7\mathrm{V}\) and \(s_H=8\mathrm{V}\), the stress levels \(7\mathrm{V}\), \(7.5\mathrm{V}\), and \(8\mathrm{V}\) correspond to \(x=0\), \(x=0.517\), and \(x=1\), respectively. Accordingly, the observed failure times are
	therefore converted into the corresponding numbers of failures occurring within successive inspection intervals. This provides an interval-based representation of the two step-stress experiments while retaining the information on the changing stress levels. The resulting data, together with the
	associated stress levels and standardized stress values, are presented in Table \ref{tab:LED_interval}.
	
	\begin{table}[!t]
		\centering
		\caption{Interval-monitored LED data used in the proposed meta-analysis setup}
		\label{tab:LED_interval}
		\footnotesize
		\setlength{\tabcolsep}{5pt}
		\renewcommand{\arraystretch}{1.15}
		\begin{tabular}{@{}ccccc@{}}
			\toprule
			\begin{tabular}[c]{@{}c@{}}Inspection\\ number\end{tabular} &
			\begin{tabular}[c]{@{}c@{}}Stress\\ (V)\end{tabular} &
			\begin{tabular}[c]{@{}c@{}}Standardized\\ stress\end{tabular} &
			\begin{tabular}[c]{@{}c@{}}Inspection\\ time (hrs)\end{tabular} &
			\begin{tabular}[c]{@{}c@{}}Failures\\ counts\end{tabular} \\
			\midrule
			\multicolumn{5}{c}{\textit{Experiment I: } \(7\mathrm{V}\rightarrow8\mathrm{V}\)} \\
			1 & 7.0 & 0   & 1.667  & 11 \\
			2 & 8.0 & 1   & 10.000 & 10 \\
			3 & 8.0 & 1   & 22.333 & 7  \\
			\midrule
			\multicolumn{5}{c}{\textit{Experiment II: } \(7.5\mathrm{V}\rightarrow8\mathrm{V}\)} \\
			1 & 7.5 & 0.517& 1.667  & 16 \\
			2 & 8.0 & 1   & 10.000 & 8  \\
			3 & 8.0 & 1   & 18.950 & 4  \\
			\bottomrule
		\end{tabular}
	\end{table}
	
	\begin{table}[!t]
		\centering
		\caption{MLEs, 95\% CIs (within brackets), and model fit statistics under the Weibull and Chen model assumptions for the LED data.}
		\label{tab:combined_real_results}
		\scriptsize
		\setlength{\tabcolsep}{5pt}
		\renewcommand{\arraystretch}{0.95}
		\begin{tabular}{@{}lcc@{}}
			\toprule
			Quantity & Weibull & Chen \\
			\midrule
			\multicolumn{3}{@{}l}{\textit{Parameter estimates}}\\
			$\beta_0$  &$-0.914$ $(-1.430,\,-0.399)$ & $-1.409$ $(-1.924,\,-0.893)$\\
			$\alpha$   &$0.504$ $(0.166,\,0.842)$ & $0.262$ $(0.129,\,0.395)$\\
			$\beta_1$  & $0.324$ $(-1.072,\,1.720)$ & $0.280$ $(-1.076,\,1.637)$\\
			\midrule
			\multicolumn{3}{@{}l}{\textit{reliability characteristics}}\\
			$R_{\text{NOC}}(t_M)$ & $0.406$ $(0.172,\,0.692)$ & $0.416$ $(0.196,\,0.676)$ \\
			MTTF       & $12.090$ $(0.789,\,185.338)$ & $8.941$ $(1.397,\,57.224)$\\
			\begin{tabular}[t]{@{}l@{}}Median\\ \end{tabular} & $2.965$ $(0.799,\,11.005)$ & $3.096$ $(0.838,\,11.446)$\\
			\midrule
			\multicolumn{3}{@{}l}{\textit{Model comparison}}\\
			MLL       & $-74.087$ & $-73.239$ \\
			AIC       & $154.174$ & $152.478$ \\
			BIC       & $160.457$ & $158.761$ \\
			$p$-value & $0.146$   & $0.382$   \\
			\bottomrule
		\end{tabular}
	\end{table}
	
	\begin{figure}[!t]
		\centering
		
		\begin{subfigure}[b]{0.32\textwidth}
			\centering
			\includegraphics[width=\textwidth]{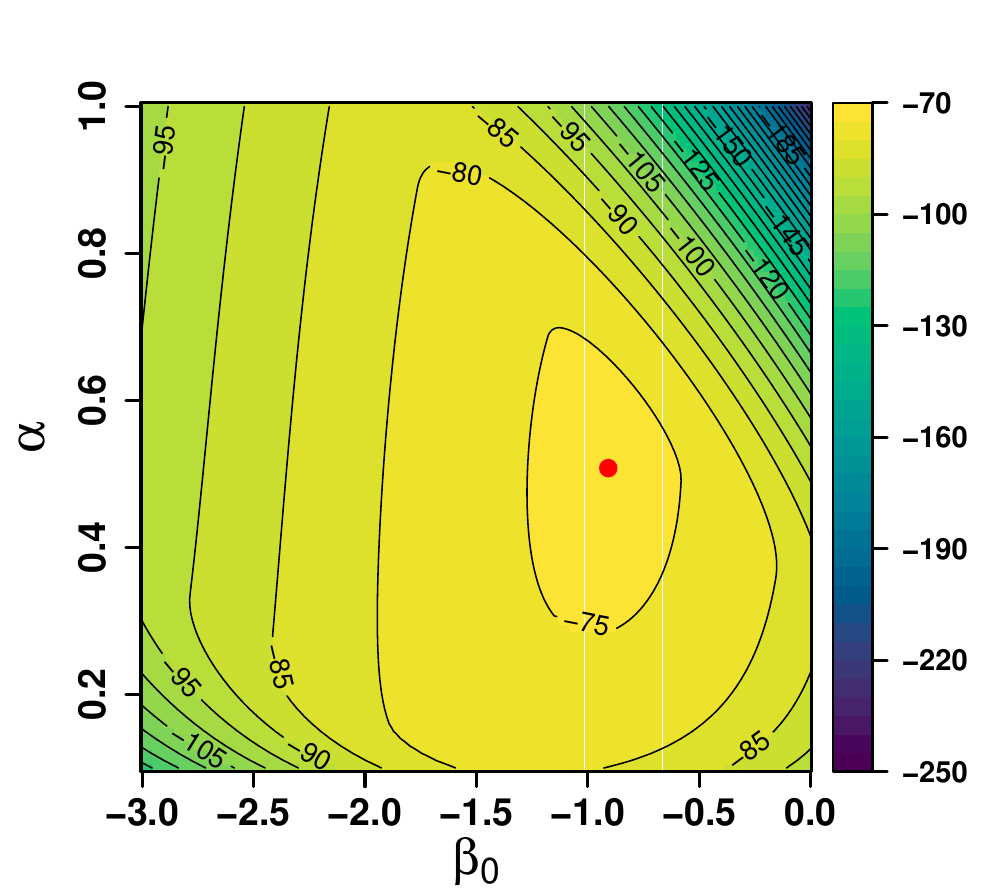}
			\caption{}
		\end{subfigure}
		\hfill
		\begin{subfigure}[b]{0.32\textwidth}
			\centering
			\includegraphics[width=\textwidth]{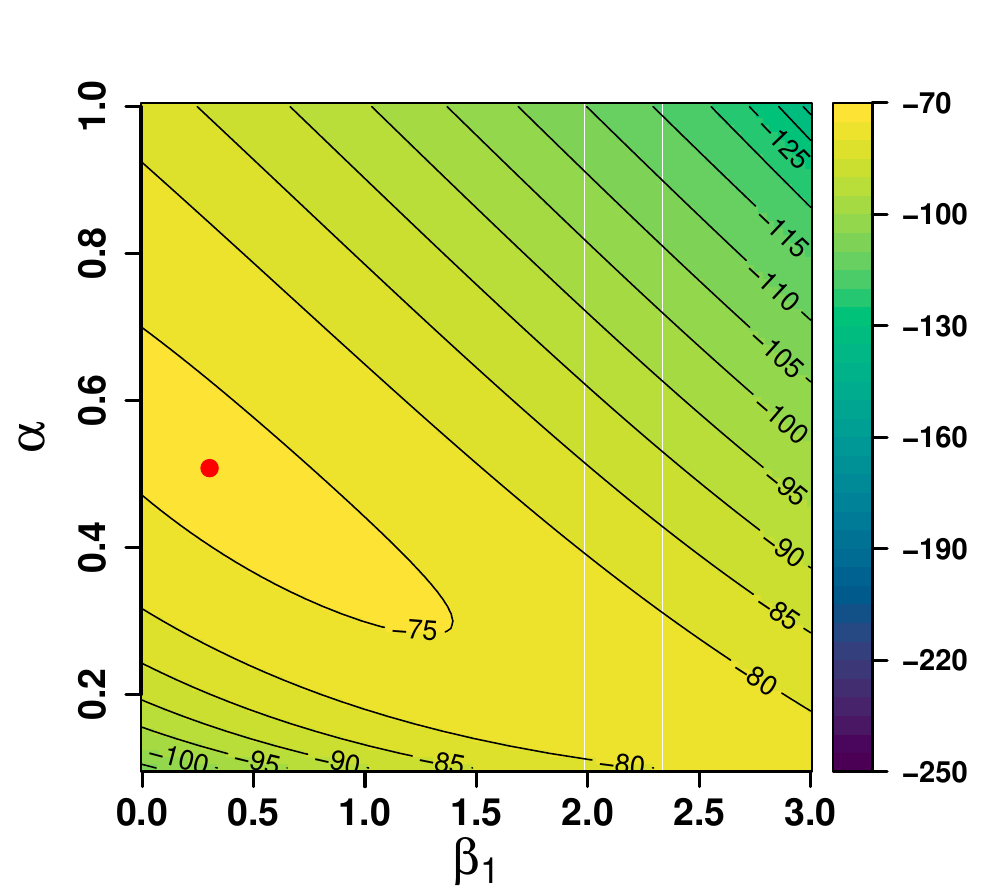}
			\caption{}
		\end{subfigure}
		\hfill
		\begin{subfigure}[b]{0.32\textwidth}
			\centering
			\includegraphics[width=\textwidth]{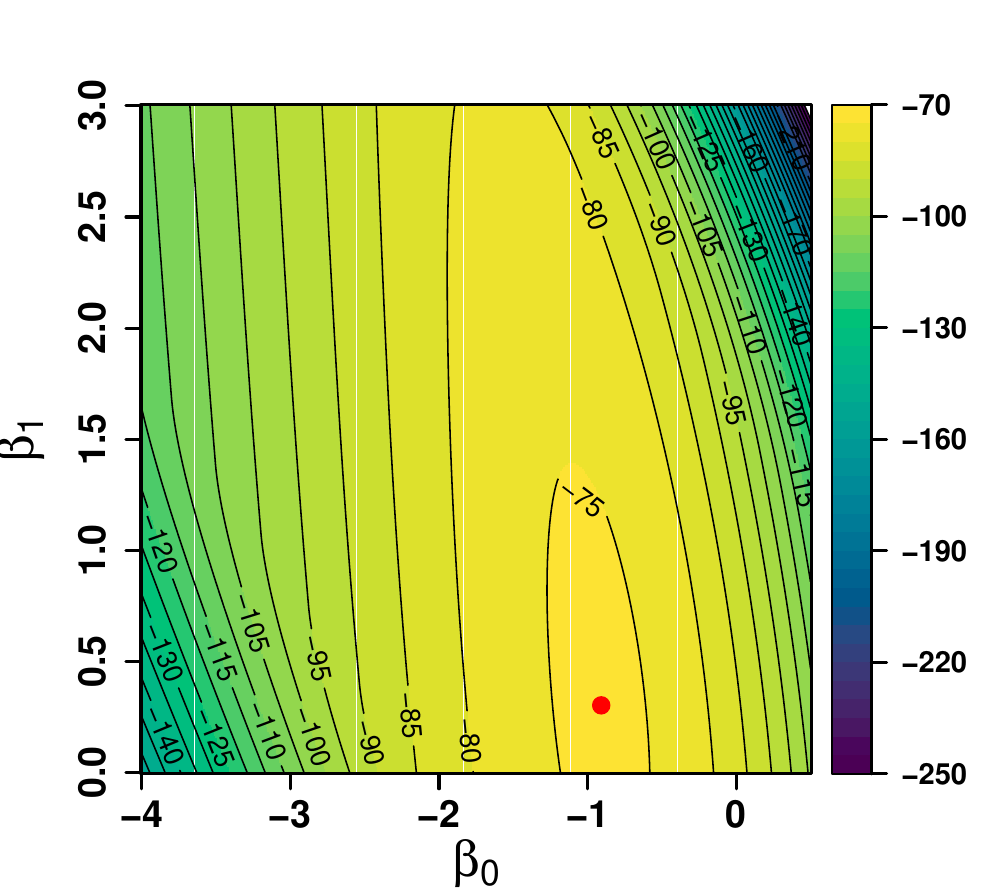}
			\caption{}
		\end{subfigure}
		
		\vspace{0.5cm}
		
		\begin{subfigure}[b]{0.32\textwidth}
			\centering
			\includegraphics[width=\textwidth]{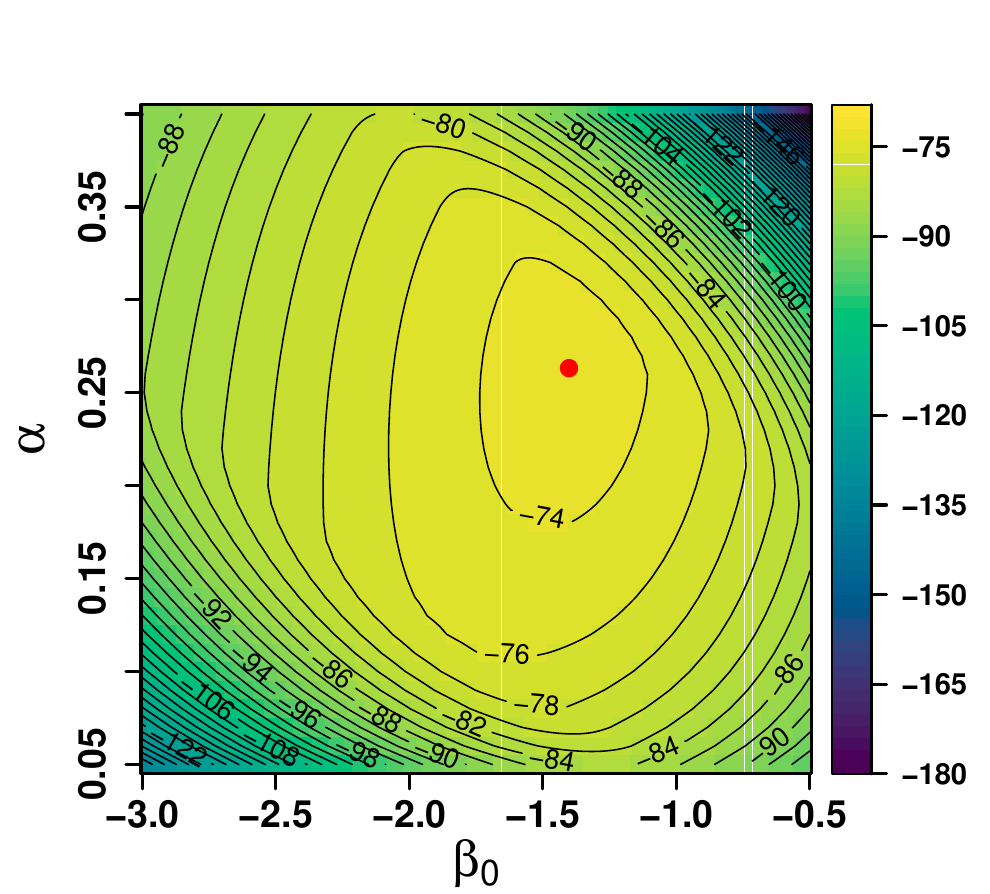}
			\caption{}
		\end{subfigure}
		\hfill
		\begin{subfigure}[b]{0.32\textwidth}
			\centering
			\includegraphics[width=\textwidth]{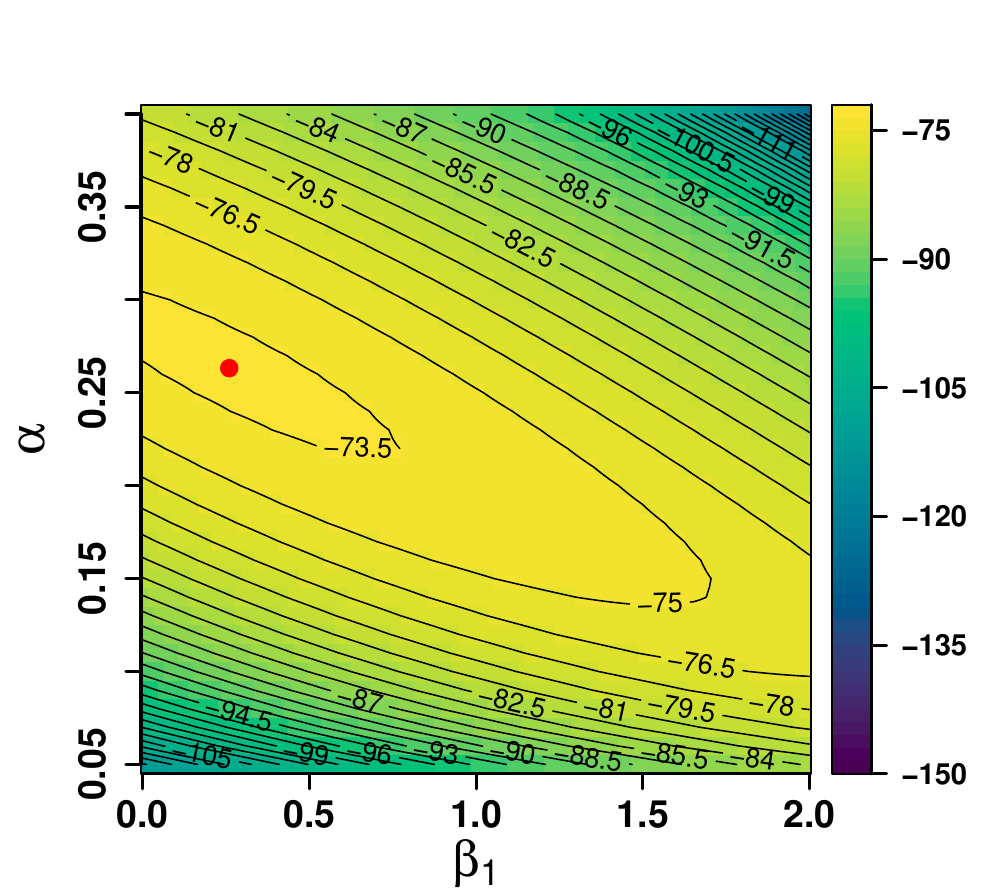}
			\caption{}
		\end{subfigure}
		\hfill
		\begin{subfigure}[b]{0.32\textwidth}
			\centering
			\includegraphics[width=\textwidth]{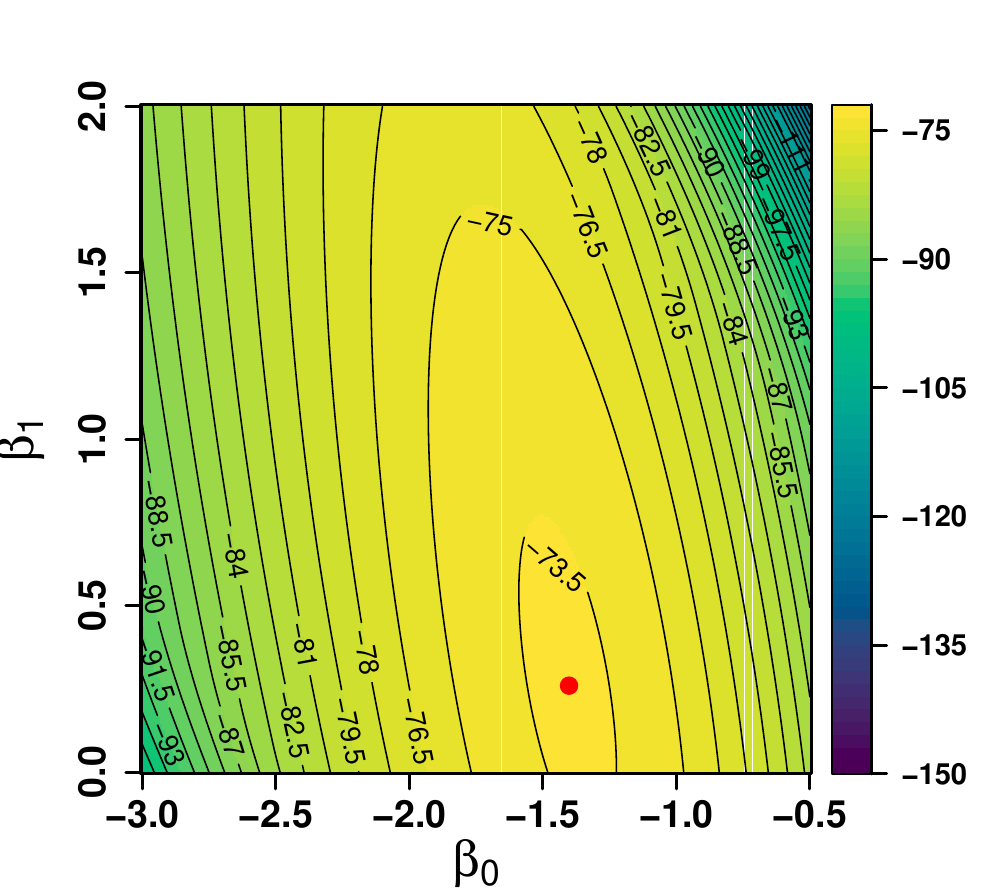}
			\caption{}
		\end{subfigure}
		
		\caption{Contour plots exhibiting the unimodality of the profile log-likelihood function for the Weibull (top row) and Chen (bottom row) models: (i) $\beta_0$ and $\alpha$, (ii) $\beta_1$ and $\alpha$, and (iii) $\beta_0$ and $\beta_1$. The red points indicate the maximum likelihood estimates.}
		\label{fig:LED_contours}
	\end{figure}
	
	The interval-monitored LED data used in the proposed meta-analysis setup are  analyzed using two important members of the PH class of distributions- (i) Weibull distribution, and (ii) Chen distribution. The contour plots of the profile log-likelihood functions for the Weibull and Chen models are provided  in Figure  \ref{fig:LED_contours}, with the red points denoting the corresponding MLEs. This graphical assessment supports the unimodality of the likelihood surface and provides a useful basis for choosing initial values for numerical maximization. The corresponding parameter estimates, reliability characteristics, and model-fit measures, along with the 95\% CIs obtained from the asymptotic normality of the MLEs  for the two time-to-failure distributions   are reported in Table \ref{tab:combined_real_results}. For the LED data, the estimated Weibull shape parameter ($\hat{\alpha}=0.504$) indicates a decreasing failure rate, whereas the Chen model with $\hat{\alpha}=0.262$ provides greater flexibility in describing the failure-rate behavior, including a possible bathtub-shaped pattern. Furthermore, both models yield a positive estimate of $\beta_1$, consistent with a higher failure intensity at higher voltage; however, the corresponding CIs include zero, so the stress effect is estimated with considerable uncertainty, owing to the narrow voltage range (7--8V) and the small number of test units. Regarding the reliability characteristics, both models provide broadly consistent estimates of reliability at the specified mission time of $5$ hours and median lifetime, while the MTTF is more sensitive to the assumed time-to-failure  distribution with estimates of $12.090$ hours for the Weibull model and $8.941$ hours for the Chen model. In terms of model fit, the Chen model achieves a higher maximized log-likelihood  value together with lower Akaike Information Criterion (AIC) and Bayesian Information Criterion (BIC) values than the Weibull model, indicating a better fit to the observed data. Nevertheless, the goodness-of-fit tests yield $p$-values of $0.146$ and $0.382$ for the Weibull and Chen models, respectively, providing no significant evidence of lack of fit for either distribution. Overall, both models provide an adequate representation of the observed LED failure data, while the Chen model offers a modest advantage in fit and greater flexibility in describing the failure-rate behavior.
	
	\subsubsection{Gain from pooling the two experiments}
	\label{subsubsec:pooling_gain}
	\begin{figure*}[!h]
		\centering
		\includegraphics[width=\textwidth]{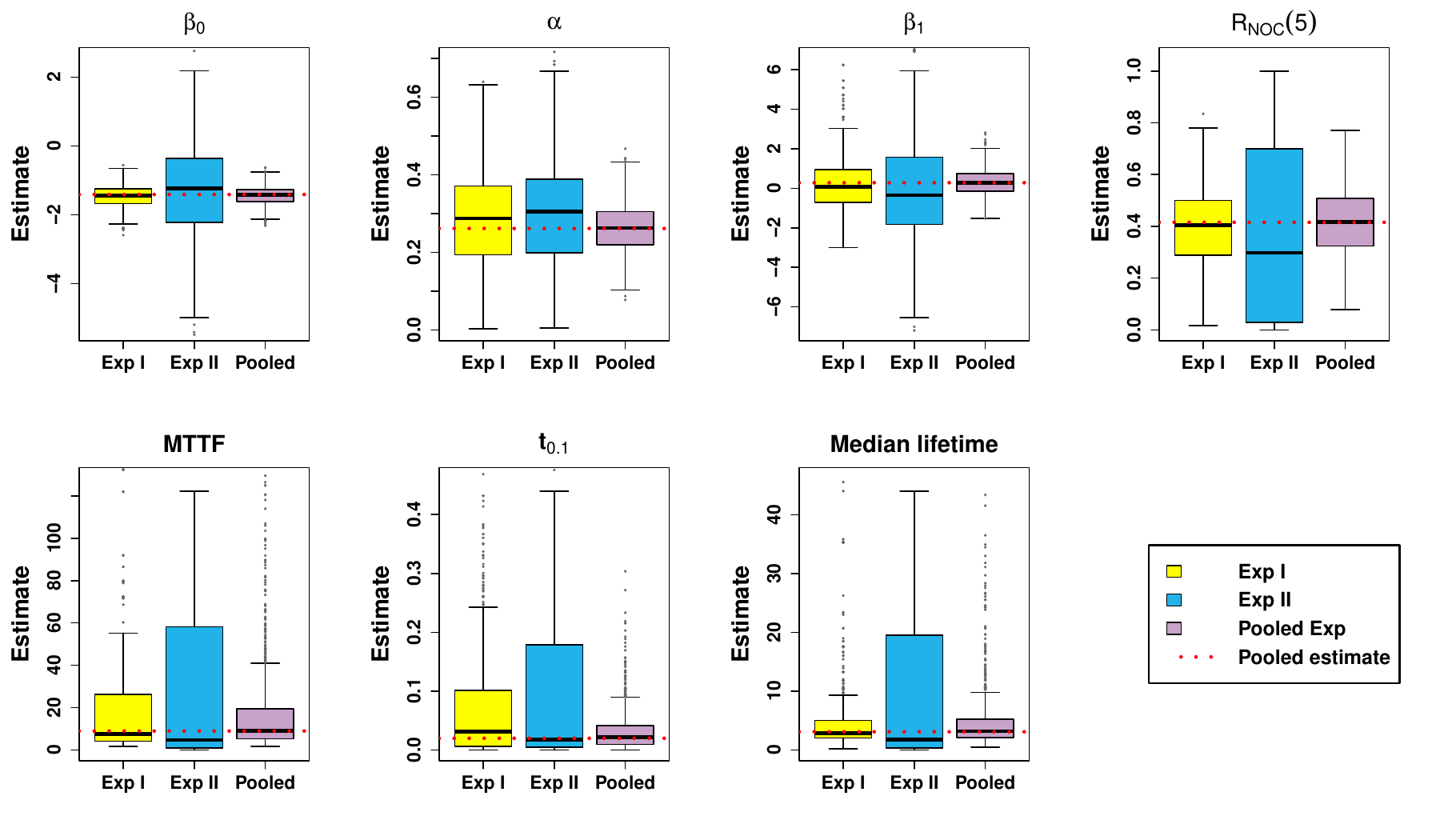}
		\caption{Bootstrap estimates for the LED data under the Chen model, obtained from Experiment I alone, Experiment II alone, and both experiments pooled ($1000$ replications). The dotted line marks the pooled estimate from the observed data. Some outlying estimates lie outside the plotted range: $15.1\%$, $21.6\%$, and $4.5\%$ for the MTTF; $1.1\%$, $18.0\%$, and $0\%$ for the $0.10$ quantile; and $5.9\%$, $19.9\%$, and $1.3\%$ for the median lifetime (Exp I, Exp II, and pooled, respectively).}
		\label{fig:LED_pooling}
	\end{figure*}
	
	To examine the benefit of combining the two LED experiments, a parametric bootstrap study is carried out under the Chen model, which gave the better fit. Using the pooled MLEs as reference values, $1000$ data sets are generated with the design of each experiment. The model is then fitted to each experiment separately (i.e., $K=1$) and to both experiments together, with standard errors obtained from the observed information matrix. All pooled fits are usable, whereas $8.4\%$ and $12.4\%$ of the fits to Experiments I and II alone are not, mostly because the estimated MTTF or lifetime quantiles become too large to be computed.
	
	Figure~\ref{fig:LED_pooling} shows that the pooled estimates are centred on the reference values and are much less spread out than those from either experiment alone. We focus on the reliability at the mission time and the $0.10$ quantile, which are the quantities most relevant for qualification and warranty decisions. For $R_{\text{NOC}}(5)$, pooling reduces the RMSE by $20.7\%$ and the AL by $22.2\%$ compared with Experiment I, while keeping the CP close to the nominal level ($0.958$, against $0.977$ for Experiment I). Experiment II alone is of little use here: its CI covers almost the whole interval $(0,1)$, with an AL of $0.990$. For the $0.10$ quantile, the RMSE is reduced by $65.8\%$ compared with Experiment I, and the CP improves from $0.911$ to $0.942$. Moreover, a few extremely large estimates of the MTTF and the lifetime quantiles occur in the single-experiment fits, which make their RMSE and AL very large; this does not happen under pooling. The complete results are provided in Table S1 of the Supplementary Material. A similar pattern is observed under the Weibull model (Figure~S4 of the Supplementary Material).  

	\FloatBarrier
	
	\FloatBarrier
	\subsection{Simulated data illustration}
	\label{subsec:simulated_data}
	
	In this subsection, we consider the  two simulated simple SSALT data by (i) \cite{xiong1998} and  (ii) \cite{katerimeta2009} to illustrate the proposed methodology.  These data sets have different sample sizes and censoring patterns thereby providing  a suitable setting for demonstrating the  usefulness of combining information from independent step-stress experiments.
	For data set $(i)$, the experiment consists of $N_1=20$ test units, with $r_1=16$ observed failures and a stress-changing time  point at $\tau_1=5$. The data were originally generated from exponential distribution under the simple SSALT model with the parameter setting $ \theta_1=e^{2.5},  \theta_2=e^{1.5},$ where $\theta_1$ and $\theta_2$ denote the corresponding stress-specific mean lifetimes. For the second experiment, 
	data set $(ii)$, consists of $N_2=35$ test units, with $r_2=26$ observed failures and the same stress-change point $\tau_2=5$. The failure-time observations corresponding to the two data sets are presented in Table  \ref{tab:simulated_failure_data}.
	
	\begin{table}[!t]
		\centering
		\caption{Simulated data: failure-time observations.}
		\label{tab:simulated_failure_data}
		\footnotesize
		\setlength{\tabcolsep}{4pt}
		\renewcommand{\arraystretch}{1.2}
		\begin{tabular}{@{}cccc>{\raggedright\arraybackslash}p{0.4\columnwidth}@{}}
			\toprule
			Dataset & $n$ & $r$ &
			$\tau$ & Failure times \\
			\midrule
			(i) & 20 & 16 & 5 &
			2.01, 3.60, 4.12, 4.34, 5.04, 5.94, 6.68, 7.09, 7.17, 7.49, 7.60, 8.23, 8.24, 8.25, 8.69, 12.05 \\
			\midrule
			(ii) & 35 & 26 & 5 &
			0.053, 0.22, 0.92, 1.58, 1.70, 2.55, 2.72, 3.12, 3.85, 4.51, 4.53, 4.90, 5.45, 5.48, 5.48, 5.70, 5.74, 5.98, 6.36, 7.26, 7.73, 7.74, 8.01, 9.54, 9.67, 9.89 \\
			\bottomrule
		\end{tabular}
	\end{table}
	
	To apply the proposed interval-monitoring framework, the original
	failure-time observations are grouped according to specified
	inspection times, and only the numbers of failures occurring within
	the successive inspection intervals are used in the likelihood.
	For data set $(i)$, the predetermined inspection times are $t$= $3$, $5$, $7$, and
	$12$. It gives the respective failure counts as  $1$, $3$,
	$3$, and $8$ with $5$ units surviving beyond the final
	inspection time. 
	For data set $(ii)$, the inspection times are $t$= $2$, $5$,
	$7$, and $10$. The corresponding interval failure counts are
	$5$, $7$, $7$, and $7$.
	
	\begin{table}[!t]
		\centering
		\caption{MLEs, 95\% CIs (within brackets),
			reliability characteristics, and model fit statistics
			under the exponential and Weibull baseline distributions.}
		\label{tab:combined_simulated_results}
		\footnotesize
		\setlength{\tabcolsep}{2pt}
		\renewcommand{\arraystretch}{1.15}
		\begin{tabular}{@{}lcc@{}}
			\toprule
			Quantity & Exponential & Weibull \\
			\midrule
			\multicolumn{3}{@{}l}{\textit{Parameter estimates}}\\
			$\beta_0$ & $-2.697$ $(-3.188,\,-2.207)$ & $-3.258$ $(-4.590,\,-1.927)$ \\
			$\alpha$  & $-$ & $1.361$ $(0.569,\,2.152)$ \\
			$\beta_1$ & $0.931$ $(0.299,\,1.562)$ & $0.459$ $(-0.672,\,1.591)$ \\
			\midrule
			\multicolumn{3}{@{}l}{\textit{Reliability characteristics}}\\
			$R_{\text{NOC}}(t_M)$  & $0.667$ $(0.525,\,0.785)$ & $0.644$ $(0.486,\,0.775)$ \\
			MTTF      & $14.838$ $(9.085,\,24.235)$ & $10.042$ $(5.051,\,19.963)$ \\
			Median    & $10.285$ $(6.297,\,16.799)$ & $8.377$ $(5.147,\,13.633)$ \\
			\midrule
			\multicolumn{3}{@{}l}{\textit{Model comparison}}\\
			MLL       & $-86.864$ & $-86.364$ \\
			AIC       & $177.728$ & $178.728$ \\
			BIC       & $181.743$ & $184.750$ \\
			$p$-value & $0.67$   & $0.64$ \\
			\bottomrule
		\end{tabular}
	\end{table}
	
	For analysis purpose, we consider the two important members of the PH class of distributions, the 
	exponential and Weibull distributions.  The contour plots of the log-likelihood function under the exponential model and Weibull model reaffirm  the unimodality of the likelihood (see Figures~S2 and S3 of the Supplementary Material in this regard). Table \ref{tab:combined_simulated_results} presents the point estimates and associated 95\% CIs for the model parameters along with the estimated reliability, reliability characteristics, and goodness-of-fit measures. Under both the models, the estimated stress-effect parameter $\beta_1$ is positive ($0.931$ for the exponential model and $0.459$ for the Weibull model), indicating that elevated stress levels are associated with increased failure intensity. Furthermore, the estimated Weibull shape parameter, $\hat{\alpha}=1.361$, indicates a monotonically increasing failure rate. At the mission time $t_M=6$, the estimated reliability is $0.667$ for the exponential model and $0.644$ for the Weibull model. The corresponding MTTF estimates are $14.838$ and $10.042$, while the median lifetime estimates are $10.285$ and $8.377$, respectively, showing that the choice of baseline distribution has a noticeable effect on the estimated lifetime measures. Among the candidate models considered, the exponential model yields AIC and BIC values of $177.728$ and $181.743$, respectively, compared with $178.728$ and $184.750$ for the Weibull model. Although the Weibull model attains a slightly higher maximized log-likelihood value of $-86.364$ than the exponential model ($-86.864$), the information criteria favor the exponential model after accounting for model complexity. The goodness-of-fit tests result in  $p$-values of $0.67$ and $0.64$ for the exponential and Weibull models, respectively, indicating no significant evidence of lack of fit for either model. Overall, both models provide an adequate fit to the simulated data, with the exponential model preferred according to the information criteria.
	
	\FloatBarrier
	
	\section{Concluding remarks}
	\label{sec:conclusion}
	
	We have proposed a way to analyze several interval-monitored SSALT experiments together, such as tests of the same product carried out in different laboratories or test campaigns, even when they use different stress levels, stress-change times, and inspection schedules. Under the FRM, each experiment contributes its own multinomial likelihood, while all experiments share the baseline distribution and the stress--life parameters. Because the score equations and the Fisher information depend on the baseline distribution only through $H_0(t;\alpha)$ and its derivative with respect to $\alpha$, the same expressions apply to any member of the PH family. The exponential, Weibull, and Burr XII distributions were worked out as special cases, and the Chen distribution was used in the misspecification study and in the LED analysis. The proposed methodology provides estimates and CIs for the model parameters and for the reliability characteristics at NOC, namely the reliability at a mission time, the MTTF, and lifetime quantiles.
	
	In the simulations, the estimators were nearly unbiased and the CIs had close to nominal coverage, but precision depended strongly on the number of units. With $50$ units per test, the 95\% CI of the MTTF was about as wide as the MTTF itself; with $150$ units per test, it was about half as wide. Misspecifying the lifetime distribution shifted the median by no more than about $6\%$, but shifted the $0.10$ and $0.95$ quantiles by up to a third, and more data did not remove this error. When a warranty period or a replacement policy depends on a tail quantile, the choice of lifetime model therefore matters as much as the size of the test. For the LED data, combining the two tests reduced the RMSE of the estimated reliability at $5$ hours by about $20\%$ and that of the $0.10$ quantile by about 66\%, compared with the better single test, while keeping the coverage probability close to the nominal level. Both the Weibull and the Chen models fitted these data adequately, with the Chen model preferred by AIC and BIC.
	
	Several directions remain open for future research within the proposed framework. One possible extension is to develop a Bayesian version of the proposed methodology to incorporate prior information into the estimation of model parameters and reliability characteristics. Investigating optimal SSALT designs under practical time and budget constraints to determine efficient sample allocations and stress paths would be a valuable extension of the present work. In practice, experiments carried out in different laboratories or at different times may also differ because of operators, equipment, or changes during testing; incorporating such experiment-specific variation through random effects would make the framework more realistic. Another promising direction is to extend the proposed meta-analysis set-up to jointly analyse observed degradation measurements and time-to-failure data, thereby exploiting complementary sources of information for reliability assessment.
	
	\section*{Data availability}
	The LED data and the simulated data sets used in this paper are taken from the cited sources \cite{shyur1999,samanta2024led,xiong1998,katerimeta2009}. The R code used for the analyses is available from the corresponding author upon request.

	\bibliographystyle{elsarticle-harv}
	\bibliography{references_meta}

	\clearpage
	\setcounter{section}{0}
	\setcounter{figure}{0}
	\setcounter{table}{0}
	\renewcommand{\thesection}{S\arabic{section}}
	\renewcommand{\thefigure}{S\arabic{figure}}
	\renewcommand{\thetable}{S\arabic{table}}
	
	\section*{Supplementary Material}
	This supplement contains additional material supporting the main paper: kernel density plots for the simulation study under the Burr XII distribution (Section~S1), log-likelihood contour plots for the simulated-data illustration (Section~S2), the complete bootstrap results comparing pooled and single-experiment analyses of the LED data (Section~S3), and the log-likelihood functions, likelihood equations, and estimation procedure for the exponential, Weibull, and Burr XII distributions (Section~S4).
	
	\subsection*{S1 ~Kernel density plots under the Burr XII distribution}
	
\begin{figure}[htbp]
		\centering
		\subfloat[$N=150$]{\includegraphics[width=0.5\textwidth, height=0.46\textwidth]{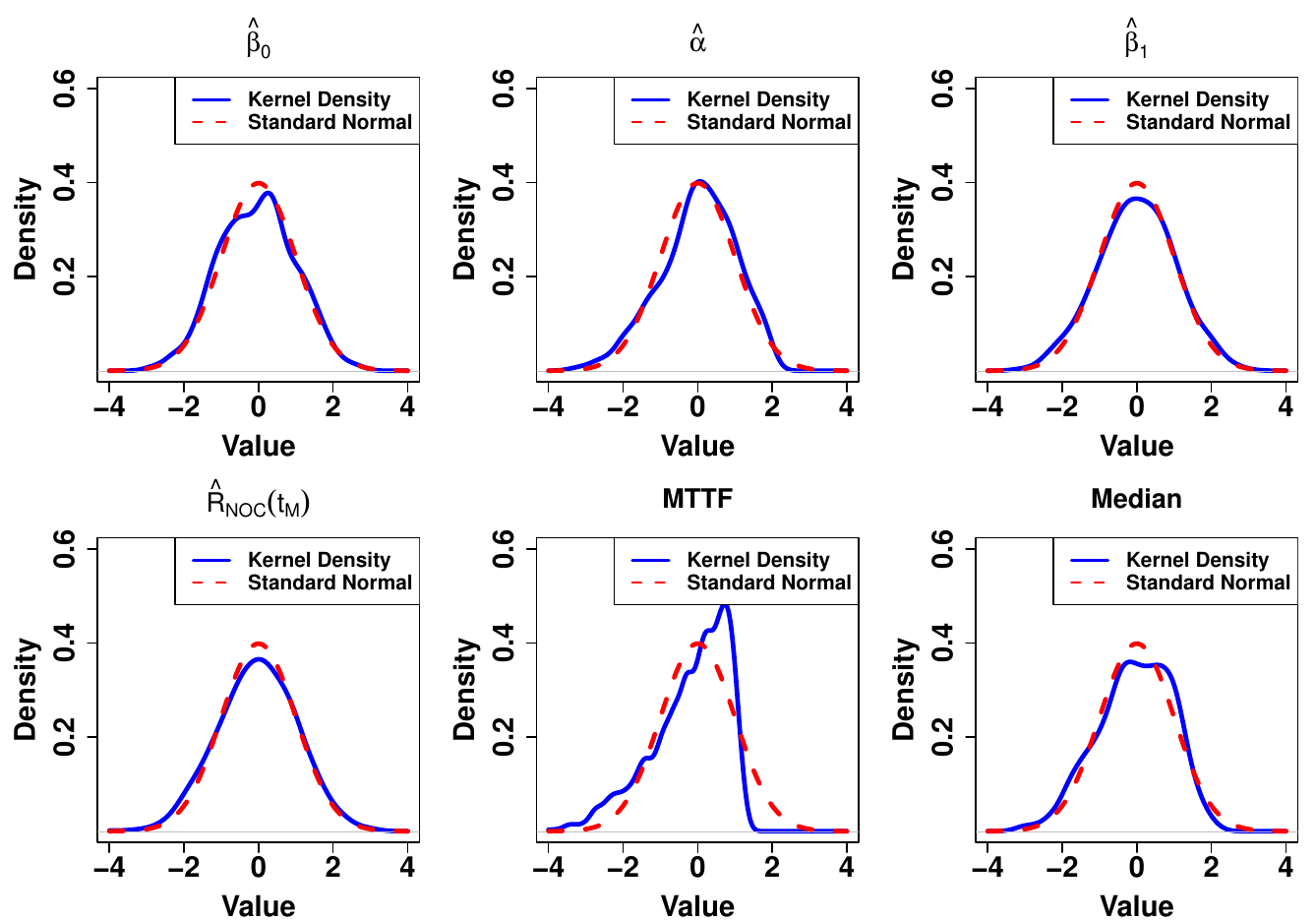}\label{fig:burr_kernel_a}}
		\hfil
		\subfloat[$N=300$]{\includegraphics[width=0.5\textwidth, height=0.46\textwidth]{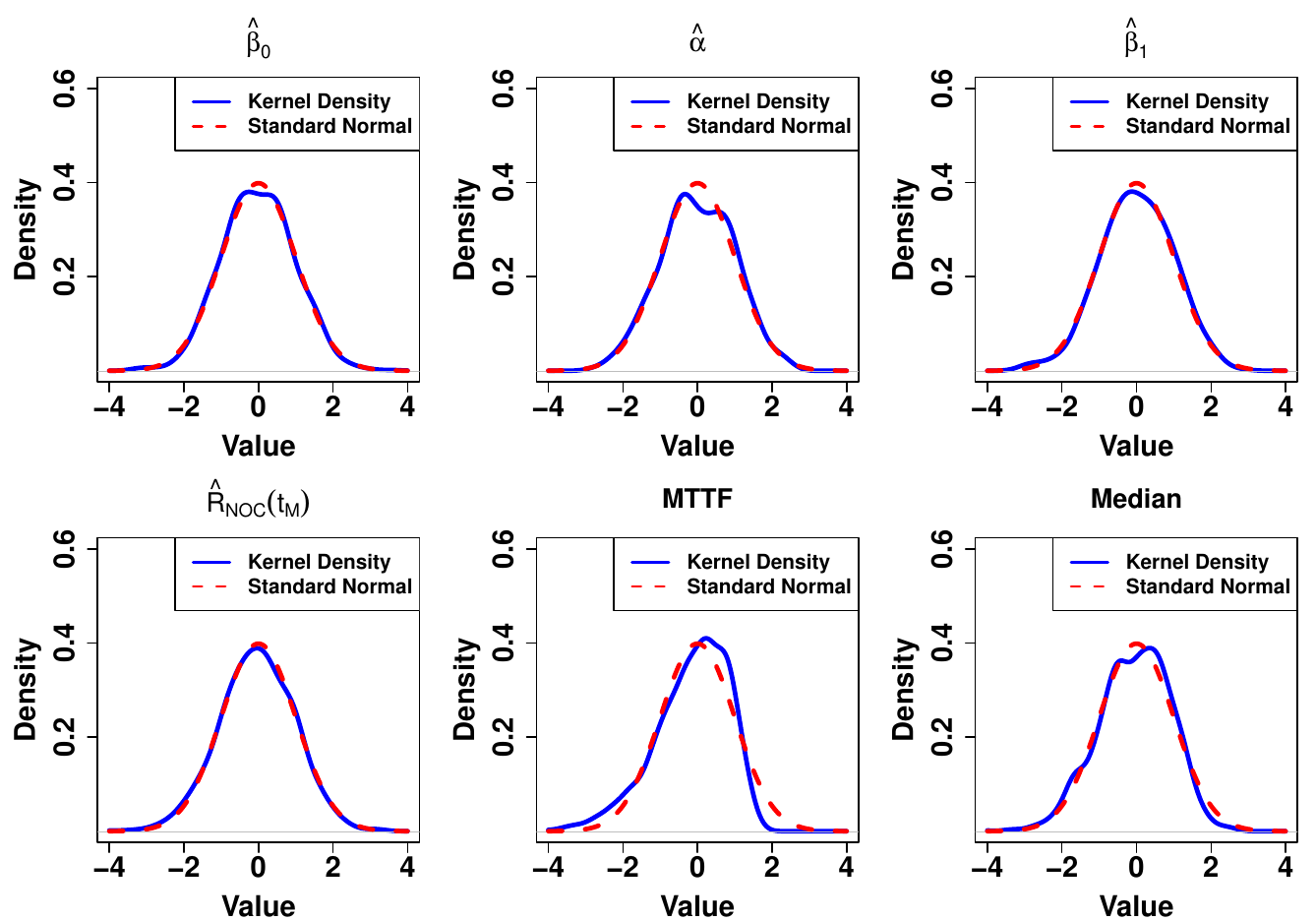}\label{fig:burr_kernel_b}}
		\caption{Kernel density plots of the standardized estimators with the standard normal curve under the Burr XII distribution for sample sizes (a) $N=150$ and (b) $N=300$.}
		\label{fig:burr_kernel}
	\end{figure}

	Figure~\ref{fig:burr_kernel} presents the kernel density estimates of the standardized estimators together with the standard normal curve for $N=150$ and $N=300$. At $N=150$, the standardized estimators of $\beta_0$, $\alpha$, and $\beta_1$ are already reasonably close to the standard normal curve. The standardized MTTF shows a somewhat larger departure from normality, with a sharper peak and a thinner right tail than the standard normal curve, while $\hat{R}_0(t)$ and the median exhibit only minor deviations from normality at this sample size. As the sample size increases to $N=300$, the kernel densities of all the reported quantities move closer to the standard normal curve. Overall, the numerical and graphical results provide evidence of satisfactory finite-sample performance and asymptotic normal behavior of the proposed estimators under the Burr XII baseline, consistent with the findings obtained under the Weibull case.
	
	\subsection*{S2 ~Log-likelihood contour plots for the simulated data illustration}
	
\begin{figure}[!ht]
		\centering
		\includegraphics[width=0.4\linewidth]{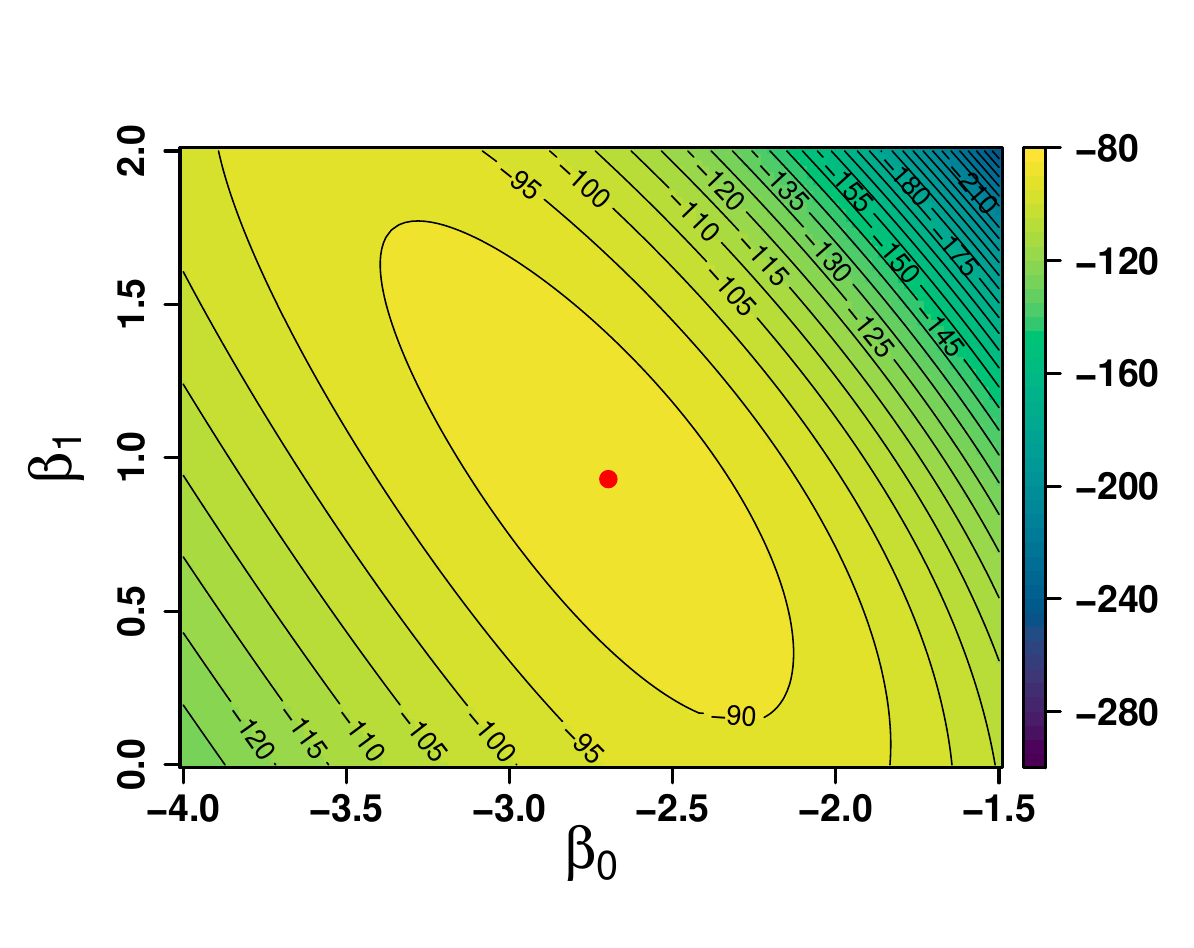}
		\caption{Contour plot exhibiting unimodality of the FR model based on log-likelihood of $\beta_0$ and $\beta_1$ under the exponential distribution.}
		\label{contour exponential}
	\end{figure}

\begin{figure}[!ht]
		\centering
		\subfloat[$\beta_0$ and $\alpha$]{\includegraphics[width=0.31\textwidth]{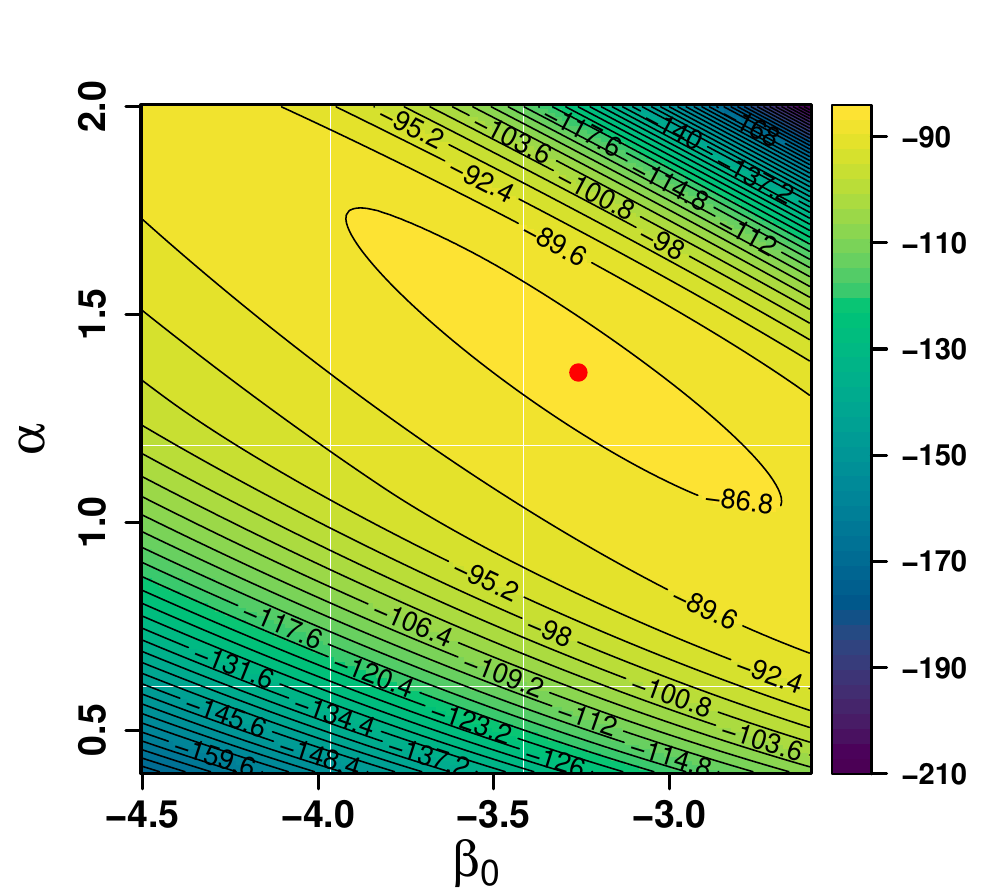}\label{fig:cw_sim_a}}
		\hfil
		\subfloat[$\beta_1$ and $\alpha$]{\includegraphics[width=0.31\textwidth]{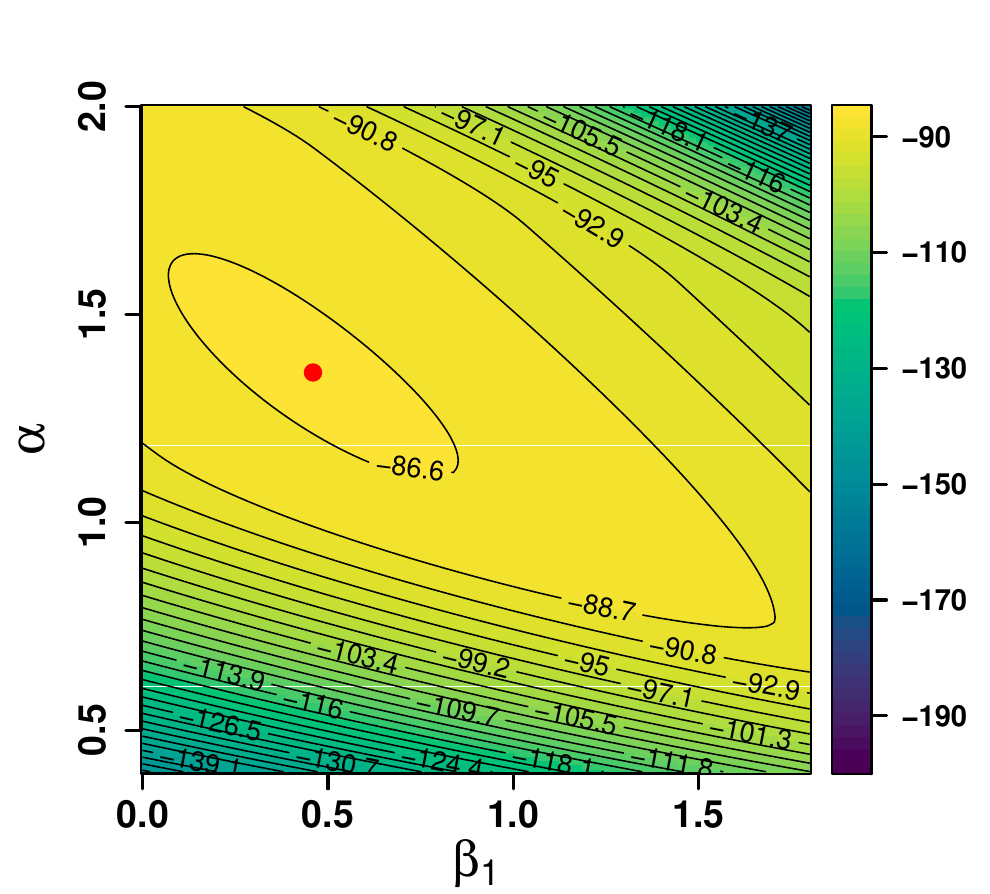}\label{fig:cw_sim_b}}
		\hfil
		\subfloat[$\beta_0$ and $\beta_1$]{\includegraphics[width=0.31\textwidth]{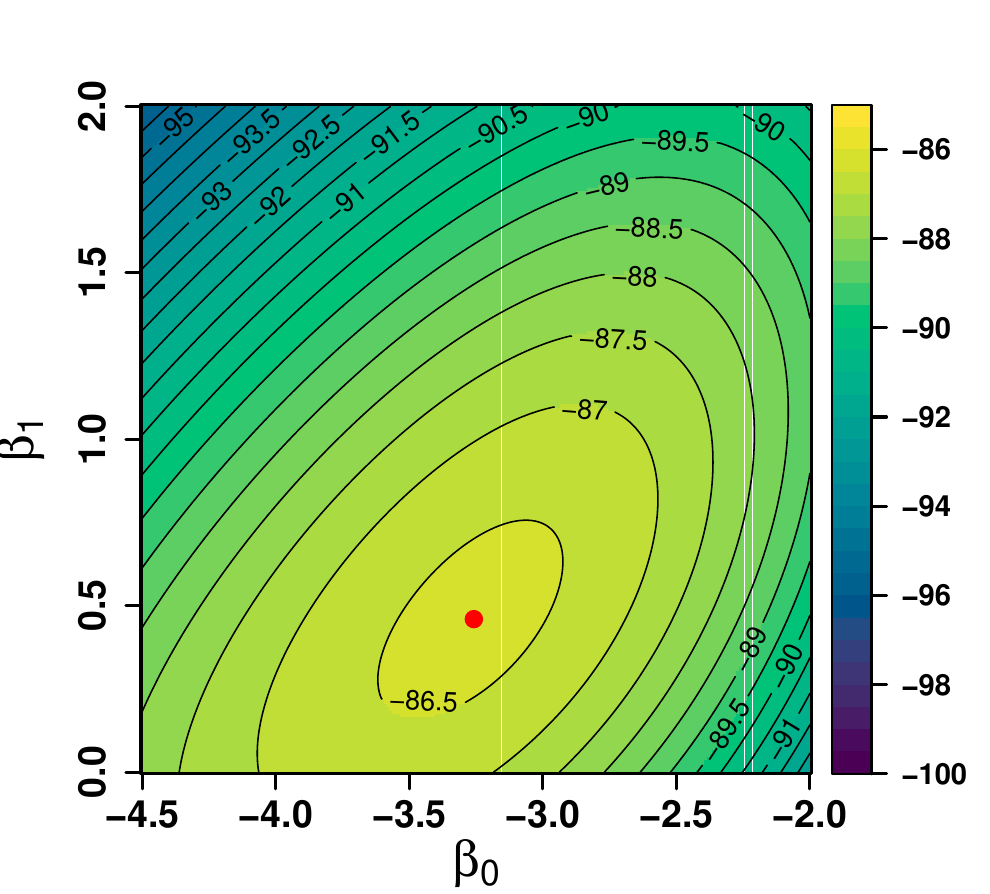}\label{fig:cw_sim_c}}
		\caption{Contour plots exhibiting the unimodality of the profile log-likelihood function for the Weibull model: (a) $\beta_0$ and $\alpha$, (b) $\beta_1$ and $\alpha$, (c)   $\beta_0$ and $\beta_1$. The red points indicate the maximum likelihood estimates.}
		\label{fig:contour_weibull_sim}
	\end{figure}

	Figures~\ref{contour exponential} and \ref{fig:contour_weibull_sim} show the contour plots of the log-likelihood function under the exponential model and of the profile log-likelihood functions under the Weibull model, respectively, for the simulated data sets considered in the main paper.
	
	\subsection*{S3 ~Bootstrap comparison of pooled and single-experiment analyses for the LED data}
	
	Table \ref{tab:S_LED_pooling} reports the complete results of the bootstrap study described in the main paper, for the model parameters and all reliability characteristics. For the single experiments, a few extremely large estimates of the MTTF and the lifetime quantiles dominate the averages; the corresponding entries are marked $>10^{6}$. These large estimates also raise the averages of the pooled MTTF and median lifetime above their reference values, although the pooled estimates are concentrated around the reference values, as seen in the boxplots in the main paper.
	
\begin{figure}[ht]
		\centering
		\includegraphics[width=\textwidth]{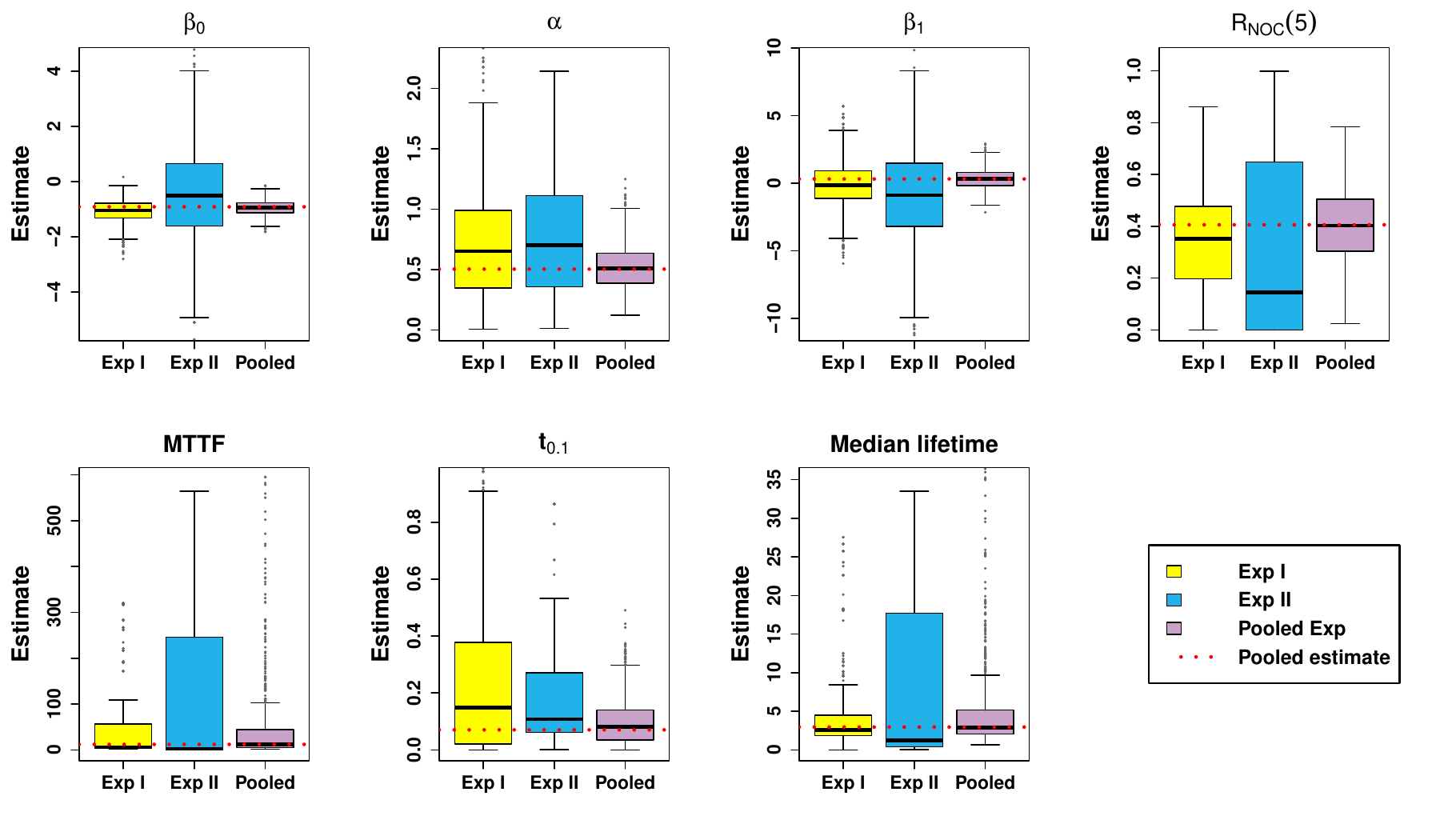}
		\caption{Bootstrap estimates for the LED data under the Weibull model, obtained from Experiment I alone, Experiment II alone, and both experiments pooled ($1000$ replications). The dotted line marks the pooled estimate from the observed data. Some outlying estimates lie outside the plotted range: $5.9\%$, $5.4\%$, and $0\%$ for $\alpha$; $17.2\%$, $23.2\%$, and $7.7\%$ for the MTTF; $2.1\%$, $17.9\%$, and $0\%$ for the $0.10$ quantile; and $8.9\%$, $22.9\%$, and $3.7\%$ for the median lifetime (Exp I, Exp II, and pooled, respectively).}
		\label{fig:S_LED_weibull}
	\end{figure}

	\begin{table}[!ht]
		\centering
		\caption{Bootstrap results for the LED data under the Chen model ($1000$ replications). Reference values are the pooled estimates from the observed data; the CI for $\alpha$ is computed on the log scale. Entries marked $>10^{6}$ are dominated by a few extremely large estimates.}
		\label{tab:S_LED_pooling}
		\scriptsize
		\setlength{\tabcolsep}{5pt}
		\begin{tabular}{@{}llccccc@{}}
			\toprule
			Design & Quantity & AE & Bias & RMSE & CP & AL \\
			\midrule
			\multirow{7}{*}{Exp I}
			& $\beta_0$ & $-1.461$ & $-0.053$ & $0.317$ & $0.972$ & $1.707$ \\
			& $\alpha$ & $0.293$ & $0.031$ & $0.145$ & $0.930$ & $>10^{6}$ \\
			& $\beta_1$ & $0.187$ & $-0.093$ & $1.430$ & $0.980$ & $13.20$ \\
			& $R_{\text{NOC}}(5)$ & $0.389$ & $-0.027$ & $0.163$ & $0.977$ & $0.585$ \\
			& MTTF & $>10^{6}$ & $>10^{6}$ & $>10^{6}$ & $0.913$ & $>10^{6}$ \\
			& $t_{0.10}$ & $0.071$ & $0.051$ & $0.109$ & $0.911$ & $>10^{6}$ \\
			& $t_{0.50}$ & $>10^{6}$ & $>10^{6}$ & $>10^{6}$ & $0.956$ & $>10^{6}$ \\
			\midrule
			\multirow{7}{*}{Exp II}
			& $\beta_0$ & $-1.343$ & $0.065$ & $1.541$ & $0.984$ & $17.65$ \\
			& $\alpha$ & $0.296$ & $0.034$ & $0.147$ & $0.901$ & $>10^{6}$ \\
			& $\beta_1$ & $0.031$ & $-0.250$ & $3.005$ & $0.981$ & $35.84$ \\
			& $R_{\text{NOC}}(5)$ & $0.376$ & $-0.039$ & $0.349$ & $0.999$ & $0.990$ \\
			& MTTF & $>10^{6}$ & $>10^{6}$ & $>10^{6}$ & $0.929$ & $>10^{6}$ \\
			& $t_{0.10}$ & $>10^{6}$ & $>10^{6}$ & $>10^{6}$ & $0.957$ & $>10^{6}$ \\
			& $t_{0.50}$ & $>10^{6}$ & $>10^{6}$ & $>10^{6}$ & $0.951$ & $>10^{6}$ \\
			\midrule
			\multirow{7}{*}{Pooled}
			& $\beta_0$ & $-1.435$ & $-0.026$ & $0.277$ & $0.948$ & $1.065$ \\
			& $\alpha$ & $0.263$ & $0.002$ & $0.064$ & $0.929$ & $0.250$ \\
			& $\beta_1$ & $0.313$ & $0.033$ & $0.681$ & $0.954$ & $2.619$ \\
			& $R_{\text{NOC}}(5)$ & $0.418$ & $0.003$ & $0.129$ & $0.958$ & $0.455$ \\
			& MTTF & $867.1$ & $858.2$ & $22870$ & $0.910$ & $>10^{6}$ \\
			& $t_{0.10}$ & $0.032$ & $0.012$ & $0.037$ & $0.942$ & $0.259$ \\
			& $t_{0.50}$ & $9.230$ & $6.133$ & $83.70$ & $0.947$ & $218600$ \\
			\bottomrule
		\end{tabular}
	\end{table} 
	Figure \ref{fig:S_LED_weibull} shows the corresponding bootstrap results under the Weibull model. The pattern is the same as under the Chen model: all pooled fits are usable, whereas $17.1\%$ and $21.1\%$ of the fits to Experiments I and II alone are not. Compared with Experiment I, pooling reduces the RMSE of $\widehat{R}_{\text{NOC}}(5)$ by $29.3\%$ and that of the $0.10$ quantile by $72.3\%$.
	
	\subsection*{S4 ~Likelihood equations for the exponential, Weibull, and Burr XII distributions}
	
	\subsubsection*{S4.1 ~Exponential distribution}
	\label{subsubsec:exponential}
	
	The exponential distribution is obtained as a special case of the PH class of distributions in case when 
	$\alpha=1.$ For the $k$-th experiment ($k=1,2,\ldots,K$), the reliability function associated with the $i$-th stress-specific time-to-failure distribution then reduces to
	\[
	R_{\text{exp}}(t)=
	\begin{cases}
		e^{{-\lambda_{k,i} t}},& t>0 \\
		1, & \text{otherwise.}
	\end{cases}
	\]
	Now, assuming the exponential time-to-failure distribution, the overall log-likelihood function can be obtained as:
	\begin{align}
		\ell_{\mathrm{exp}}(\beta_0,\beta_1 \mid\mathcal{D})
		&=
		\sum_{k=1}^{K}\sum_{i=1}^{I_k}
		n_{k,i}
		\ln
		\Bigg\{
		e^{-\sum_{j=1}^{i-1}
			e^{\beta_0+\beta_1x_{k,j}}
			(\tau_{k,j}-\tau_{k,j-1})}
		-
		e^{-\sum_{j=1}^{i}
			e^{\beta_0+\beta_1x_{k,j}}
			(\tau_{k,j}-\tau_{k,j-1})}
		\Bigg\}
		\nonumber\\
		&\quad
		-
		\sum_{k=1}^{K}
		n_{k,I_k+1}
		\sum_{j=1}^{I_k}
		e^{\beta_0+\beta_1x_{k,j}}
		(\tau_{k,j}-\tau_{k,j-1}).
		\label{eq:loglikelihood_expo}
	\end{align}
	Differentiating the log-likelihood function $\ell_{\mathrm{exp}}(\beta_0,\beta_1 \mid\mathcal{D})$ with respect to $\beta_0$ and $\beta_1,$ the score  equations are obtained as follows:
	\begin{align}
		\frac{\partial\ell_{\mathrm{exp}}}{\partial\beta_0}
		&=
		\sum_{k=1}^{K}\sum_{i=1}^{I_k}
		\frac{n_{k,i}}{D_{k,i}^{\mathrm{exp}}}
		\left[
		R_{k,i}^{\mathrm{exp}}H_{k,i}^{\mathrm{exp}}
		-
		R_{k,i-1}^{\mathrm{exp}}H_{k,i-1}^{\mathrm{exp}}
		\right]
		-
		\sum_{k=1}^{K}
		n_{k,I_k+1}H_{k,I_k}^{\mathrm{exp}}
		=0,
		\label{eq:exp_score_beta0}
	\end{align}
	\begin{align}
		\frac{\partial\ell_{\mathrm{exp}}}{\partial\beta_1}
		&=
		\sum_{k=1}^{K}\sum_{i=1}^{I_k}
		\frac{n_{k,i}}{D_{k,i}^{\mathrm{exp}}}
		\left[
		R_{k,i}^{\mathrm{exp}}H_{k,i}^{\mathrm{exp}}B_{k,i}^{\mathrm{exp}}
		-
		R_{k,i-1}^{\mathrm{exp}}H_{k,i-1}^{\mathrm{exp}}B_{k,i-1}^{\mathrm{exp}}
		\right]
		-
		\sum_{k=1}^{K}
		n_{k,I_k+1}H_{k,I_k}^{\mathrm{exp}}B_{k,I_k}^{\mathrm{exp}}
		=0,
		\label{eq:exp_score_beta1}
	\end{align}
	where, for $i=1,\dots,I_k$,  we have 
	\begin{align*}
		H_{k,i}^{\mathrm{exp}}
		&=
		\sum_{j=1}^{i}
		e^{\beta_0+\beta_1x_{k,j}}
		(\tau_{k,j}-\tau_{k,j-1}),
		\\
		R_{k,i}^{\mathrm{exp}}
		&=
		e^{-H_{k,i}^{\mathrm{exp}}}
		\\
		D_{k,i}^{\mathrm{exp}}
		&=
		e^{-H_{k,i-1}^{\mathrm{exp}}}
		-
		e^{-H_{k,i}^{\mathrm{exp}}},
		\\
		B_{k,i}^{\mathrm{exp}}
		&=
		\frac{
			\displaystyle
			\sum_{j=1}^{i}
			x_{k,j}e^{\beta_0+\beta_1x_{k,j}}
			(\tau_{k,j}-\tau_{k,j-1})
		}{
			H_{k,i}^{\mathrm{exp}}
		}.
	\end{align*}
	\noindent Since the resulting  score equations given by \eqref{eq:exp_score_beta0} and \eqref{eq:exp_score_beta1} are nonlinear in $\boldsymbol{\theta_{\text{exp}}}=(\beta_0,\beta_1),$  their solutions cannot be obtained  in explicit  form. The MLE of $\boldsymbol{\theta_{\text{exp}}}$ (denoted by $\boldsymbol{\widehat{\theta}_{\text{exp}}}$)  is then obtained by numerically maximizing the log-likelihood function $\ell_{\mathrm{exp}}(\beta_0,\beta_1 \mid\mathcal{D})$ provided in \eqref{eq:loglikelihood_expo}. However, the choice  of initial values turns out often to be a non-trivial problem in initiating any numerical optimization routine. We address this problem at first  by evaluating the log-likelihood function over a suitably defined two-dimensional grid of $(\beta_0,\beta_1)$ and then  the corresponding log-likelihood surface is visualized using a contour plot to assess its unimodality within the parameter region under consideration. The grid point yielding the maximum log-likelihood is subsequently selected as the initial guess,  denoted by \((\beta_0^{(0)},\beta_1^{(0)})\).  An iterative numerical optimization algorithm is  employed  to maximize the log-likelihood function until the convergence, yielding thus \(\boldsymbol{\widehat{\theta}_{\text{exp}}}=(\widehat{\beta}_0,\widehat{\beta}_1)\).
	
	Based on  $\boldsymbol{\widehat{\theta}_{\text{exp}}}$,  the estimated  metrics  for assessing the lifetime characteristics under  the NOC are given by 
	\begin{align*}
		\widehat{R}_{\text{NOC}}(t_M)
		&=
		e^{-e^{\widehat{\beta}_0}t_M},
		&
		\widehat{\mu}_{\text{NOC}}
		&=
		e^{-\widehat{\beta}_0},
		&
		\widehat{t}_{0.5,\text{NOC}} 
		&=
		(\ln 2)e^{-\widehat{\beta}_0},
	\end{align*}
	where $t_M$, $\widehat{\mu}_{\text{NOC}}$ and   $\widehat{t}_{0.5,\text{NOC}} $ denote the specified mission time, the
	estimated MTTF, and  the estimated median time-to-failure respectively.
	
	\subsubsection*{S4.2 ~Weibull distribution}
	\label{subsubsec:weibull}
	
	The Weibull distribution arises as a widely employed time-to-failure distribution belonging to the PH  family of distributions. For the $k$-th experiment ($k=1,2,\ldots,K$), the reliability function associated with the $i$-th stress-specific Weibull time-to-failure distribution is given by
	\[
	R_{\text{Weib}}(t)=
	\begin{cases}
		e^{{-\lambda_{k,i} t^{\alpha}}},& t>0 \\
		1, & \text{otherwise.}
	\end{cases}
	\]
	Considering  the Weibull  time-to-failure distribution, the overall log-likelihood function  can be obtained as:
	\begin{align}
		\ell_{\mathrm{Weib}}(\beta_0,\alpha,\beta_1\mid \mathcal{D})
		&=
		\sum_{k=1}^{K}\sum_{i=1}^{I_k}
		n_{k,i}
		\ln
		\Bigg\{
		e^{-\sum_{j=1}^{i-1}
			e^{\beta_0+\beta_1x_{k,j}}
			\left(\tau_{k,j}^{\alpha}-\tau_{k,j-1}^{\alpha}\right)}
		-
		e^{-\sum_{j=1}^{i}
			e^{\beta_0+\beta_1x_{k,j}}
			\left(\tau_{k,j}^{\alpha}-\tau_{k,j-1}^{\alpha}\right)}
		\Bigg\}
		\nonumber\\
		&\quad
		-
		\sum_{k=1}^{K}
		n_{k,I_k+1}
		\sum_{j=1}^{I_k}
		e^{\beta_0+\beta_1x_{k,j}}
		\left(\tau_{k,j}^{\alpha}-\tau_{k,j-1}^{\alpha}\right).
		\label{eq:weibull_loglikelihood}
	\end{align}
	Differentiating the log-likelihood function $\ell_{\mathrm{Weib}}(\beta_0,\alpha,\beta_1\mid \mathcal{D})$ with respect to
	$\beta_0$, $\alpha$ and $\beta_1,$  the score equations are obtained as follows:
	\begin{align}
		\frac{\partial\ell_{\mathrm{Weib}}}
		{\partial\beta_0}
		&=
		\sum_{k=1}^{K}\sum_{i=1}^{I_k}
		\frac{n_{k,i}}{D_{k,i}^{\mathrm{Weib}}}
		\left[
		R_{k,i}^{\mathrm{Weib}}H_{k,i}^{\mathrm{Weib}}
		-
		R_{k,i-1}^{\mathrm{Weib}}H_{k,i-1}^{\mathrm{Weib}}
		\right]
		-
		\sum_{k=1}^{K}
		n_{k,I_k+1}H_{k,I_k}^{\mathrm{Weib}}
		=0,
		\\
		\frac{\partial\ell_{\mathrm{Weib}}}
		{\partial\alpha}
		&=
		\sum_{k=1}^{K}\sum_{i=1}^{I_k}
		\frac{n_{k,i}}{D_{k,i}^{\mathrm{Weib}}}
		\left[
		R_{k,i}^{\mathrm{Weib}}H_{k,i}^{\mathrm{Weib}}C_{k,i}^{\mathrm{Weib}} 
		-
		R_{k,i-1}^{\mathrm{Weib}}
		H_{k,i-1}^{\mathrm{Weib}}C_{k,i-1}^{\mathrm{Weib}}
		\right]
		-
		\sum_{k=1}^{K}
		n_{k,I_k+1}
		H_{k,I_k}^{\mathrm{Weib}}C_{k,I_k}^{\mathrm{Weib}}
		=0,
		\\
		\frac{\partial\ell_{\mathrm{Weib}}}
		{\partial\beta_1}
		&=
		\sum_{k=1}^{K}\sum_{i=1}^{I_k}
		\frac{n_{k,i}}{D_{k,i}^{\mathrm{Weib}}}
		\Big[
		R_{k,i}^{\mathrm{Weib}}H_{k,i}^{\mathrm{Weib}}B_{k,i}^{\mathrm{Weib}}
		-
		R_{k,i-1}^{\mathrm{Weib}}
		H_{k,i-1}^{\mathrm{Weib}}B_{k,i-1}^{\mathrm{Weib}}
		\Big]
		-
		\sum_{k=1}^{K}
		n_{k,I_k+1}
		H_{k,I_k}^{\mathrm{Weib}}B_{k,I_k}^{\mathrm{Weib}}
		=0,   
	\end{align}
	where
	\begin{align*}
		H_{k,i}^{\mathrm{Weib}}
		&=
		\sum_{j=1}^{i}
		e^{\beta_0+\beta_1x_{k,j}}
		\left(
		\tau_{k,j}^{\alpha}
		-
		\tau_{k,j-1}^{\alpha}
		\right),\\
		R_{k,i}^{\mathrm{Weib}}
		&=
		e^{-H_{k,i}^{\mathrm{Weib}}},		
		\\
		D_{k,i}^{\mathrm{Weib}}
		&=
		e^{-H_{k,i-1}^{\mathrm{Weib}}}
		-
		e^{-H_{k,i}^{\mathrm{Weib}}},\\
		B_{k,i}^{\mathrm{Weib}}
		&=
		\frac{
			\displaystyle
			\sum_{j=1}^{i}
			x_{k,j}e^{\beta_0+\beta_1x_{k,j}}
			\left(
			\tau_{k,j}^{\alpha}
			-
			\tau_{k,j-1}^{\alpha}
			\right)
		}{
			H_{k,i}^{\mathrm{Weib}}
		}, \\
		C_{k,i}^{\mathrm{Weib}}
		&=
		\frac{
			\displaystyle
			\sum_{j=1}^{i}
			e^{\beta_0+\beta_1x_{k,j}}
			\left[
			\tau_{k,j}^{\alpha}\ln(\tau_{k,j})
			-
			\tau_{k,j-1}^{\alpha}\ln(\tau_{k,j-1})
			\right]
		}{
			H_{k,i}^{\mathrm{Weib}}
		}.
	\end{align*}
	Here, $\tau_{k,0}^{\alpha} \ln(\tau_{k,0}) = 0$ for $j = 1$ subject to the initial condition $H_0(0; \alpha) = 0$.
	
	However, the likelihood equations do not admit explicit solutions for the model parameters. In particular, the nonlinear dependence of the likelihood function on $\alpha$, $\beta_0$, and $\beta_1$ precludes closed-form expressions for their maximum likelihood estimators. Accordingly, the profile log-likelihood functions are employed to investigate the likelihood structure over each of the three possible parameter pairs, with the remaining parameter treated as a nuisance parameter and maximized out.
	
	The three profile log-likelihood functions are expressed as
	\begin{align*}
		\ell_p(\alpha,\beta_0)
		&=
		\max_{\beta_1>0}
		\ell_{\mathrm{Weib}}(\beta_0,\alpha,\beta_1\mid\mathcal{D}),\\
		\ell_p(\alpha,\beta_1)
		&=
		\max_{\beta_0}
		\ell_{\mathrm{Weib}}(\beta_0,\alpha,\beta_1\mid\mathcal{D}),\\
		\ell_p(\beta_0,\beta_1)
		&=
		\max_{\alpha>0}
		\ell_{\mathrm{Weib}}(\beta_0,\alpha,\beta_1\mid\mathcal{D}).
	\end{align*}
	
	The resulting two-dimensional contour plots provide a graphical assessment of the unimodal behaviour and identify the regions containing the profile maxima. The parameter values corresponding to these maxima
	are then subsequently used to specify the initial guess for the joint numerical maximization of the full three-dimensional log-likelihood,
	$
	\boldsymbol{\theta}^{(0)}
	=
	\left(\beta_0^{(0)},\alpha^{(0)},\beta_1^{(0)}\right).
	$
	Starting from these initial values, an iterative numerical optimization procedure is employed until convergence to obtain the maximum likelihood estimates (MLEs),
	$
	\widehat{\boldsymbol{\theta}}
	=
	\left(\widehat{\beta}_0,\widehat{\alpha},\widehat{\beta}_1\right).
	$
	Based on these MLEs, the corresponding reliability characteristics at the normal operating condition (NOC) are  obtained as
	\begin{align*}
		\widehat{R}_{\text{NOC}}(t_M)
		&=
		e^{-e^{\widehat{\beta}_0}t_M^{\widehat{\alpha}}},
		&
		\widehat{\mu}_{\text{NOC}}
		&=
		e^{-\widehat{\beta}_0/\widehat{\alpha}}
		\Gamma\left(1+\frac{1}{\widehat{\alpha}}\right),
		&
		\widehat{t}_{0.5,\text{NOC}} 
		&=
		\left(\frac{\ln 2}{e^{\widehat{\beta}_0}}\right)^{1/\widehat{\alpha}},
	\end{align*}
	where $t_M$ denotes the specified mission time, $\widehat{\mu}_{\text{NOC}}$ is the estimated MTTF, and $\widehat{t}_{0.5,\text{NOC}} $ is the estimated median lifetime of the Weibull distribution.
	
	\subsubsection*{S4.3 ~Burr XII distribution}
	\label{subsubsec:burr}
	
	For the Burr XII distribution, the reliability function is specified as $ R(t;\alpha,\lambda_{k,i}) = \left(1+t^\alpha\right)^{-\lambda_{k,i}},$ $t>0,$ with shape parameters $\alpha>0$ and $\lambda_{k,i}>0$. The corresponding log-likelihood function is given by
	\begin{align}
		\ell_{\mathrm{Burr}}(\beta_0,\alpha,\beta_1\mid\mathcal{D})
		&=
		\sum_{k=1}^{K}\sum_{i=1}^{I_k}
		n_{k,i}
		\ln
		\Bigg\{
		e^{-\sum_{j=1}^{i-1}
			e^{\beta_0+\beta_1x_{k,j}}
			\left[
			\ln(1+\tau_{k,j}^{\alpha})
			-
			\ln(1+\tau_{k,j-1}^{\alpha})
			\right]}
		\nonumber \\
		&\quad-
		e^{-\sum_{j=1}^{i}
			e^{\beta_0+\beta_1x_{k,j}}
			\left[
			\ln(1+\tau_{k,j}^{\alpha})
			-
			\ln(1+\tau_{k,j-1}^{\alpha})
			\right]}
		\Bigg\}
		\nonumber\end{align}
	\begin{align}\quad-
		\sum_{k=1}^{K}
		n_{k,I_k+1}
		\sum_{j=1}^{I_k}
		e^{\beta_0+\beta_1x_{k,j}}
		\left[
		\ln(1+\tau_{k,j}^{\alpha})
		-
		\ln(1+\tau_{k,j-1}^{\alpha})
		\right].
		\label{eq:burr_loglikelihood}
	\end{align}
	Differentiation of the log-likelihood with respect to
	$\beta_0$, $\beta_1$, and $\alpha$ gives the likelihood equations
	\begin{align}
		\frac{\partial\ell_{\mathrm{Burr}}}
		{\partial\beta_0}
		&=
		\sum_{k=1}^{K}\sum_{i=1}^{I_k}
		\frac{n_{k,i}}{D_{k,i}^{\mathrm{Burr}}}
		\left[
		R_{k,i}^{\mathrm{Burr}}H_{k,i}^{\mathrm{Burr}}-
		R_{k,i-1}^{\mathrm{Burr}}H_{k,i-1}^{\mathrm{Burr}}
		\right]
		-
		\sum_{k=1}^{K}
		n_{k,I_k+1}H_{k,I_k}^{\mathrm{Burr}}
		=0,
		\\
		\frac{\partial\ell_{\mathrm{Burr}}}
		{\partial\beta_1}
		&=
		\sum_{k=1}^{K}\sum_{i=1}^{I_k}
		\frac{n_{k,i}}{D_{k,i}^{\mathrm{Burr}}}
		\Big[
		R_{k,i}^{\mathrm{Burr}}H_{k,i}^{\mathrm{Burr}}B_{k,i}^{\mathrm{Burr}}
		-
		R_{k,i-1}^{\mathrm{Burr}}
		H_{k,i-1}^{\mathrm{Burr}}B_{k,i-1}^{\mathrm{Burr}}
		\Big]
		-
		\sum_{k=1}^{K}
		n_{k,I_k+1}
		H_{k,I_k}^{\mathrm{Burr}}B_{k,I_k}^{\mathrm{Burr}}
		=0,
		\\
		\frac{\partial\ell_{\mathrm{Burr}}}
		{\partial\alpha}
		&=
		\sum_{k=1}^{K}\sum_{i=1}^{I_k}
		\frac{n_{k,i}}{D_{k,i}^{\mathrm{Burr}}}
		\left[
		R_{k,i}^{\mathrm{Burr}}H_{k,i}^{\mathrm{Burr}}C_{k,i}^{\mathrm{Burr}}
		-
		R_{k,i-1}^{\mathrm{Burr}}
		H_{k,i-1}^{\mathrm{Burr}}C_{k,i-1}^{\mathrm{Burr}}
		\right]
		-
		\sum_{k=1}^{K}
		n_{k,I_k+1}
		H_{k,I_k}^{\mathrm{Burr}}C_{k,I_k}^{\mathrm{Burr}}
		=0,
	\end{align}
	where
	\begin{align*}
		H_{k,i}^{\mathrm{Burr}}
		&=
		\sum_{j=1}^{i}
		e^{\beta_0+\beta_1x_{k,j}}
		\left[
		\ln(1+\tau_{k,j}^{\alpha})
		-
		\ln(1+\tau_{k,j-1}^{\alpha})
		\right],
		\\
		R_{k,i}^{\mathrm{Burr}}
		&=
		e^{-H_{k,i}^{\mathrm{Burr}}},
		\\
		D_{k,i}^{\mathrm{Burr}}
		&=
		e^{-H_{k,i-1}^{\mathrm{Burr}}}-e^{-H_{k,i}^{\mathrm{Burr}}},
		\\
		B_{k,i}^{\mathrm{Burr}}
		&=
		\frac{
			\displaystyle
			\sum_{j=1}^{i}
			x_{k,j}e^{\beta_0+\beta_1x_{k,j}}
			\left[
			\ln(1+\tau_{k,j}^{\alpha})
			-
			\ln(1+\tau_{k,j-1}^{\alpha})
			\right]
		}{
			\displaystyle
			H_{k,i}^{\mathrm{Burr}}
		},\\
		C_{k,i}^{\mathrm{Burr}}
		&=
		\frac{
			\displaystyle
			\sum_{j=1}^{i}
			e^{\beta_0+\beta_1x_{k,j}}
			\left[
			\frac{\tau_{k,j}^{\alpha}\ln(\tau_{k,j})}
			{1+\tau_{k,j}^{\alpha}}
			-
			\frac{\tau_{k,j-1}^{\alpha}\ln(\tau_{k,j-1})}
			{1+\tau_{k,j-1}^{\alpha}}
			\right]
		}{
			\displaystyle
			H_{k,i}^{\mathrm{Burr}}
		}.
	\end{align*}
	Note that $\tau_{k,0}^{\alpha} \ln(\tau_{k,0}) = 0$ for $j = 1$ subject to  $H_0(0; \alpha) = 0$.
	
	For the Burr XII model, the likelihood equations depend nonlinearly on the model parameters and cannot be solved explicitly.
	The likelihood
	surface is examined through the profile log-likelihoods for the three
	parameter pairs, $(\alpha,\beta_0)$, $(\alpha,\beta_1)$, and
	$(\beta_0,\beta_1)$. The associated contour plots are examined to determine the neighborhood of the profile maxima, from which the starting point
	$
	\boldsymbol{\theta}^{(0)}
	=
	(\beta_0^{(0)},\alpha^{(0)},\beta_1^{(0)})
	$
	is determined. The full log-likelihood is subsequently maximized
	numerically from this starting point to obtain the maximum likelihood estimates (MLEs).
	
	Based on these estimates, the reliability characteristics at the normal operating condition (NOC) are obtained as
	\begin{align*}
		\widehat{R}_{\text{NOC}}(t_M)
		&=
		\left(1+t_M^{\widehat{\alpha}}\right)^{-e^{\widehat{\beta}_0}},
		&
		\widehat{\mu}_{\text{NOC}}
		&=
		\frac{
			\Gamma\left(1+\frac{1}{\widehat{\alpha}}\right)
			\Gamma\left(e^{\widehat{\beta}_0}
			-\frac{1}{\widehat{\alpha}}\right)
		}{
			\Gamma\left(e^{\widehat{\beta}_0}\right)
		},
		&
		\widehat{t}_{0.5,\text{NOC}} 
		&=
		\left[
		2^{e^{-\widehat{\beta}_0}}-1
		\right]^{1/\widehat{\alpha}}.
	\end{align*}
	where $t_M$ denotes the specified mission time, $\widehat{\mu}_{\text{NOC}}$ is the estimated MTTF, and $\widehat{t}_{0.5,\text{NOC}} $ is the estimated median lifetime of the Burr XII distribution.
	
	\noindent \textbf{Remark.} The existence of a finite MTTF under NOC ($\mu_{\text{NOC}}$) depends on the underlying lifetime distribution. Specifically, $\mu_{\text{NOC}}$ is finite for the exponential, Weibull, Gompertz, and Chen distributions under admissible parameter values, provided that $e^{\beta_0} > 0$. For the Pareto and Lomax distributions, finiteness requires $e^{\beta_0} > 1$, whereas for the Burr XII distribution, it requires $\alpha e^{\beta_0} > 1$.
	
\end{document}